\documentclass{emulateapj}

\newcommand{\tb}{\textbf}

\newcommand{\tx}{\textit}
\newcommand{\beq}{\begin{equation}}
\newcommand{\eeq}{\end{equation}}
\newcommand{\bdi}{\begin{displaymath}}
\newcommand{\edi}{\end{displaymath}}
\newcommand{\beqn}{\begin{eqnarray}}
\newcommand{\eeqn}{\end{eqnarray}}
\newcommand{\f}{\frac}
\newcommand{\bay}{\begin{array}{c}}
\newcommand{\eay}{\end{array}}
\newcommand{\ben}{\begin{enumerate}}
\newcommand{\een}{\end{enumerate}}

\newcommand{\M}{M_{\odot}}

\shorttitle{Dark Stars: Improved Models and Pulsations}
\shortauthors{Rindler-Daller et al.}

\begin{document}

\title{Dark Stars: Improved Models and First Pulsation Results}


\author{T. Rindler-Daller$^{(1)}$, M.H. Montgomery$^{(2)}$, K. Freese$^{(1)}$, D.E. Winget$^{(2)}$, B. Paxton$^{(3)}$}
\affil{$(1)$ Department of Physics and Michigan Center for Theoretical Physics, \\University of Michigan, 
    Ann Arbor, MI 48109, USA\\
    $(2)$ Department of Astronomy, McDonald Observatory and Texas Cosmology Center, \\University of Texas, Austin, TX 78712, USA\\
    $(3)$ Kavli Insitute for Theoretical Physics, \\University of California,
    Santa Barbara, CA 93106, USA}

\begin{abstract}
We use the stellar evolution code MESA to study dark stars.  Dark stars (DSs), which are powered by dark matter (DM) self-annihilation
rather than by nuclear fusion, may be the first stars to form in the Universe. We compute stellar models for accreting DSs 
with masses up to $10^6$ $M_{\odot}$. The heating due to DM annihilation is self-consistently included, assuming extended adiabatic 
contraction of DM within the minihalos in which DSs form. We find remarkably good overall agreement with 
previous models, which assumed polytropic interiors.  
There are some differences in the details, with positive implications for observability. We found that, 
 in the mass range of $10^4 -10^5 M_\odot$, our DSs are hotter by a factor of 1.5 than those in Freese et al.(2010), 
 are smaller in radius by a factor of 0.6, denser by a factor of 3~--~4, and more luminous by a factor of 2. Our models also 
confirm previous results, according to which supermassive DSs are very well approximated by
$(n=3)$-polytropes. We also perform a first study of dark star pulsations.
Our DS models have pulsation modes with timescales ranging from less than a day to more than two years in 
their rest frames,
at $z \sim 15$, depending on DM particle mass and overtone number. 
Such pulsations may someday be used to identify bright, cool objects
uniquely as DSs; if properly calibrated, they might, in principle, also supply novel standard candles for
cosmological studies. 
\end{abstract}


\section{Introduction}\label{Intro}


The first stars are thought to form at redshifts of $z \sim 15-30$ inside dark matter (DM) minihalos of mass $\sim 10^{6} - 10^{8} M_{\sun}$ 
which consist of $85 \%$ DM and $15 \%$ baryons, mostly in the form of hydrogen and helium from Big Bang
nucleosynthesis.  The formation of the first stars is currently a hot topic in astrophysical
cosmology
\citep[for more on the formation of the first stars, see,
e.g.][]{BLoeb, ABN, Yoshida03, BL, ABS, AS, TAO09, Hosokawa11,
  Greif12, Smithetal12}. Up-to-date reviews can be found, e.g., 
in \cite{Bromm13} and \cite{Glover13}.

Early and most subsequent investigations have largely neglected the impact of self-annihilating DM on the chemistry and physics of primordial protostellar
clouds, and hence on first star formation. 
The canonical DM candidates, weakly-interacting massive particles
(WIMPs), in many theories, are their own antiparticles and able to
annihilate with one another.  This annihilation process in the early
Universe can leave the correct relic abundance today. In
  addition, annihilation can be important wherever the DM density
$\rho_{\chi}$ is high, since the annihilation rate scales as
$\rho_{\chi}^2$.

The formation of the first stars is expected to be particularly
affected by this process, since they form at high redshifts
($\rho_{\chi} \sim (1+z)^3$) and in the high-density centers of DM
minihalos. \cite{Spolyar08} were the first to consider the effect of
DM particles on the first stars during their formation. They found
that, above a certain baryonic density threshold (the value of which
depends on the DM particle mass) heating by DM annihilation will come
to dominate over all cooling mechanisms. Subsequently, the
protostellar cloud will continue to contract, albeit at a slower rate,
and eventually, above a certain baryonic density threshold (again
depending on the DM particle mass), DM annihilation products remain
trapped in the star, thermalize and provide a heat source for
hydrostatic equilibrium: a dark star (DS) is born.  These first DSs are made 
primarily of hydrogen and helium, with less than $ 0.1
\%$ of the mass in form of DM.  Nevertheless, they shine due to DM
heating, not fusion, and so the term ``dark'' refers to the power
source, and not the appearance or the primary matter
constituent. Subsequently, \cite{Freese08} and \cite{Spolyar09}
studied the evolution of DSs from their birth at $\sim 1
M_{\sun}$, as they accreted material from the surrounding halo, up to
$\sim 1000 M_{\sun}$ which is approximately the Jeans mass of the
  collapsing molecular cloud. They showed that DSs are giant, puffy
($\sim 10$ AU), cool ($T_{\rm{eff}} < 10,000$ K), and bright ($> 10^6
L_{\sun}$) objects.  Since their surface temperatures never exceed
values that are high enough to trigger feedback mechanisms that
would shut off further accretion \citep[see][]{TM}, dark stars can in
principle grow as long as there is a supply of DM fuel. Indeed,
\cite{Freese10} followed the growth of DSs to become supermassive with
masses in excess of $10^5 M_{\sun}$. These supermassive DSs are
extraordinarily luminous, $L \sim 10^9-10^{11} L_{\sun}$, and may be
observable with upcoming facilities \citep[see][]{Freese10, Ilie12}.

The WIMP annihilation rate is $n_\chi^2 \langle \sigma v \rangle$ where
$n_\chi$ is the WIMP number density and we take the standard annihilation cross
section
\begin{equation}
\label{eq:sigmav}
\langle \sigma v \rangle = 
3 \times 10^{-26} {\rm cm}^3/{\rm s},
\end{equation}
and WIMP masses in the range $m_{\chi} = 10 ~\rm{GeV} -1 ~\rm{TeV}$.  
WIMP annihilation produces energy at a rate per unit volume 
\begin{equation} \label{DMheat}
\hat Q_{DM} = n_\chi^2 \langle \sigma v \rangle m_\chi =
\langle \sigma v \rangle \rho_\chi^2/m_\chi,
\end{equation}
where $\rho_\chi$ is the WIMP mass density.  We note the dependence of the DM heating 
$\hat Q_{DM} \propto  \langle \sigma v \rangle/m_\chi,$
so that by studying a wide range of WIMP masses we are effectively studying a comparable range of annihilation
cross sections. 

The annihilation products
typically are electrons, photons, and neutrinos. The neutrinos escape
the star, while the other annihilation products are trapped in the
dark star, thermalize with the star, and heat it up.  The luminosity
from the DM heating is
\begin{equation}
\label{DMheating}
L \sim f_Q \int \hat Q_{DM} dV 
\end{equation}
where $f_Q$ is the fraction of the annihilation energy deposited in
the star (not lost to neutrinos) and $dV$ is the volume element. We
take $f_Q=2/3$ as is typical for WIMPs, see \cite{Spolyar08}.

\cite{Spolyar09} and \cite{Freese10} consider two mechanisms for
supplying DM ``fuel''. One is gravitational contraction in
which DM is supplied by the gravitational attraction of baryons in the star. We label this mechanism AC for 
adiabatic contraction (see \cite{Blumenthal}), the
technique that allows us to calculate the resultant DM density inside the star.  This is a generic process,
and is expected to occur in the halos in which DSs form. It has been shown in \cite{Freese09} that the 
choice of initial DM profile, as well as details in the assumption of the DM orbits involved, are not crucial, and the 
effect prevails.
 \cite{Spolyar09} found the following approximation on how the DM density follows the
(baryonic) gas density $n_h$, namely
\begin{equation}
\label{eq:simple}
\rho_\chi \sim 5 ~{\rm (GeV/cm}^{-3}) (n_h/{\rm cm}^3)^{0.81}.
\end{equation}
In this paper, we implement AC, following \cite{Spolyar09} and \cite{Freese10}, using the 
Blumenthal method\footnote{We do not use the simple formula given in Eqn.(\ref{eq:simple}).
}

The second mechanism refers to replenishing of DM inside the star by
capture of WIMP DM from the surroundings as it scatters elastically off of nuclei in the star, see \cite{Freese08b, Iocco08}. This elastic scattering is the same
mechanism that is searched for in direct WIMP detection experiments. In both fueling mechanisms, the final stellar
mass is driven to be very high (with and without capture), and while DM reigns, the star remains bright but cool.  The DSs growing
via WIMPs captured via elastic scattering are hotter and denser than the ones formed via AC alone.  In this paper, we restrict our 
studies to WIMPs gravitationally brought in via AC alone and do not consider captured DM.

The main focus of this paper is to improve upon the modeling of the stellar evolution of dark stars.    
The calculations of \cite{Freese08,Spolyar09,Freese10} were based on the assumption of polytropic stellar interiors.
More precisely, as soon as the conditions were ripe for dark star formation $(M \sim 1-10 M_{\sun})$, 
dark stars were built up by accretion of their surrounding material, while an iterative procedure ensured that 
polytropic equilibrium configurations were found along the evolutionary sequence. 

Instead of polytropic models, our current work employs the
  fully-fledged 1D stellar evolution code
  MESA\footnote{http://mesa.sourceforge.net/}, which allows us to
  solve the stellar structure equations self-consistently, without
  restrictive assumptions on the equation of state or overall
  structure of the stellar models. This is accomplished through an
  additional module in MESA that locally adds the energy
  due to dark matter heating.  

Using these models, we are also able to study deviations from
equilibrium, which allows us to explore the question of DS
pulsations. We accomplish this at the expense of neglecting some
physical effects in this work. While we do implement extended AC and
DM heating self-consistently, we do not include DM capture, nor
nuclear burning, in contrast to \cite{Spolyar09} and \cite{Freese10}.
However, now that we are comfortable that our newly implemented module
in MESA is working successfully, i.e. giving robust results,
  regardless of initial stellar mass, WIMP mass, or halo environment,
work is in progress to include the above effects in a future
publication.

There have been some recent critiques of the idea of dark
  stars. While a full response is not appropriate here, we will
  mention a few points; a more detailed account can be found in a
  response published on the arXiv \cite[see][]{Gondolo13}.  

First, previous simulations such as those in \cite{Ripa10}
  and \cite{Smith12} have studied collapsing protostellar clouds and
  noted that the collapse continues past a hydrogen density that is
  higher than the one quoted in \cite{Spolyar08} (The exact value of
  this threshold depends on the adopted WIMP mass). This fact led to the incorrect conclusion 
  that DM heating is not potent enough for the establishment
  of a hydrostatic equilibrium of a DM heating-powered dark
  star. However, subsequent work by \cite{Freese08} and
  \cite{Spolyar09} has shown that the expected dark matter densities
  in models of fully-formed dark stars, which \emph{are} supported by
  dark matter annihilation, are indeed much higher.  Due to resolution
  limits, the aforementioned simulations are unable to reach densities
  this high, and are therefore unable to directly address the dark
  star regime. 

Second, a growing body of literature finds that the central
  accretion disk around the protostar can fragment
  \citep[see][]{Stacy10, Clark11, Greif12, Stacy14}. This could lead
  to the formation of multiple protostars, and the removal of a
  central object from the DM cusp. Furthermore, these protostars
  \citep[see][]{Stacy12} could gravitationally scatter the dark matter
  and thereby remove it from the central DM cusp. Of course, this
  latter study pre-supposes the presence of multiple protostars, and
  therefore assumes that DM heating does not prevent fragmentation
  of the disk. However, \cite{Smith12} have found that the inclusion
  of DM heating can actually help stabilize the protostellar disk,
  preventing fragmentation around the central protostar within a
  radius of about $1000 AU \sim 2\cdot 10^5 R_{\odot}$ within $\sim
  6100$ years after formation of the primary protostar in the
  center. This in turn could prevent the formation of multiple
  protostars and their subsequent gravitational scattering
  of DM from the center.

  We note that the simulations can only follow the evolution
    for so long by adopting sink particles to those regions which
    approach stellar densities. Hence, it is plausible that a
    dark star could form in the central sink particle of the
    \cite{Smith12} simulations, at densities that are much higher and
    at radii that are much smaller than can be resolved, as noted by
    \cite{Smith12}. However, the DM is ``locked in'' to the central
    protostar in \cite{Smith12}, and does not dynamically follow the
    movement of the baryons. Future studies with high enough
    resolution to capture the mutual dynamical effects of baryons and
    DM in those very central regions will be necessary in order to
    assess the feasibility of the formation of dark stars.  In the
    meantime, we have good reasons for assuming that dark stars can
    form, and the final arbiter for their existence will be any
    distinct observational signatures that they possess. To this end,
    in this paper we present improved, state-of-the-art calculations of the
    observable parameters of dark stars. As with nearby stars,
    studying their evolution is a much simpler problem than describing
    the formation process itself.

\section{Treatment of stellar evolution} \label{SE}

The stellar evolution calculations in this paper were performed using the open source software package 
MESA (``Modules for Experiments in Stellar Astrophysics'') \citep{Paxton11}. 
This flexible and state-of-the-art 1D evolution code can be used to treat a wide variety of problems, from main sequence evolution,
to the red giant and asymptotic giant phases, and finally to the white dwarf phase. Additionally, it can be applied to models
which are accreting or losing mass. Lately, MESA has been significantly updated and advanced in its capabilities to model the 
evolution of giant planets, low-mass stars, massive stars, along with additional stellar features such as rotation and 
asteroseismology \citep{Paxton13}. MESA is regularly upgraded, as a result of continuous feedback by the community of 
MESA users and their differing scientific needs. In fact, the distinctive physics of rapidly growing dark stars in a primordial
environment is one example of a problem that can easily challenge a stellar evolution code, which may otherwise be optimized 
to the more standard evolution of stars in our local Universe.  

In MESA, material which is accreted is set to have the same entropy as
the surface layers of the model.  Thus, accretion does not directly
heat the surface. This is consistent with a physical picture in which
material passes through an accretion disk, gradually radiating away
the gravitational energy of its infall. By contrast, spherical
accretion involves the formation of a shock front at the stellar
surface, which increases the entropy of the accreted material and
heats the surface layers. By examining both mechanisms for the growth
of massive protostars, \citet{Hosokawa10} find that their models
converge to the same radii and temperatures as a function of mass for
masses $\gtrsim 40 \,\M$. In addition, given the larger radii of our
models, the gravitational energy release of any infalling matter
should be even less important. For these reasons, we use the default
prescription in MESA for accretion.

In order to study the effect of DM self-annihilation heating
  in stars, we have used MESA's \verb+``other_energy_implicit''+ interface to
  add a DM heating module to MESA (see Appendix B for details).
DM annihilation provides a powerful heat source in the first stars.
As long as the central temperature of a star in equilibrium is below
the onset of nuclear fusion, DM heating will be the only heat source
in that star, solely responsible for its luminosity. We shall clarify
here some terms: In the initial contracting phase of a protostellar
cloud, heat is released by means of gravitational contraction which is
given by the Kelvin-Helmholtz time scale of the protostar. Usually,
this whole process by which the newborn star settles into a
quasi-equilibrium up to the final onset of nuclear burning in the
star's center is called the pre-main-sequence phase. In the absence of
dark matter heating, this phase usually lasts much shorter than the
main-sequence phase of nuclear burning.

In the presence of DM heating, however, dark stars can accrete
substantial amounts of mass and yet stay cool enough for nuclear fusion to be delayed; we are confirming this scenario
in this paper. In fact, in accordance with \cite{Freese10}, we find that dark stars residing in the centers of their host halos can 
grow to supermassive size of $10^{4-6} M_{\sun}$, and this process can take $10^{5-9}$ years, depending on the accretion rate,
and assuming a continuous fuel of DM is provided.
Once the DM fuel runs out, the supermassive dark star may run through a rapid sequence of changes, whereby it shrinks, seeking for a new
equilibrium, until the central temperature is high enough for fusion to start. A fusion-powered star that massive, however, can not  
survive for long, and the dark star may collapse to form a massive black hole soon after.  Indeed, this way, DSs could
provide a compelling cause for the early formation of supermassive black holes, which have been observed in the centers of 
galaxies in the local, as well as in the high-redshift Universe. Thus, massive dark stars spend most of their lives
in the pre-fusion phase, and hence the term 'pre-main-sequence' phase is a misnomer in the case of dark stars. Therefore, we will 
try to avoid this term, noting, though, that the preparation of the initial conditions in MESA are accomplished with a module of
that name (see also the next subsection).

\subsection{Initial conditions}

In MESA a new evolution can be started by creating a 'pre-main-sequence model' upon specifying the mass, a uniform composition,
a luminosity, and a central temperature $T_c$ low enough to prevent nuclear burning. For a fixed $T_c$ and composition, the total
mass depends only on the central density, $\rho_c$. An initial guess for $\rho_c$ is made by using an $n=3/2$ polytrope, appropriate
for a fully convective star, although MESA does not assume that the star is fully convective during the subsequent search for a
converged pre-main-sequence model. Instead, MESA uses its routines for solving the equations of stellar structure, equation of 
state,
and 'mixing-length-theory' (MLT) for the treatment of convection in order to search for a $\rho_c$ that gives the model of the desired mass. 
That initial guess may not be optimum for dark stars, but our MESA models tend to converge quickly towards equilibrium sequences.  

In light of the comparison of our results with previous polytropic models of supermassive dark stars, we use the same parameters for
the halo environment
as in \cite{Freese10}. According to their choice, we consider models of dark stars which are accreting matter at a 
(constant) rate of $\dot M = 10^{-3} M_{\odot}$yr$^{-1}$ in a host minihalo of $10^{6} M_{\odot}$, forming at a redshift of $z=20$,
as well as at a higher rate of $\dot M = 10^{-1} M_{\odot}$yr$^{-1}$ in a larger host halo of $10^{8} M_{\odot}$ with
a formation redshift of $z=15$, respectively. 
As our terminology throughout the paper, we use
\begin{equation}
\label{eq:defineSMH}
{\rm SMH:  \dot{M} = 10^{-3} M_{\odot}/{\rm yr,}~ ({\rm host~ halo
    ~of}~ 10^{6}~ M_{\odot}) } \, 
\end{equation}
\begin{equation}
\label{eq:LMH}
{\rm LMH:  \dot{M} = 10^{-1} M_{\odot}/{\rm yr,}~ ({\rm host~ halo~
    of}~ 10^{8}~ M_{\odot}) } \, .
\end{equation}
For both halo masses, we choose a fraction of $15 \%$ baryons and $85
\%$ DM, a primordial metallicity of $Z=0$, and a hydrogen-to-helium
fraction of $0.76$. For each halo, we assume that initially
  both the baryons and the DM can be described with the same NFW
  density profile \cite[see][]{NFW},
\begin{equation} \label{iniprofile}
\rho(r)=\frac{\rho_0}{r/r_s(1+r/r_s)^2},
\end{equation}
where $\rho_0$ is the central density and $r_s$ is the scale radius. At any point of the profile, baryons will only make up 
15\% of the mass. The density scale, $\rho_0$ can be re-expressed in terms of the critical density of the Universe at a 
given redshift, $\rho_c(z)$ via
\begin{equation}
\rho_0=\rho_c(z)\frac{178}{3}\frac{c^3}{\rm{ln}(1+c)-c/(c+1)},
\end{equation}
where $c\equiv r_{vir}/r_s$ is the concentration parameter and
$r_{vir}$ is the virial radius of the halo.  We choose a fiducial
value of $c=3.5$. In fact, the properties of DSs stay roughly the same for
  concentration parameters $c=2-5$, as has been shown in
  \cite{Ilie11}. While cosmological N-body simulations suggest
    that the profile in Equ.(\ref{iniprofile}) is a good fit to most
    DM halos, there is still uncertainty about the exact inner density
    slope of DM halos. It has been shown in \cite{Freese09}, however, that
    DSs result, regardless of the inner density profile; that paper considered even the extreme case of a cored Burkert
    profile and found the resulting DSs. We assume a flat $\Lambda$CDM universe, using the cosmological parameters from the 
    3-year data of the WMAP satellite (\cite{WMAP3}), i.e. a present matter density $\Omega_m = 0.24$, Hubble parameter 
    $h = 0.732$, and hence a dark energy density of $\Omega_{\Lambda} = 0.76$. These are the parameters used in previous
     works on DS evolution, and for the sake of comparison,
we will use these parameters for our MESA calculations, as well.\footnote{We performed test runs to see how DS properties
are affected by using more recent WMAP or Planck satellite data (\cite{Planck}), respectively. The changes are too small to be 
noticable on the plots presented in this paper. A change of the values for the cosmological parameters of the order of 
$\sim 10 \%$ will thus not affect our basic results.}

For models with WIMP masses higher than $m_{\chi} = 10$ GeV, we choose
an initial stellar mass of $2 M_{\sun}$. For models with $m_{\chi} =
10$ GeV, we need higher initial masses for the initial models to
converge. In those cases, we chose an initial mass of $5 M_{\sun}$.

\section{Evolution of supermassive dark stars}

We explore three different values of the mass of the dark matter particles, 10, 100, and 1000 GeV, and for each we compute
a sequence of DS models starting at $2 \,\M$ (or $5 \,\M$) and ending at
over $10^5 \,\M$ for the minihalo 
(``SMH''), and $10^6 \,\M$ for the larger halo (``LMH''), 
respectively\footnote{For notational convenience, we may call the
  $10^8 \,\M$ halo (LMH) sometimes a 'minihalo', as well, even though
this runs counter to the definition of minihalos as objects with virial temperatures lower than about $10000$ K, which sets
the boundary above which atomic cooling prevails over molecular cooling. The mass range of minihalos is redshift-dependent:
for our chosen redshifts, the maximum mass of minihalos amounts to
about $1.3\cdot 10^7 \,\M$ at $z=20$ and 
$2\cdot 10^7 \,\M$ at $z=15$, respectively, according to equation (2) in \cite{Shapiro04}.}. 
  Some comments are in order here before we proceed to 
show the results. The protostellar accretion rate can be estimated from the free-fall timescale
of the Jeans mass of the gas, as
 \beq \label{acc}
  \dot M \simeq \f{M_J}{t_{\rm{ff}}} \simeq \f{c_s^3}{G} \propto T^{3/2},
  \eeq
  i.e. accretion rates are substantially higher in the hotter primordial star formation clouds, compared to those in the present 
  Universe, which have more efficient ways of cooling.
 The above choice of values for $\dot M$, Equ.(\ref{eq:defineSMH}-\ref{eq:LMH}), is related to the adopted size of minihalo, since the ambient temperature is
 expected to be higher in larger minihalos, leading to higher accretion rates. It shall 
be emphasized that values of $\dot M$ in excess of $\gtrsim 10^{-3} M_{\odot}/$yr are actually very high, and present a challenge to the numerical
capabilities of stellar evolution codes. In fact, after initial difficulties, we were able to calculate models with accretion 
rates up to $10^{-1} M_{\odot}/$yr, using an upgraded version of MESA. All calculations in this paper
have been performed using release 5596 of MESA.

\begin{figure*} 
\begin{minipage}{0.5\linewidth}
     \centering
     \includegraphics[width=8.5cm]{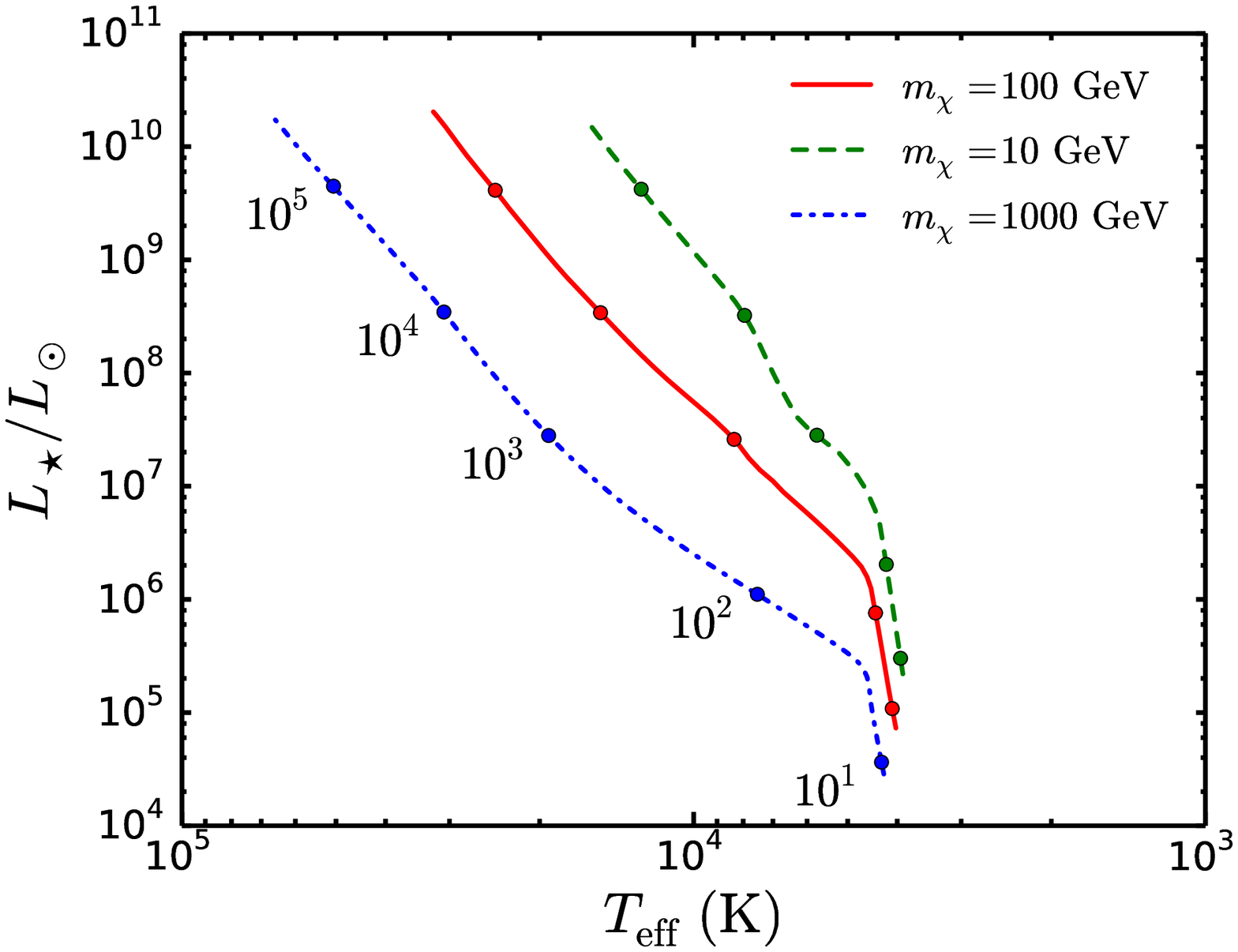}
     \vspace{0.5cm}
    \end{minipage}%
    \begin{minipage}{0.5\linewidth}
      \centering
      \includegraphics[width=8.5cm]{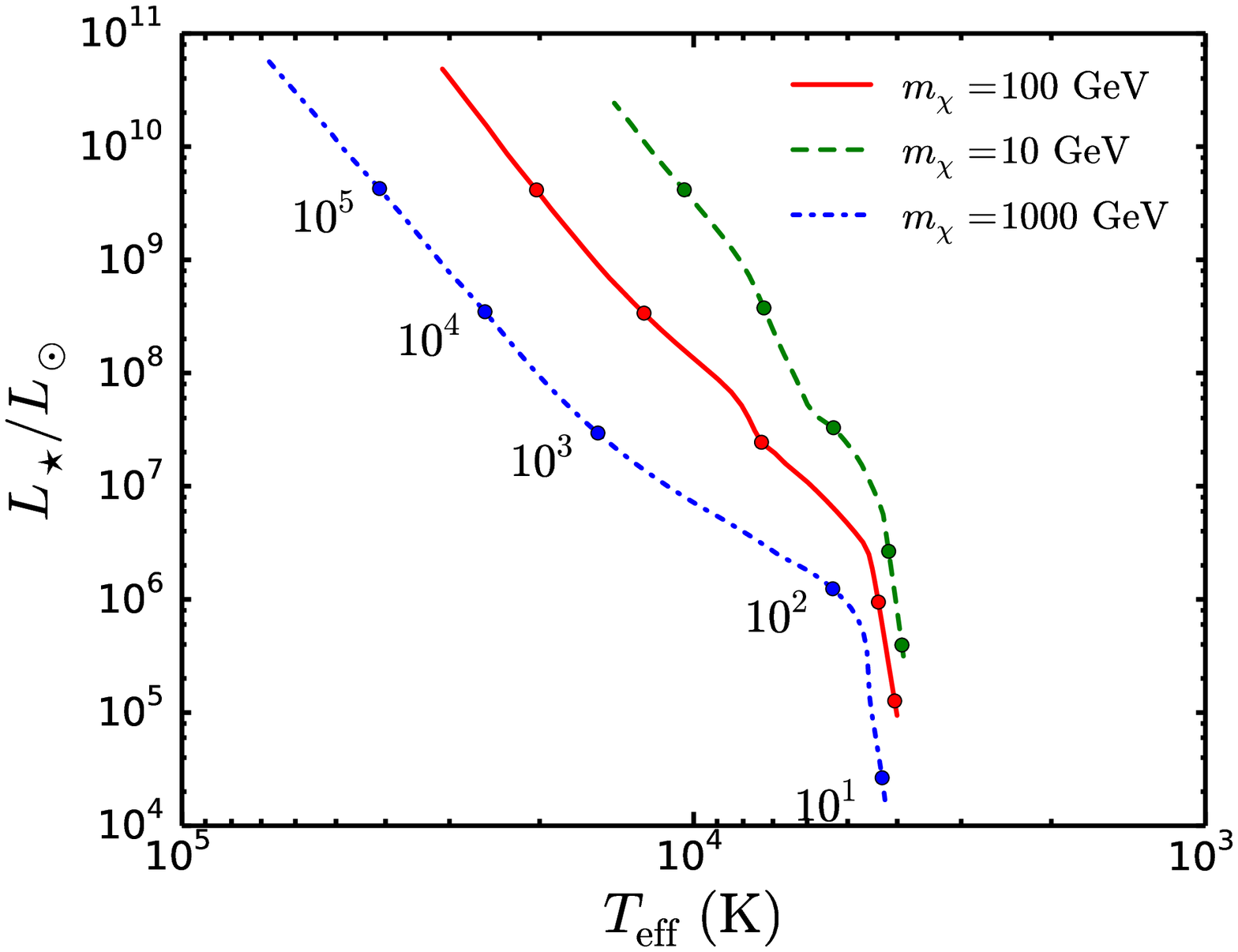}
     \vspace{0.5cm}
    \end{minipage}
 \caption{Evolution of dark stars in a host minihalo of mass $10^6~ M_{\odot}$ (``small minihalo'' - SMH) with an accretion rate of
 $\dot M = 10^{-3} M_{\odot}/$yr (\tx{left-hand plot}), and evolution of dark stars in a host halo of mass $10^8~ M_{\odot}$ 
 (``large minihalo'' - LMH) with an accretion rate of $\dot M = 10^{-1} M_{\odot}/$yr (\tx{right-hand plot}), respectively, 
 for WIMP masses of 10, 100 and 1000 GeV. The highlighted dots on the tracks correspond to the 
 benchmark values of the dark star mass, according to Table \ref{tab1} and Table \ref{tab2}, respectively. 
 The calculations assume extended adiabatic contraction and no significant depletion of dark matter due to annihilation. 
 Dark matter capture is not considered.}
 \label{HR}
\end{figure*}

In Fig.~\ref{HR}, we show the location of our model sequences in the Hertzsprung-Russell diagram for each
halo environment. The tracks are monotonic, once the star settles into
a quasi-static equilibrium.
Since the dark matter heating is proportional
to $1/m_{\chi}$ (see Equ.(\ref{DMheat})), we obtain different tracks for different particle masses. The heating in the $m_{\chi} = 10$ GeV models is the largest,
so these models have more pressure support and are therefore larger: at constant luminosity these models have cooler surface
temperatures. In contrast, the $m_{\chi} = 1000$ GeV models have less heating and are therefore smaller: these models are hotter
at fixed luminosity. We summarize some key stellar properties for every decade of mass growth in Table \ref{tab1} and 
Table \ref{tab2}, respectively.

\begin{figure*} 
\begin{minipage}{0.5\linewidth}
     \centering
     \includegraphics[width=7cm]{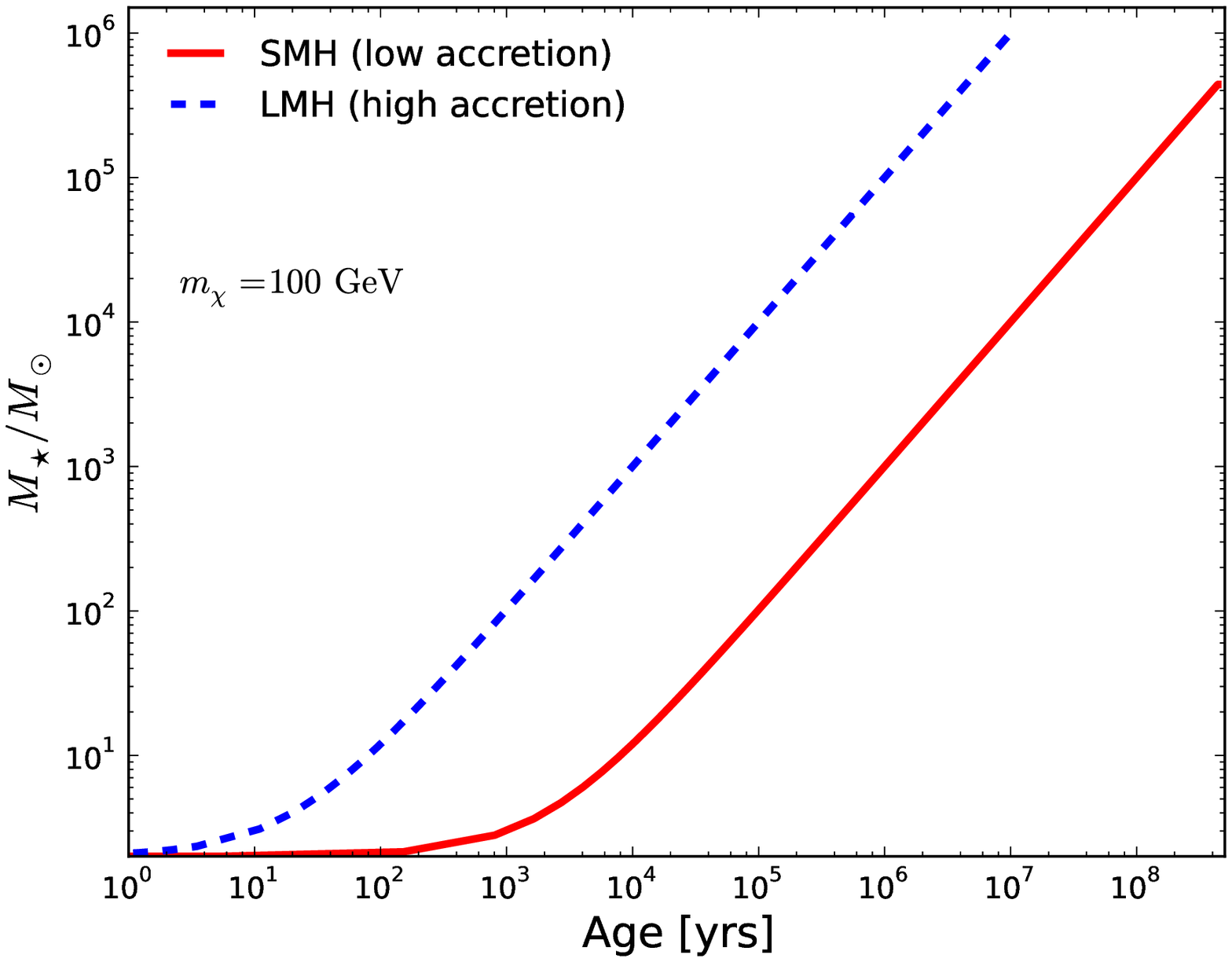}
     \vspace{0.05cm}
    \end{minipage}
    \begin{minipage}{0.5\linewidth}
      \centering\includegraphics[width=7cm]{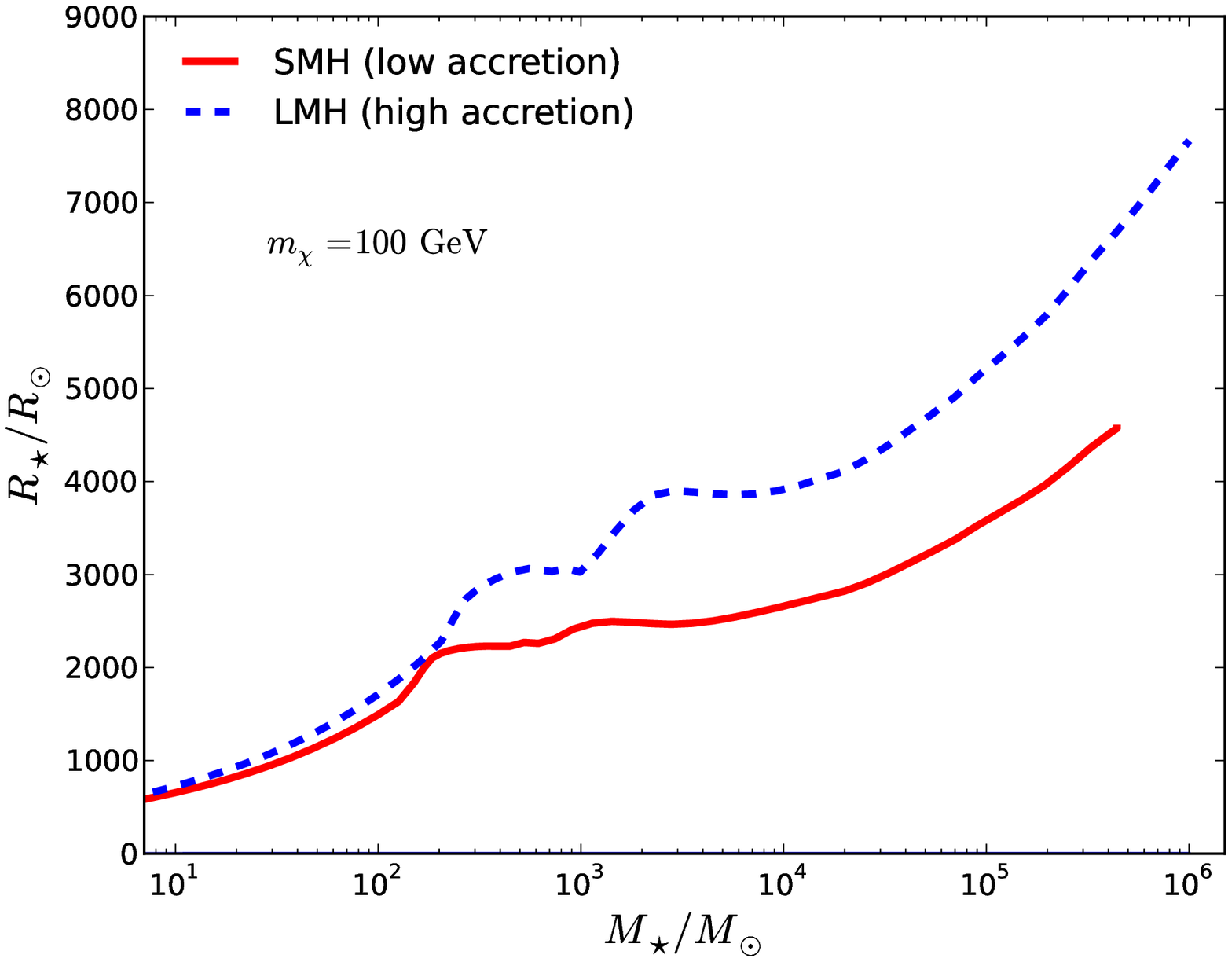}
     \hspace{0.05cm}
    \end{minipage}
 \begin{minipage}{0.5\linewidth}
     \centering
     \includegraphics[width=7cm]{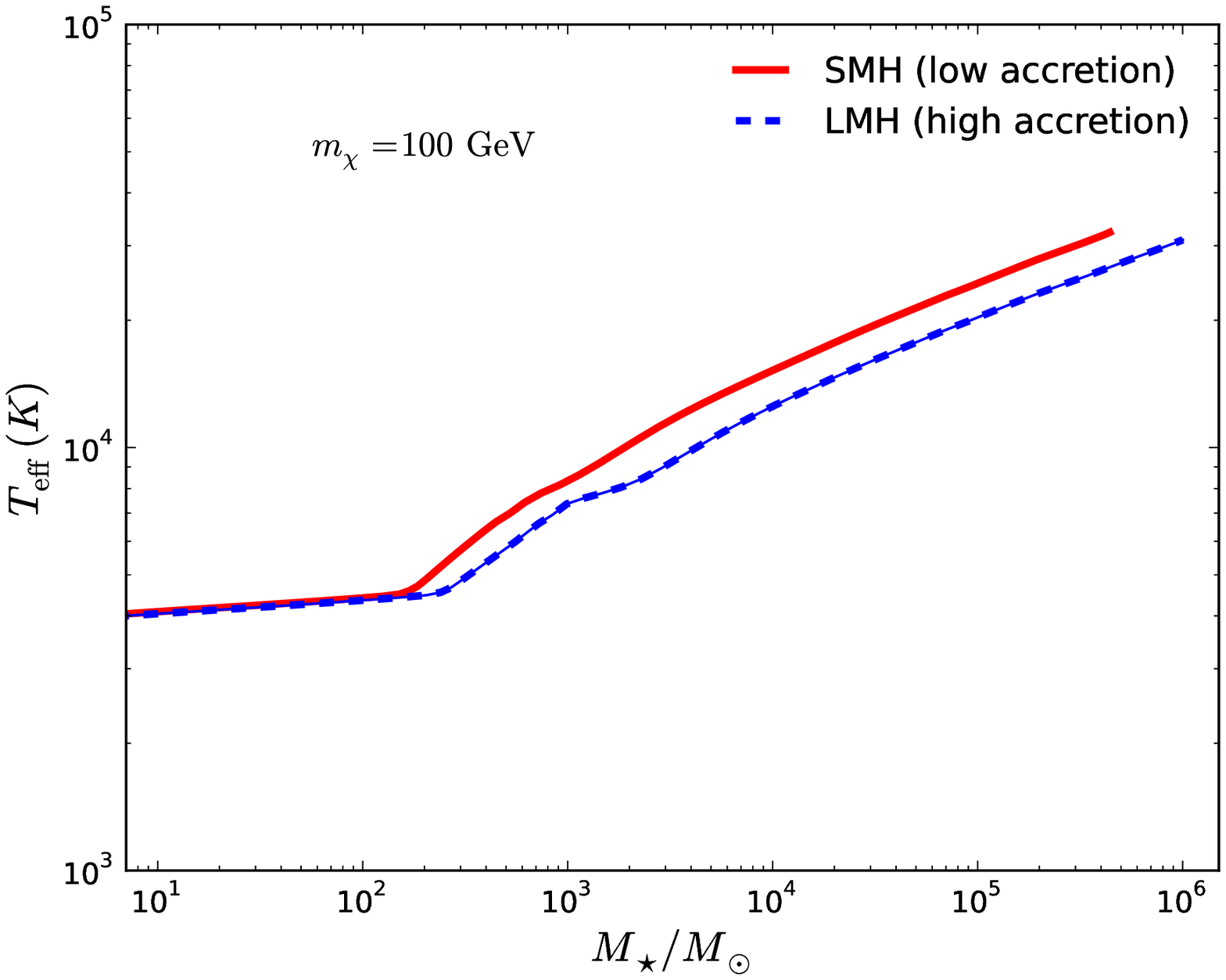}
     \vspace{0.05cm}
    \end{minipage}%
    \begin{minipage}{0.5\linewidth}
      \centering\includegraphics[width=7cm]{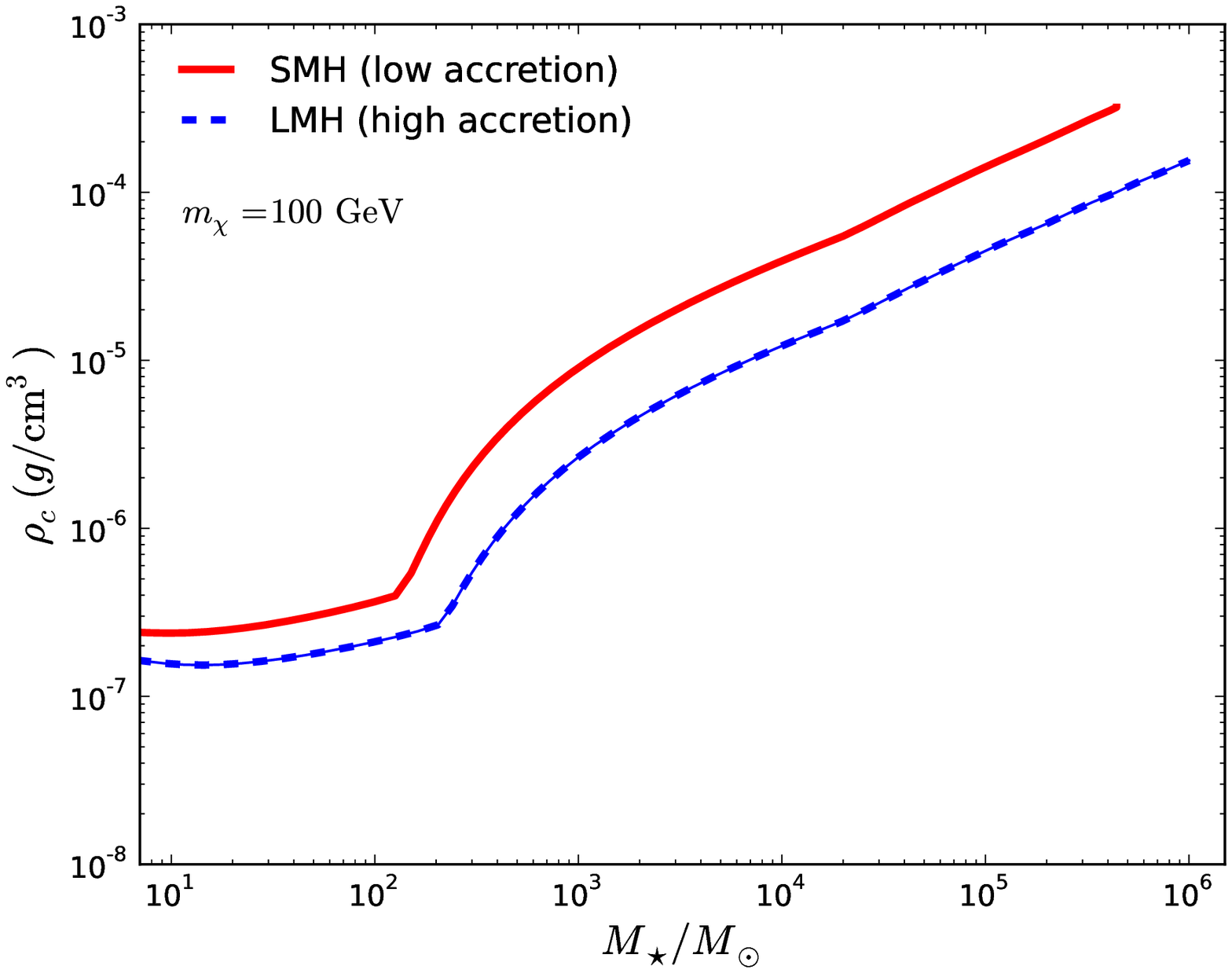}
     \hspace{0.05cm}
    \end{minipage}
 \caption{Evolution of dark stars, illustrating the dependence on accretion rate (SMH: $\dot M = 10^{-3}~M_{\odot}/$yr; 
 LMH: $\dot M = 10^{-1}~M_{\odot}/$yr). The quantities plotted are (a) stellar mass (upper left), (b) stellar radius (upper right),
 (c) surface temperature (bottom left), and (d) central density (bottom right). In each case, the DM particle
 mass is $m_{\chi} = 100$ GeV. 
 The same plots for different WIMP masses can be found in Appendix A, Fig.\ref{evol2} and \ref{evol3}.}
 \label{evol}
\end{figure*}

Figures~\ref{evol}, \ref{evol2} and \ref{evol3} show a comparison of
the evolution of different stellar characteristics, depending on the
accretion rate (Figs.~\ref{evol2} and \ref{evol3} are
  collected in Appendix A).  By assumption of a constant accretion
rate, the $M_{\star}$-Age relationship is simply linear and a given
age DS is more massive at higher accretion rates. For a given DS mass,
the low-accretion environment (SMH) will produce stars of smaller
radius, higher density and higher surface temperatures. This is true,
regardless of the value for the WIMP mass.

As our dark stars grow, they acquire radiation-dominated, weakly
convective envelopes. In general, such regions often have
large superadiabatic gradients, which can lead to slow convergence. 
These issues occur
  even for ``normal'' massive stars that are crossing the asymptotic
  giant branch. In this radiation-dominated regime, stellar evolution
  codes experience severe numerical difficulties due to the extremely
  small timescales required. 
It is an open question whether additional physical instabilities occur
in radiation-dominated stars which act to limit this
superadiabaticity.

In our calculations, we encounter these difficulties as the total mass
of our models approaches $10^4 M_{\sun}$.  Our approach is to use
MESA's MLT++ routines to partially suppress this superadiabaticity
\citep[see][for more details]{Paxton13}.  We found a reasonable range
of parameters that not only allowed our models to grow beyond
  $10^4 M_{\sun}$ but also reduced the impact caused by
superadiabaticity (see also Sec.~4). While using MLT++ can
  change the details of the mass-luminosity relationship, it does not
  fundamentally or qualitatively change the behavior of the models.
For instance, the ``bump'' in the radius at $\sim 100 M_\odot$ for the
1000 GeV WIMP case in Fig.~\ref{evol3} is caused by the sharp
transition of the envelope from subadiabatic to superadiabatic. The
MLT++ prescription limits this superadiabaticity, and this reduction
in the temperature gradient leads to a decrease in radius. Once the
model settles into a new equilibrium, the radius continues its growth. Without MLT++, this feature would have been
more pronounced. Similar behavior is found for all models in the mass
range of 100--1000 $M_{\sun}$.  This transition, however, becomes less
dramatic for smaller DM mass, as can be seen in Figs.~\ref{evol} and
\ref{evol2}.  In general, the rough features in these figures are
similar signs of the onset of superadiabaticity.

\begin{table}
\caption{Properties of dark stars along the evolutionary sequence for $m_{\chi}=100$ GeV and 
$\dot M = 10^{-3} M_{\odot}/$yr (SMH): mass $M_{\star}$, luminosity $L_{\star}$, radius $R_{\star}$, effective temperature $T_{\rm{eff}}$,
central temperature $T_c$ and total central density $\rho_c$}
\label{tab1}
\begin{center}
\begin{tabular}{r|r|r|r|r|r|r}
\hline
          & $M_{\star}$      & $L_{\star}$           &  $R_{\star}$    & $T_{\rm{eff}}$    & $T_c$             & $\rho_c$ \\
          & $[M_{\odot}]$    & $[10^6~ L_{\odot}]$   &  $[R_{\odot}]$  & $[10^3~ \rm{K}]$  & $[10^5~ \rm{K}]$  & [g cm$^{-3}$] \\           
\hline

          & $  10$ &     0.11 &  655.7 &  4.1 & $ 1.0 $  & $ 2.4 \times 10^{-7} $ \\ 
          & $ 100$ &     0.76 & 1493.3 &  4.4 & $ 2.6  $ & $ 3.7 \times 10^{-7} $ \\ 
          & $ 500$ &    10.41 & 2257.2 &  6.9 & $ 8.6  $ & $ 4.6 \times 10^{-6} $ \\ 
          & $10^3$ &    25.94 & 2437.0 &  8.3 & $ 12.4 $ & $ 9.0 \times 10^{-6} $ \\ 
          & $10^4$ &   341.21 & 2659.2 & 15.2 & $ 30.6 $ & $ 3.9 \times 10^{-5} $ \\ 
          & $10^5$ &  4121.02 & 3578.6 & 24.4 & $ 69.5 $ & $ 1.4 \times 10^{-4} $ \\
\hline
\end{tabular}
\end{center}
\end{table}

\begin{table}
\caption{Properties of dark stars along the evolutionary sequence for $m_{\chi}=100$ GeV and 
$\dot M = 10^{-1} M_{\odot}/$yr (LMH): stellar quantities as in Table \ref{tab1}} 
\label{tab2}
\begin{center}
\begin{tabular}{r|r|r|r|r|r|r}
\hline
          & $M_{\star}$      & $L_{\star}$           &  $R_{\star}$    & $T_{\rm{eff}}$    & $T_c$             & $\rho_c$ \\
          & $[M_{\odot}]$    & $[10^6~ L_{\odot}]$   &  $[R_{\odot}]$  & $[10^3~ \rm{K}]$  & $[10^5~ \rm{K}]$  & [g cm$^{-3}$] \\      
\hline
          & $  10$ &     0.13 & 724.1  &  4.0 & $ 0.8 $   & $ 1.6 \times 10^{-7} $ \\ 
          & $ 100$ &     0.95 & 1711.0 &  4.4 & $ 2.2 $   & $ 2.1 \times 10^{-7} $ \\ 
          & $ 500$ &     9.28 & 3044.2 &  5.8 & $ 5.5 $   & $ 1.2 \times 10^{-6} $ \\ 
          & $10^3$ &    24.45 & 3036.4 &  7.4 & $ 8.2 $   & $ 2.7 \times 10^{-6} $ \\ 
          & $10^4$ &   338.41 & 3916.5 & 12.5 & $ 20.7 $  & $ 1.2 \times 10^{-5} $ \\ 
          & $10^5$ &  4149.34 & 5205.5 & 20.3 & $ 47.3 $  & $ 4.5 \times 10^{-5} $ \\ 
          & $10^6$ & 48203.79 & 7797.4 & 31.4 & $ 106.6 $ & $ 1.6 \times 10^{-4} $ \\
           
\hline
\end{tabular}
\end{center}
\end{table}

\begin{figure*} 
\begin{minipage}{0.5\linewidth}
     \centering
     \includegraphics[width=7.5cm]{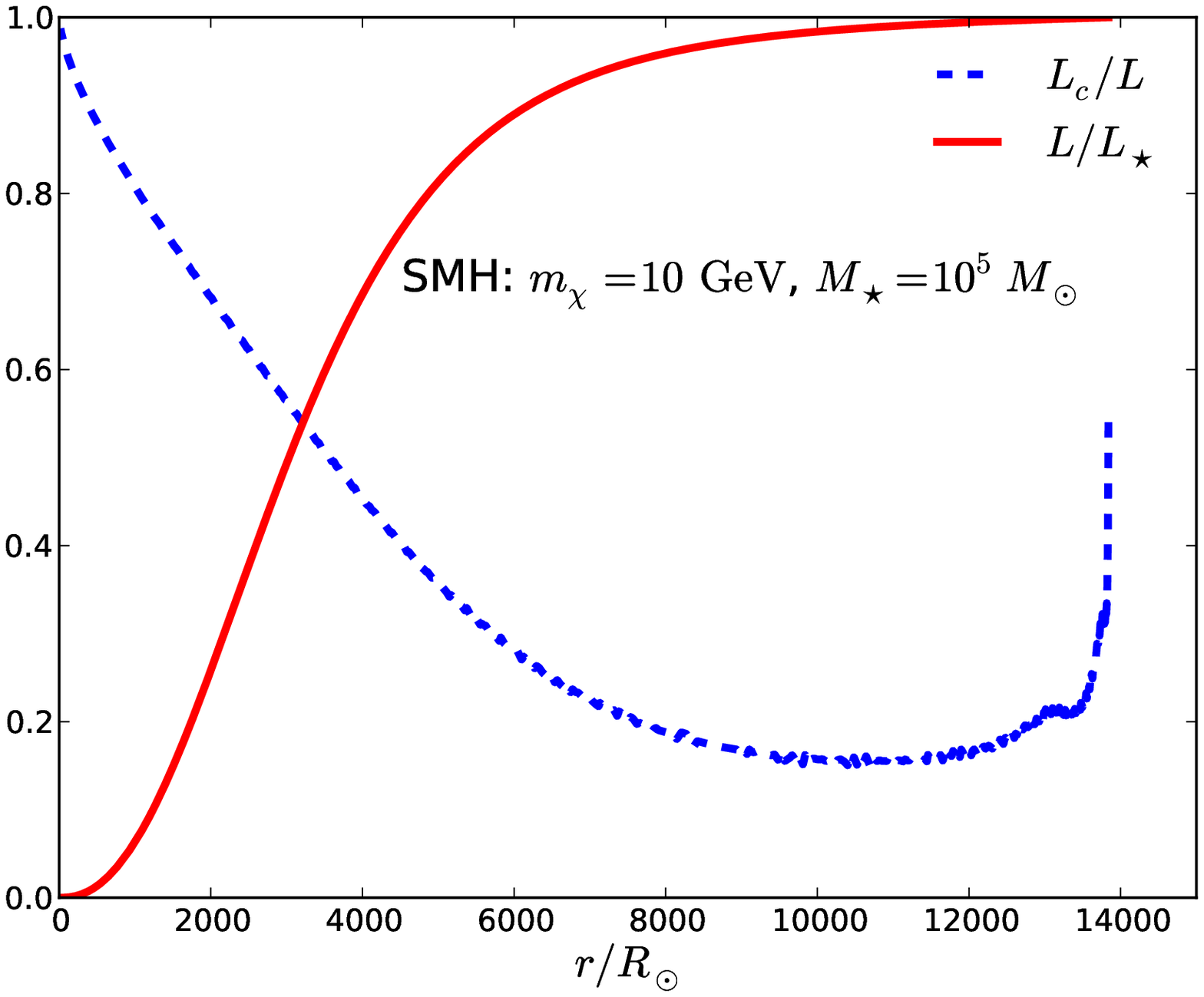}
     \vspace{0.05cm}
    \end{minipage}
    \begin{minipage}{0.5\linewidth}
      \centering\includegraphics[width=7.5cm]{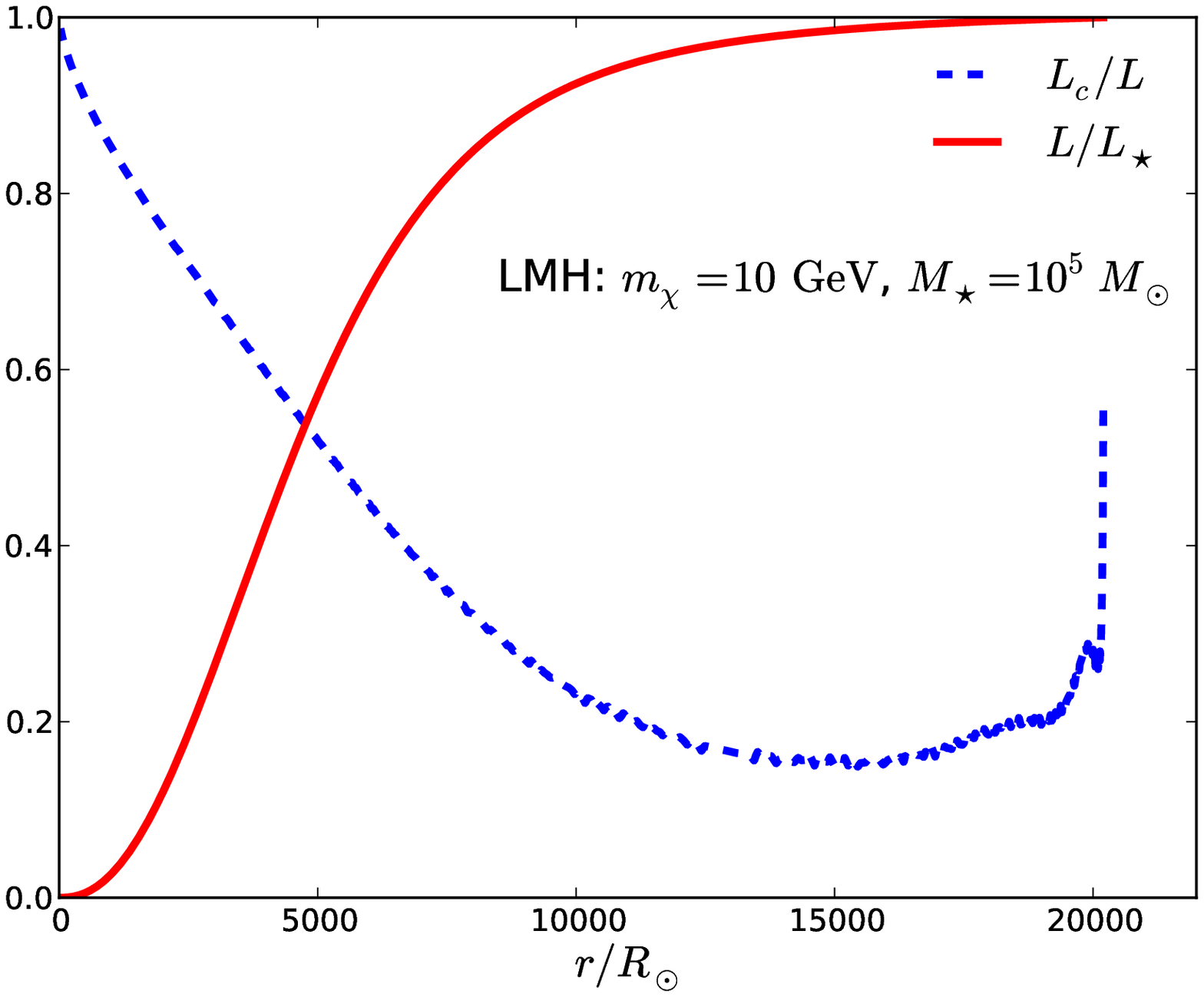}
     \hspace{0.05cm}
    \end{minipage}
    \begin{minipage}{0.5\linewidth}
     \centering
     \includegraphics[width=7.5cm]{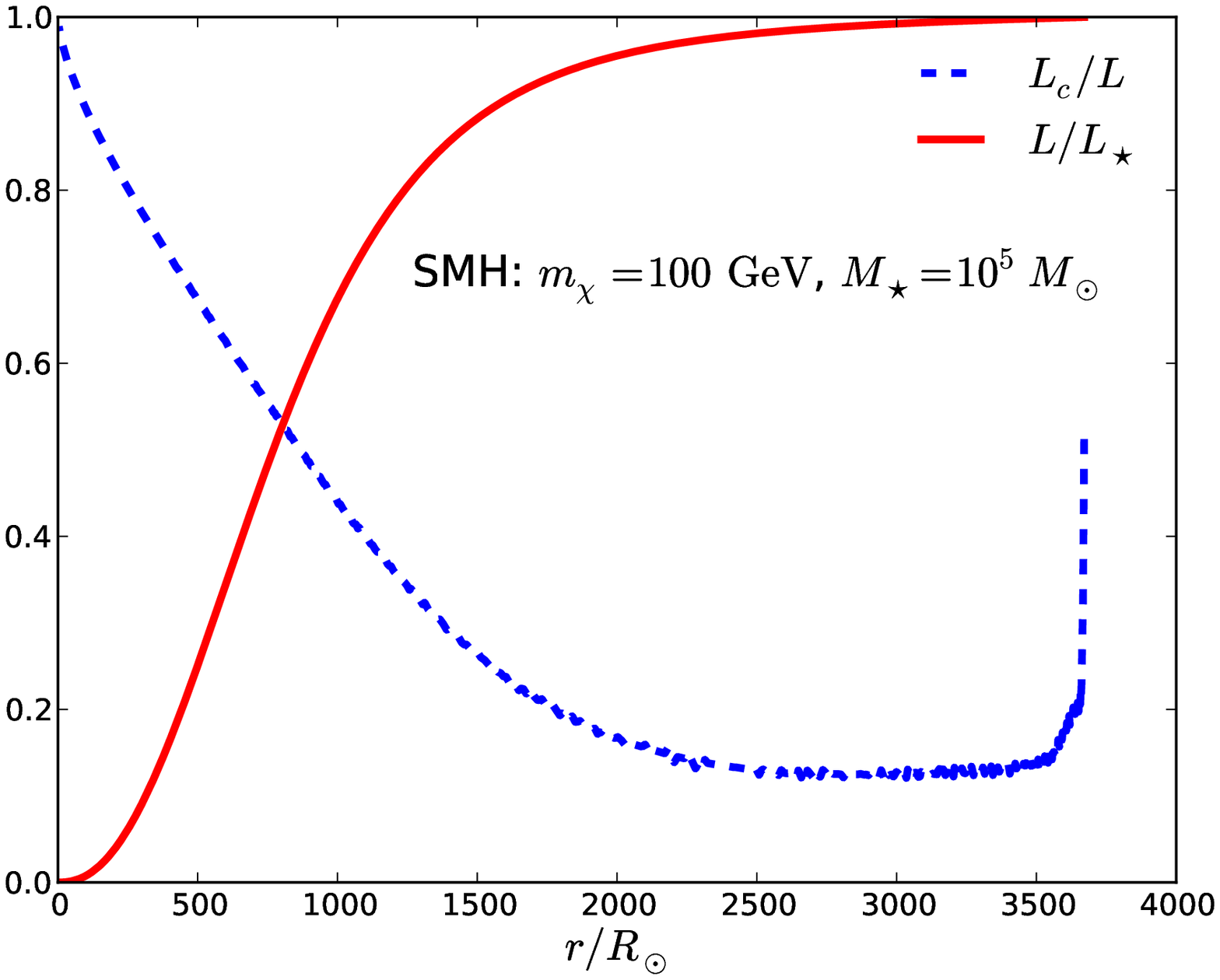}
     \vspace{0.05cm}
    \end{minipage}
    \begin{minipage}{0.5\linewidth}
      \centering\includegraphics[width=7.5cm]{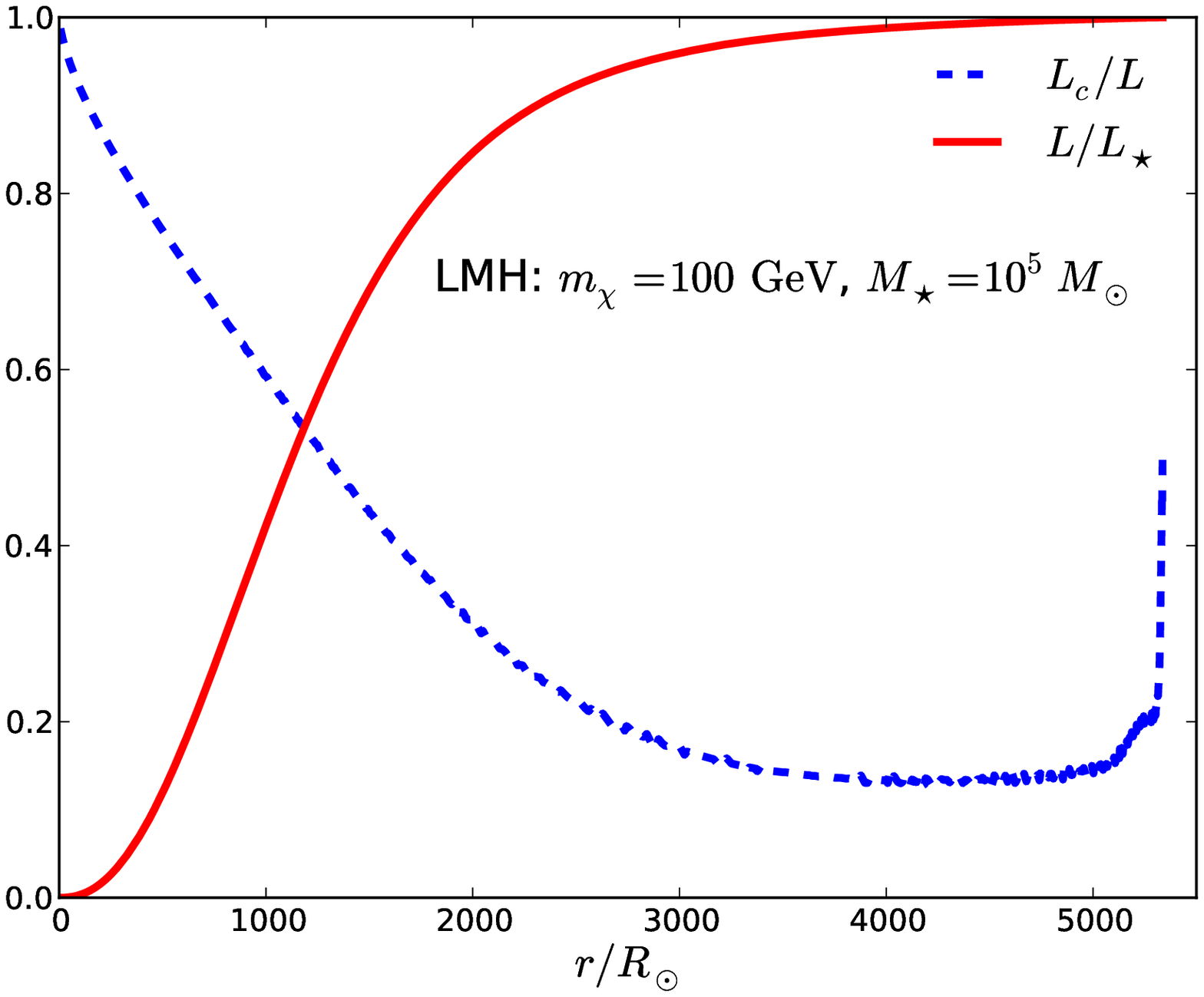}
     \hspace{0.05cm}
    \end{minipage}
    \begin{minipage}{0.5\linewidth}
     \centering
     \includegraphics[width=7.5cm]{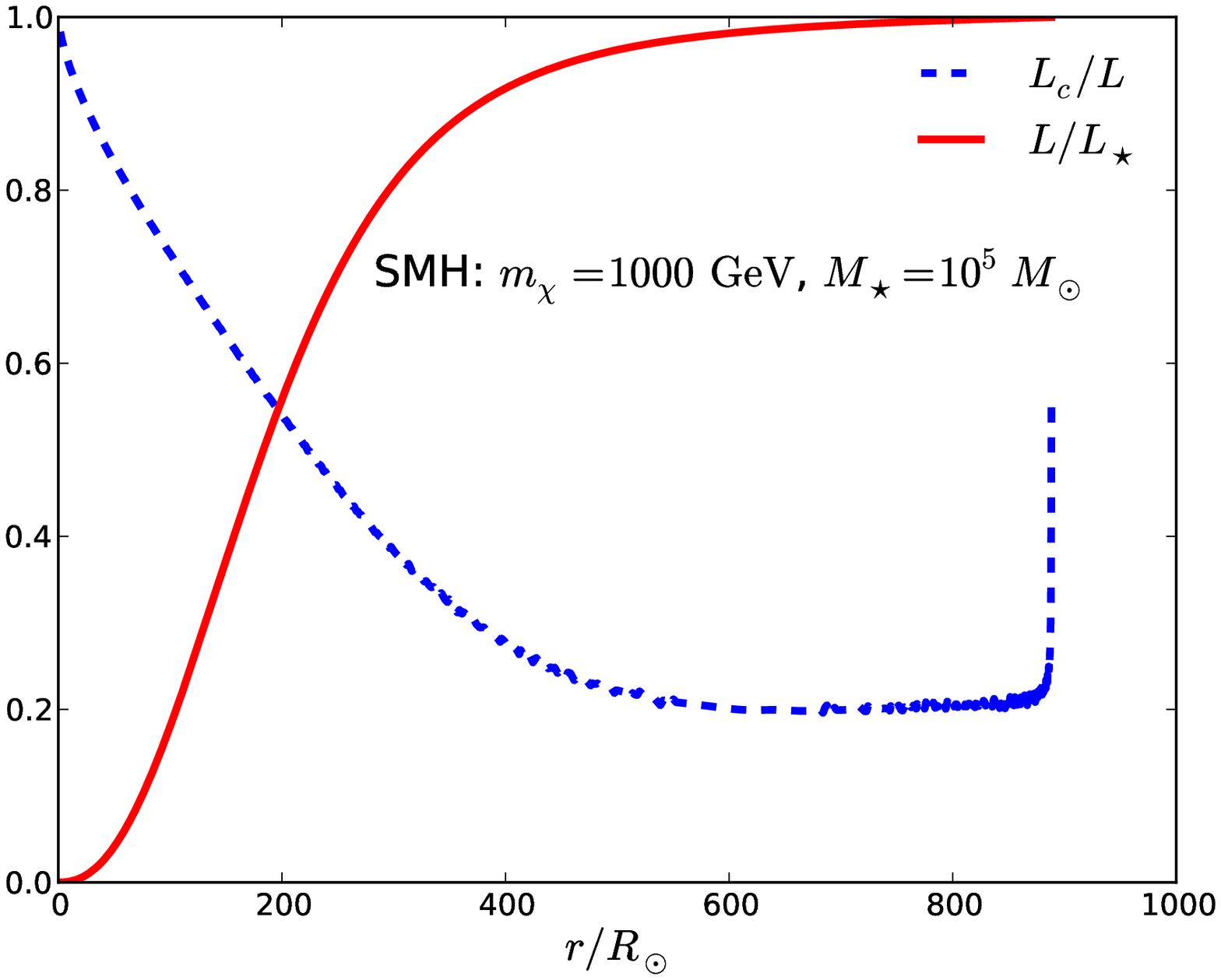}
     \vspace{0.05cm}
    \end{minipage}
    \begin{minipage}{0.5\linewidth}
      \centering\includegraphics[width=7.5cm]{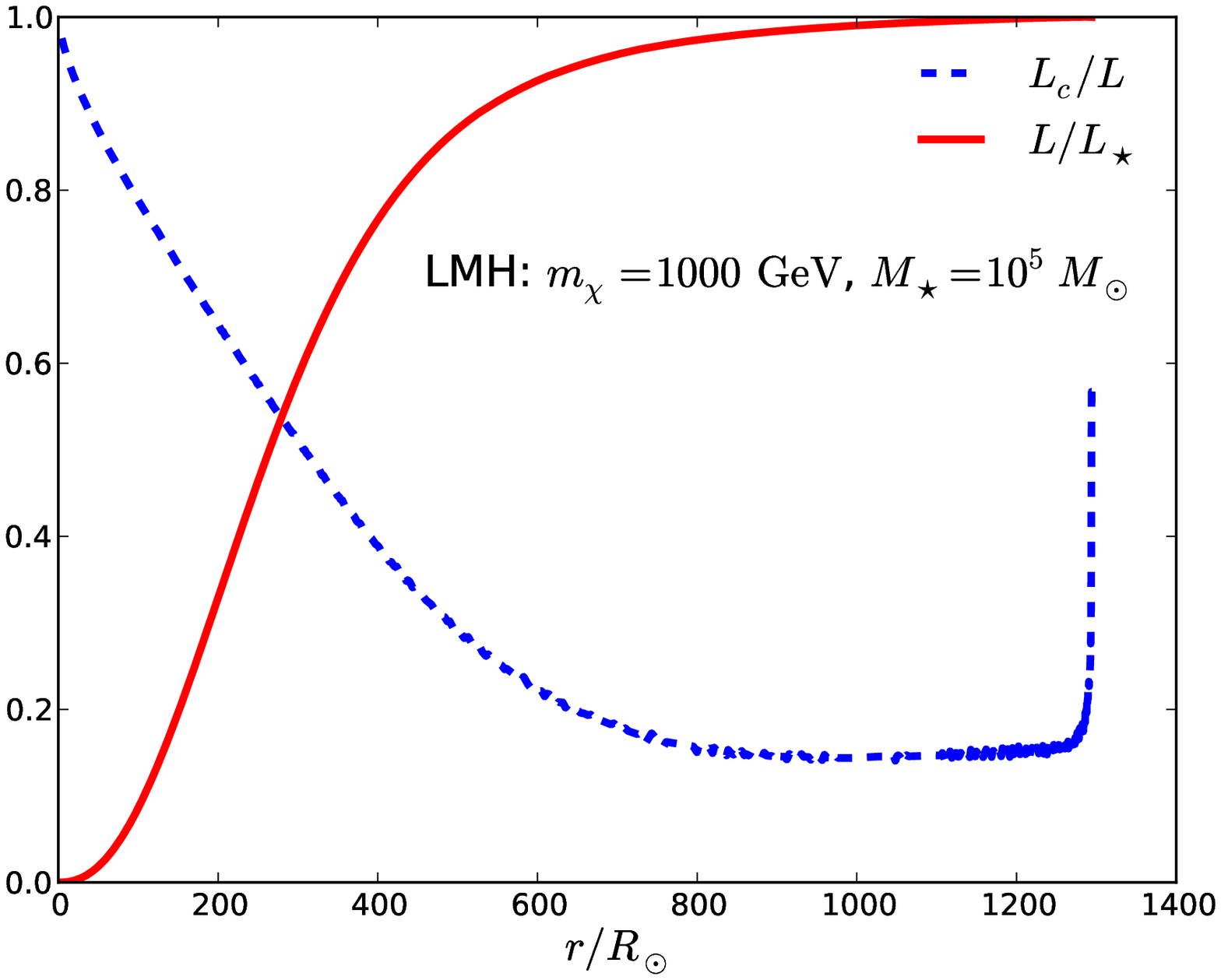}
     \hspace{0.05cm}
    \end{minipage}
 \caption{Luminosity at a given radius $L(r)$, as well as fraction of convective
 luminosity $L_c(r)$, as a function of radius. $L(r)$ is normalized
 over its value at the photosphere, $L_{\star}$, $L_c(r)$ is normalized over $L(r)$. \tx{Left column:} Dark star with $M_{\star} = 10^5~M_{\odot}$, 
 forming in SMH with accretion rate 
 $\dot M = 10^{-3}~M_{\odot}/$yr.   
 \tx{Right column:} Dark star with mass $M_{\star} = 10^5 M_{\sun}$ but forming in LMH with
 $\dot M = 10^{-1}~M_{\odot}/$yr. \tx{Upper row:} DM particle mass $m_{\chi} = 10$ GeV; 
 \tx{Middle row:} $m_{\chi} = 100$ GeV; \tx{Bottom row:} $m_{\chi} =
 1000$ GeV. In all cases, radiation dominates as the energy transport
 mechanism throughout most of the interiors of the models.}
 \label{lum1}
\end{figure*}

In Figure~\ref{lum1}, we plot the luminosity at a given radius $L(r)$ and fraction of convective luminosity $L_c(r)$, as a function of stellar radius, for
dark stars with $10^5 ~ M_{\sun}$. $L_c(r)$ is the luminosity at a given radius within the star contributed by convection, i.e.
\beq
L_c(r) = L(r) - L_{\rm{rad}}(r),
\eeq
where $L_{\rm{rad}}(r)$ reflects the part due to diffusive radiation transfer. 
We can see that the fraction due to $L_c(r)$ is important in the centers and at the very edge of the stars,
while $L_{\rm{rad}}$ dominates throughout most of the interior. Thus,
supermassive dark stars are mostly dominated by the radiative transfer of
energy, 
regardless of WIMP mass or accretion rate (see also Sec.~4). The
importance of $L_{\rm{rad}}$ increases with DS mass.

\begin{figure*} 
\begin{minipage}{0.5\linewidth}
     \centering
     \includegraphics[width=7.5cm]{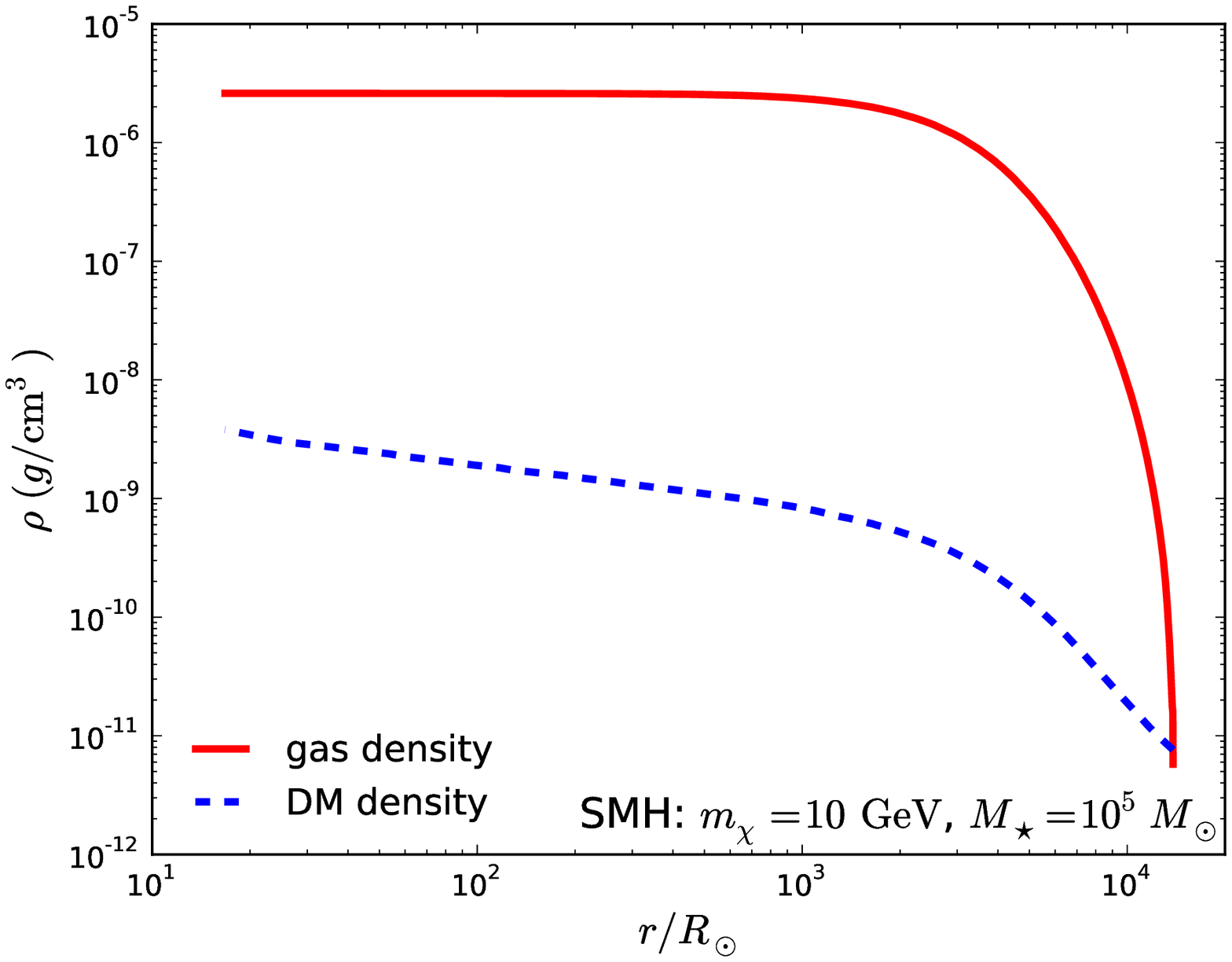}
     \vspace{0.1cm}
    \end{minipage}
    \begin{minipage}{0.5\linewidth}
     \centering\includegraphics[width=7.5cm]{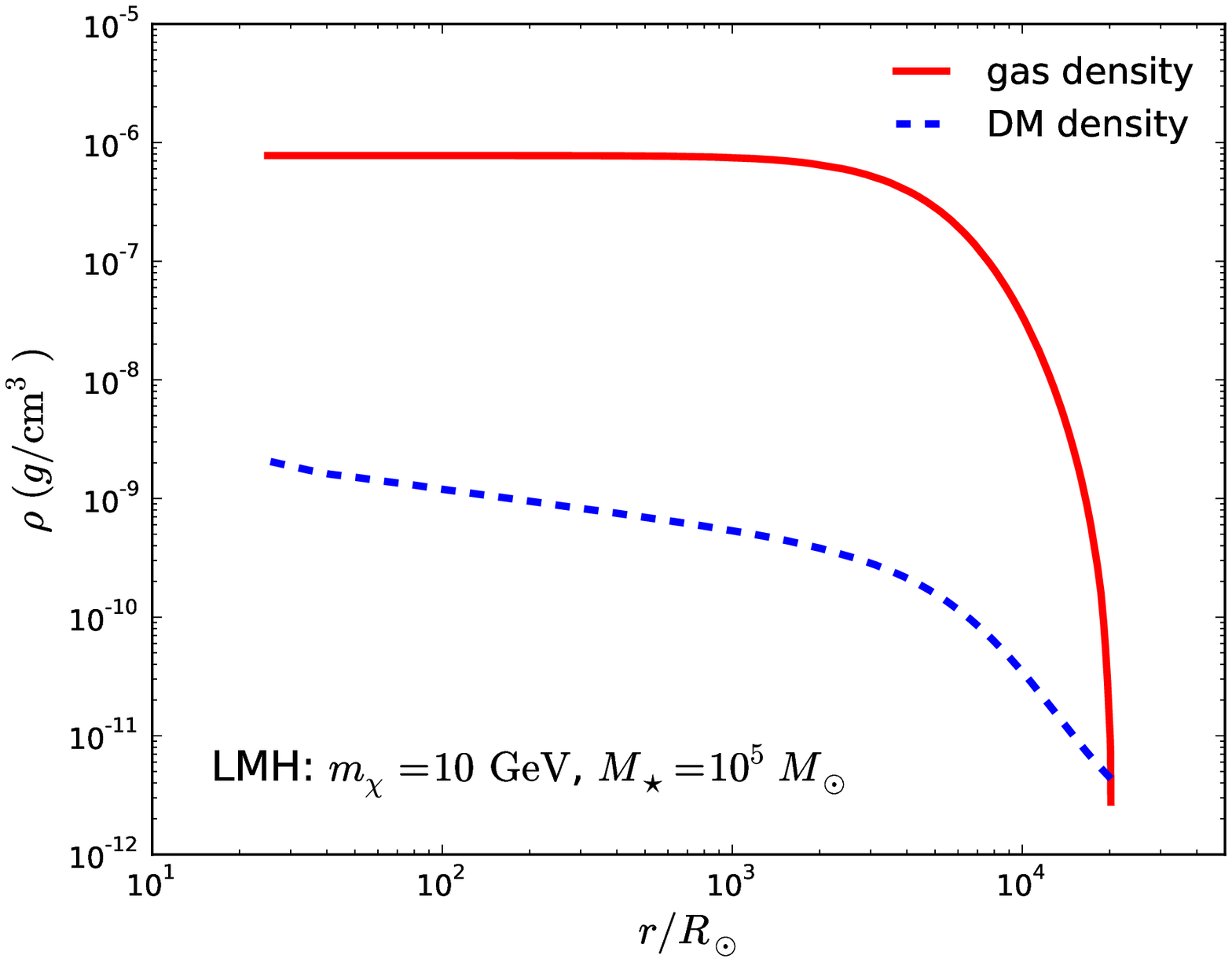}
     \hspace{0.1cm}
    \end{minipage}
\begin{minipage}{0.5\linewidth}
     \centering
     \includegraphics[width=7.5cm]{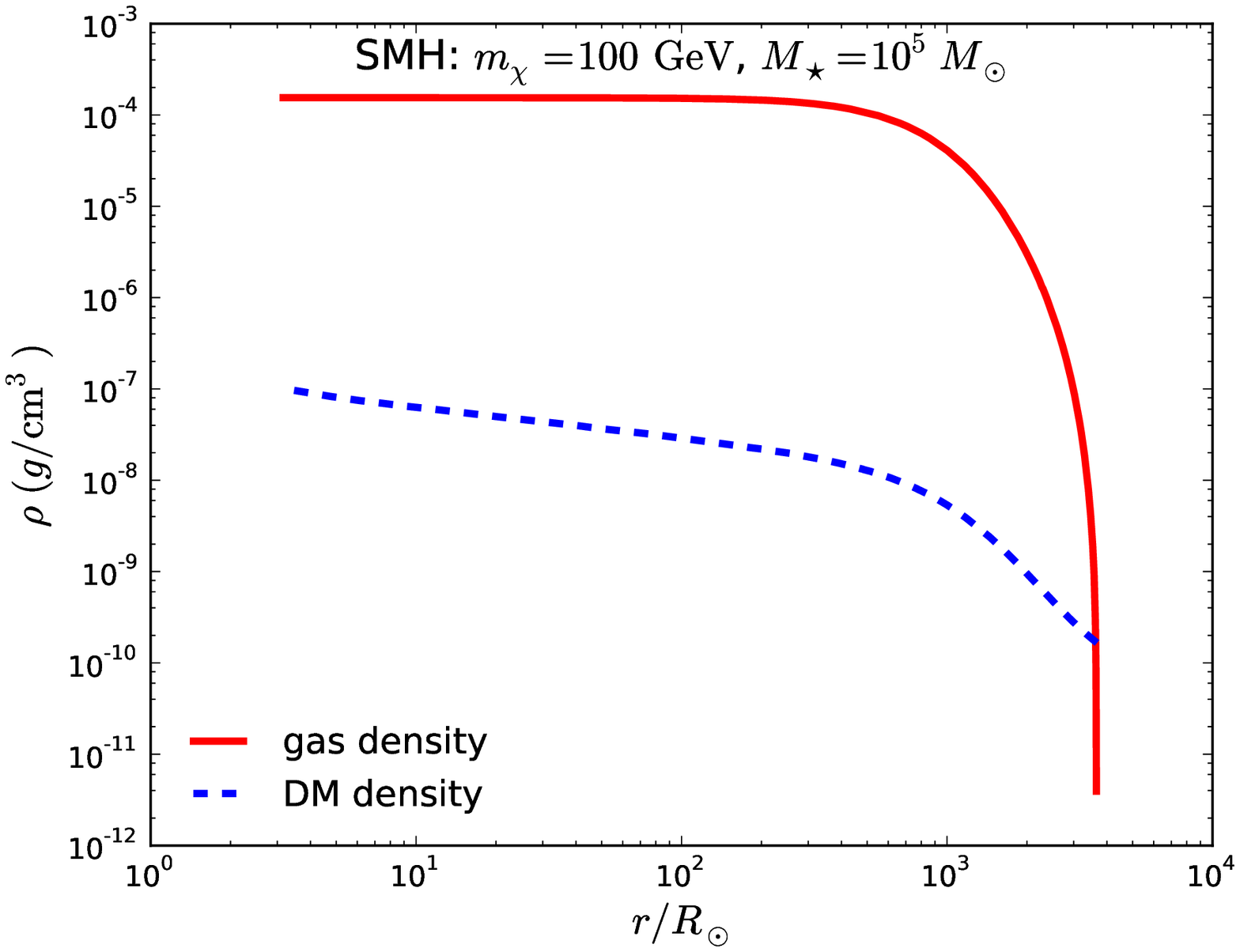}
     \vspace{0.1cm}
    \end{minipage}
    \begin{minipage}{0.5\linewidth}
     \centering\includegraphics[width=7.5cm]{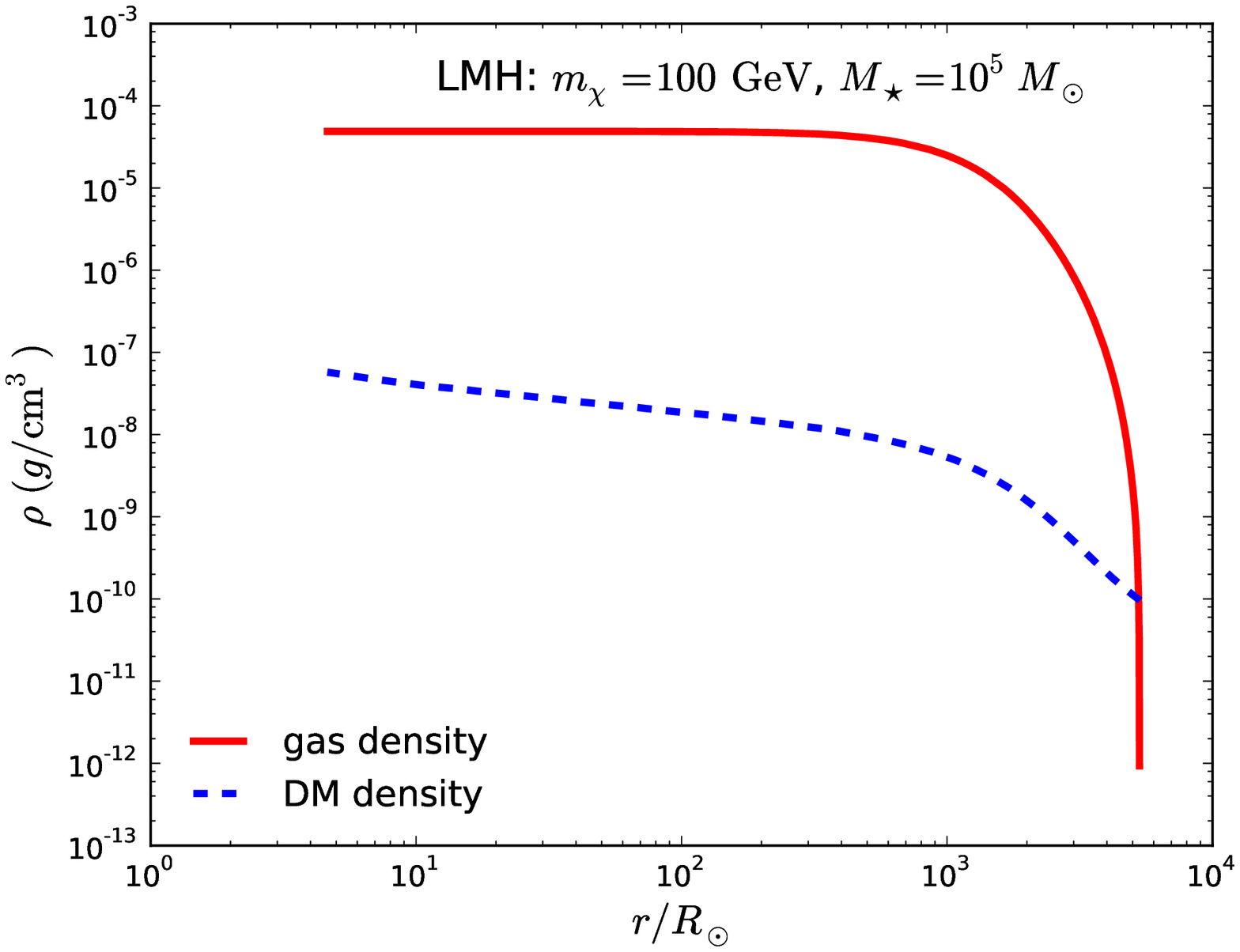}
     \hspace{0.1cm}
    \end{minipage}
    \begin{minipage}{0.5\linewidth}
     \centering
     \includegraphics[width=7.5cm]{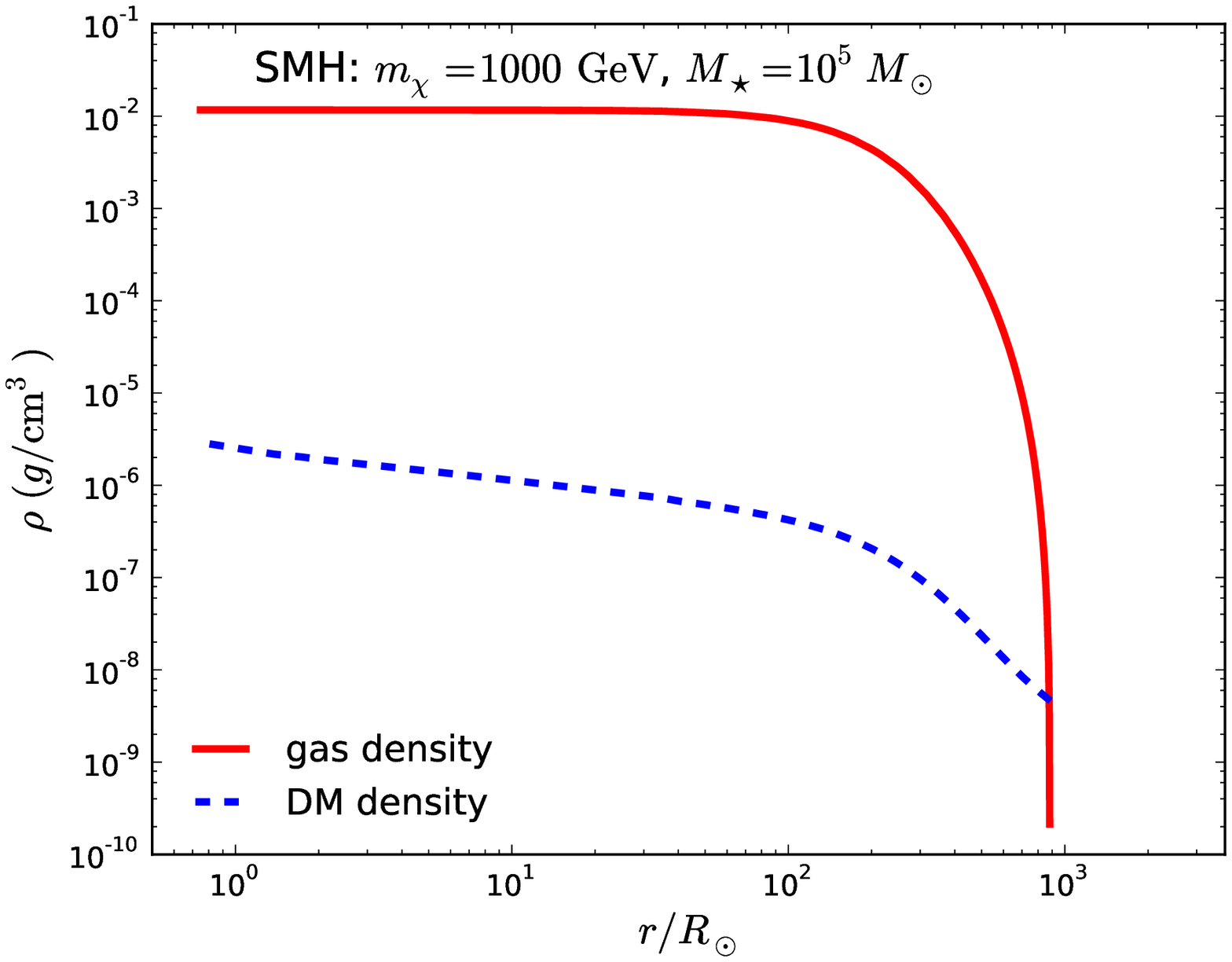}
     \vspace{0.1cm}
    \end{minipage}
    \begin{minipage}{0.5\linewidth}
     \centering\includegraphics[width=7.5cm]{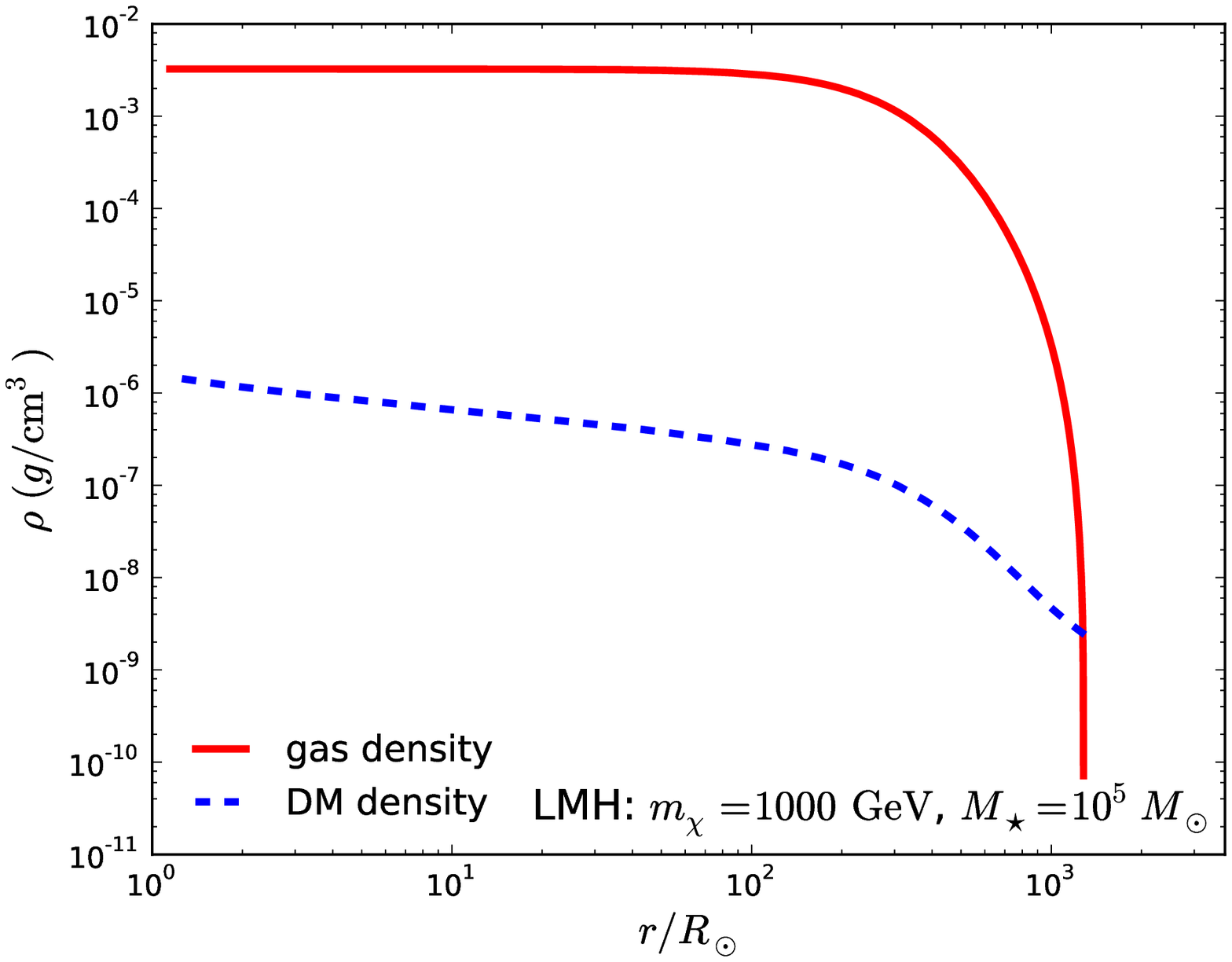}
     \hspace{0.1cm}
    \end{minipage}
 \caption{Profiles of gas and DM density within a DS of $10^5 \,\M$ for different halo environments and DM particle mass, as indicated
 in the legends of the plots. In all cases, the DM density is roughly three orders of magnitude below the gas density.}
 \label{DSdens}
\end{figure*}   

\begin{figure*} 
\begin{minipage}{0.5\linewidth}
     \centering
     \includegraphics[width=7.5cm]{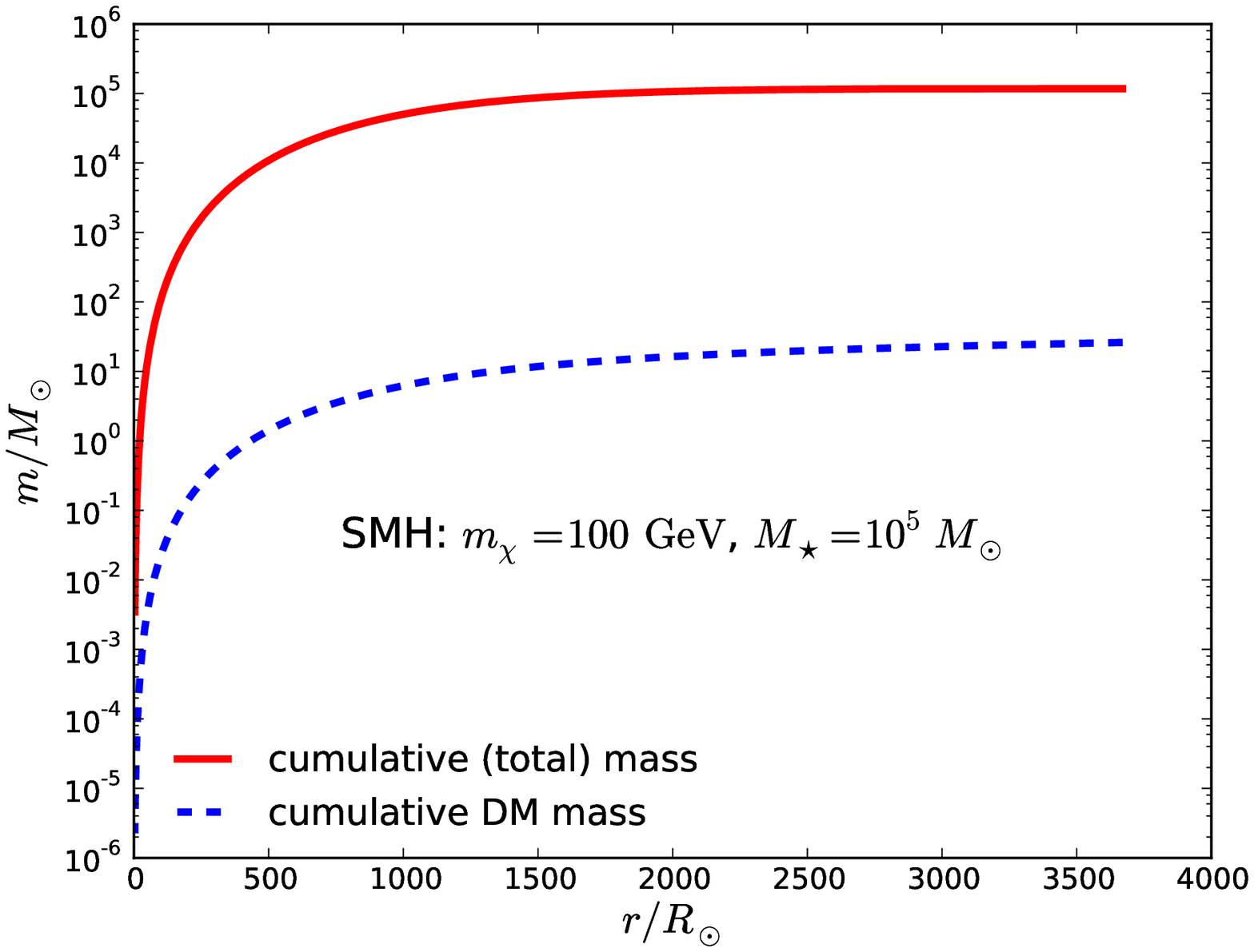}
     \vspace{0.1cm}
    \end{minipage}
    \begin{minipage}{0.5\linewidth}
      \centering\includegraphics[width=7.5cm]{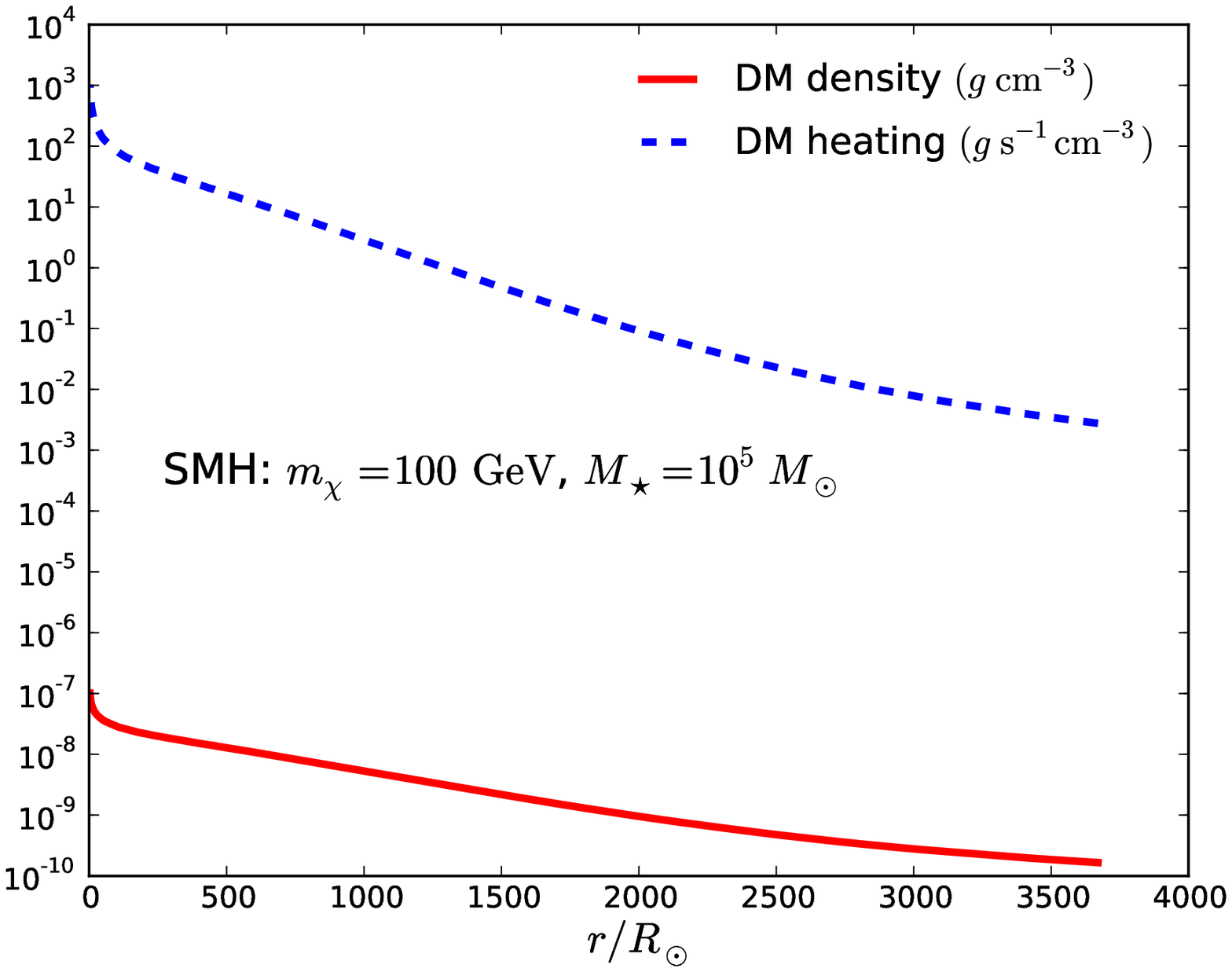}
     \hspace{0.1cm}
    \end{minipage}
 \caption{Cumulative mass profiles (\tx{left-hand plot}) and DM
   heating (\tx{right-hand plot}) for a DS of $10^5 \,\M$, residing
 in SMH and $m_{\chi} = 100$ GeV.}
 \label{DScumul}
\end{figure*}

We also look at the distribution and amount of dark matter
within dark stars. To this end, we compile plots of the gas (i.e. baryonic) and dark matter density profiles for our fiducial
$10^5 \,\M$ DS, for different halo environments and WIMP masses
in Fig.~\ref{DSdens}. The mass density in DM is roughly
three orders of magnitude below the one for the baryonic mass density, showing how subdominant DM is compared to baryonic
matter. The shape of both density profiles as well as their absolute
magnitude agrees excellently with the results in Fig.~3, case 1, of
\cite{Spolyar09}, for DSs solely powered by DM heating (i.e. no fusion included), as is the case for our models.
The shape of our density profiles is independent of the other parameters (halo and WIMP mass), while the densities
are higher for the low-accretion environment (SMH) at fixed WIMP mass, or for higher WIMP mass at fixed halo environment
(in agreement with the plots of the central density in Figs.~\ref{evol}, \ref{evol2}, \ref{evol3}).
As an illustrative corollary to these results, we show the cumulative
mass profiles for SMH and $m_{\chi} = 100$ GeV in the left-hand plot
of Fig.~\ref{DScumul}.
We can see that, for a DS with $10^5 \,\M$, the mass in DM only amounts
to roughly $20 \,\M$, or $0.02 \%$ of the total stellar mass. The right-hand plot
of this same figure, on the other hand, shows how the amount of DM heating, as defined in Eqn.(\ref{DMheat}), 
follows the DM density profile within the DS.

\section{Comparison to polytropic models} \label{evol1}

In this section, we present a detailed comparison of our results obtained using MESA with the polytropic models of \cite{Freese10}.
We first point out that polytropes do indeed provide rather good approximations to true stellar models. However, there are some 
important
differences to previous results that may impact observability. 

First, we examine the case of SMH as defined in Eq.(\ref{eq:defineSMH}).
A comparison of the main stellar characteristics summarized in Table \ref{tab1} with those in Table 1 of \cite{Freese10} 
shows that, in the mass range of $10^4-10^5 ~M_{\odot}$, our DSs are hotter by a factor of $1.41 - 1.51$ than those in \cite{Freese10},
are smaller in radius by a factor of $0.64-0.69$, denser by a factor of $3.00-3.41$, and more luminous by a factor of $1.87-1.96$.
Thus, the overall colors of our DSs are not very different from the previous models, while our luminosities and central densities 
are significantly higher.

Next we examine the case of LMH as defined in Eq.(\ref{eq:LMH}).  
Dark stars in larger minihalos are able to accrete more baryons and dark matter, and hence can grow more massive 
than in small minihalos. A comparison of the main stellar characteristics of our Table \ref{tab2} with Table 3 
from \cite{Freese10} reveals that the change in those parameters is consistent with the previous results for SMH: 
in the mass range of $10^5-10^6~M_{\odot}$, our DSs are hotter by a factor of $1.45 - 1.67$ than 
those in \cite{Freese10}, are smaller in radius by a factor of $0.60-0.62$, denser by a factor of $3.46 - 4.85$, 
and more luminous by a factor of $1.98 - 2.19$.

\begin{figure*}[t]
\begin{minipage}{0.5\linewidth}
     \centering
     \includegraphics[width=7.5cm]{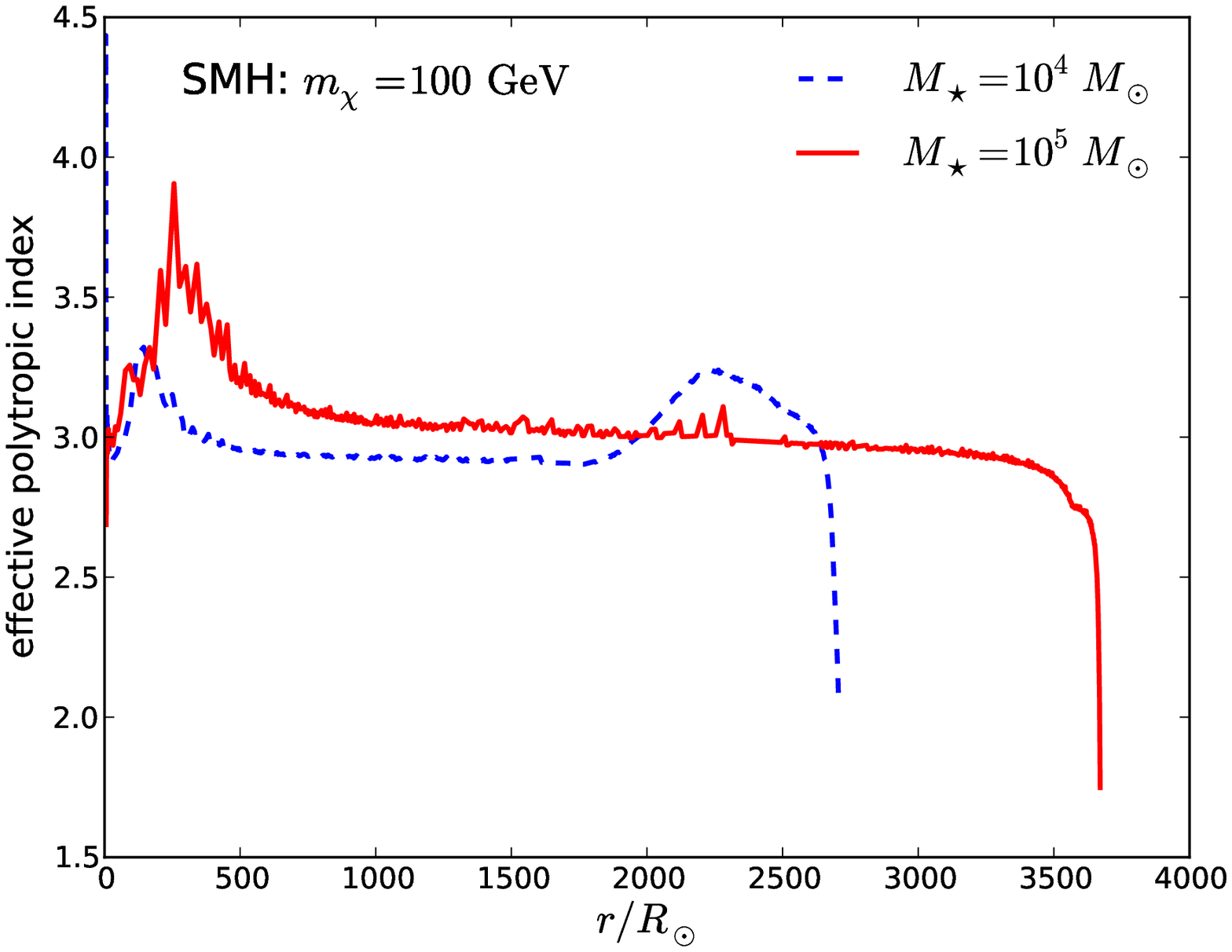}
     \vspace{0.05cm}
    \end{minipage}
    \begin{minipage}{0.5\linewidth}
      \centering\includegraphics[width=7.5cm]{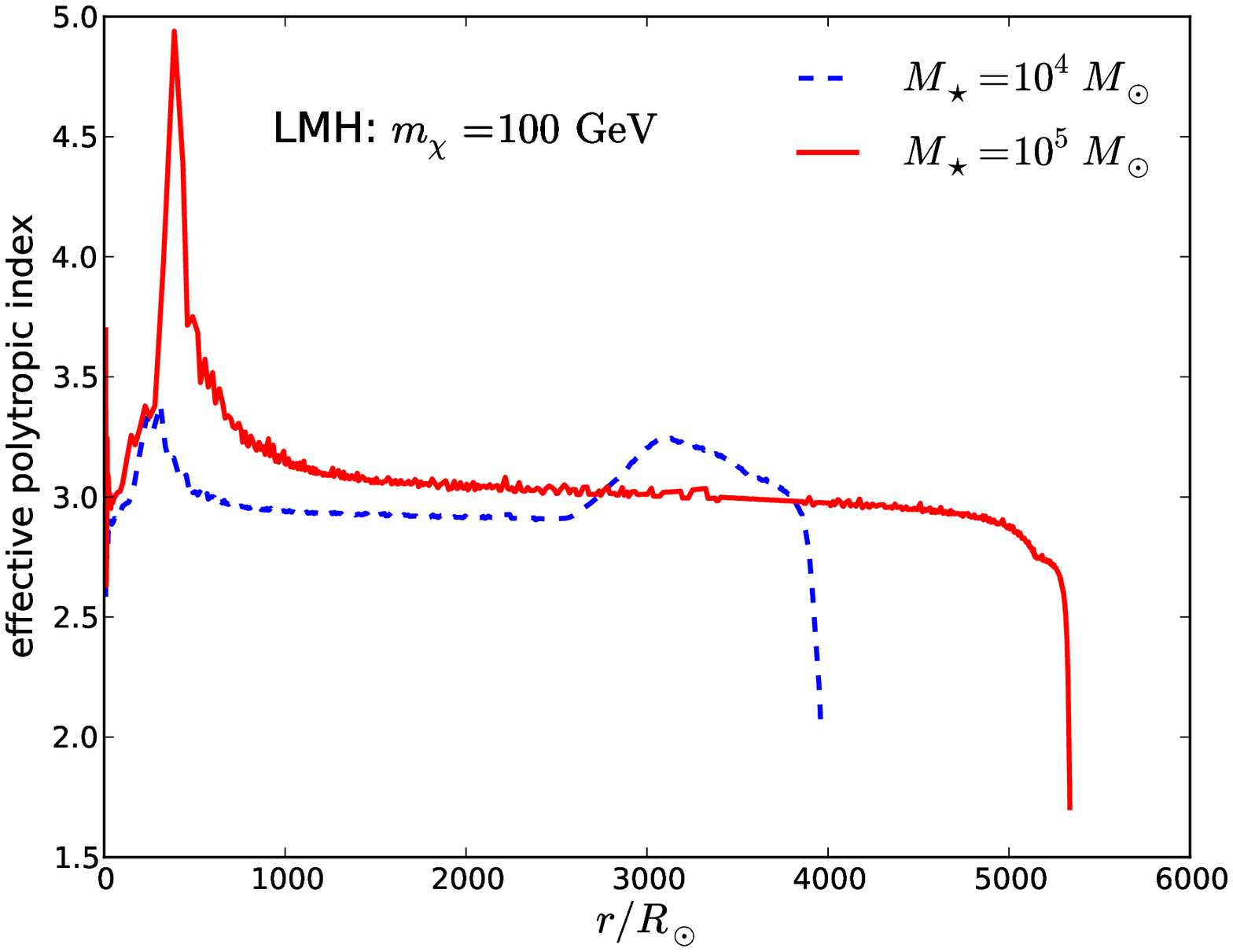}
     \hspace{0.05cm}
    \end{minipage}
\begin{minipage}{0.5\linewidth}
     \centering
     \includegraphics[width=7.5cm]{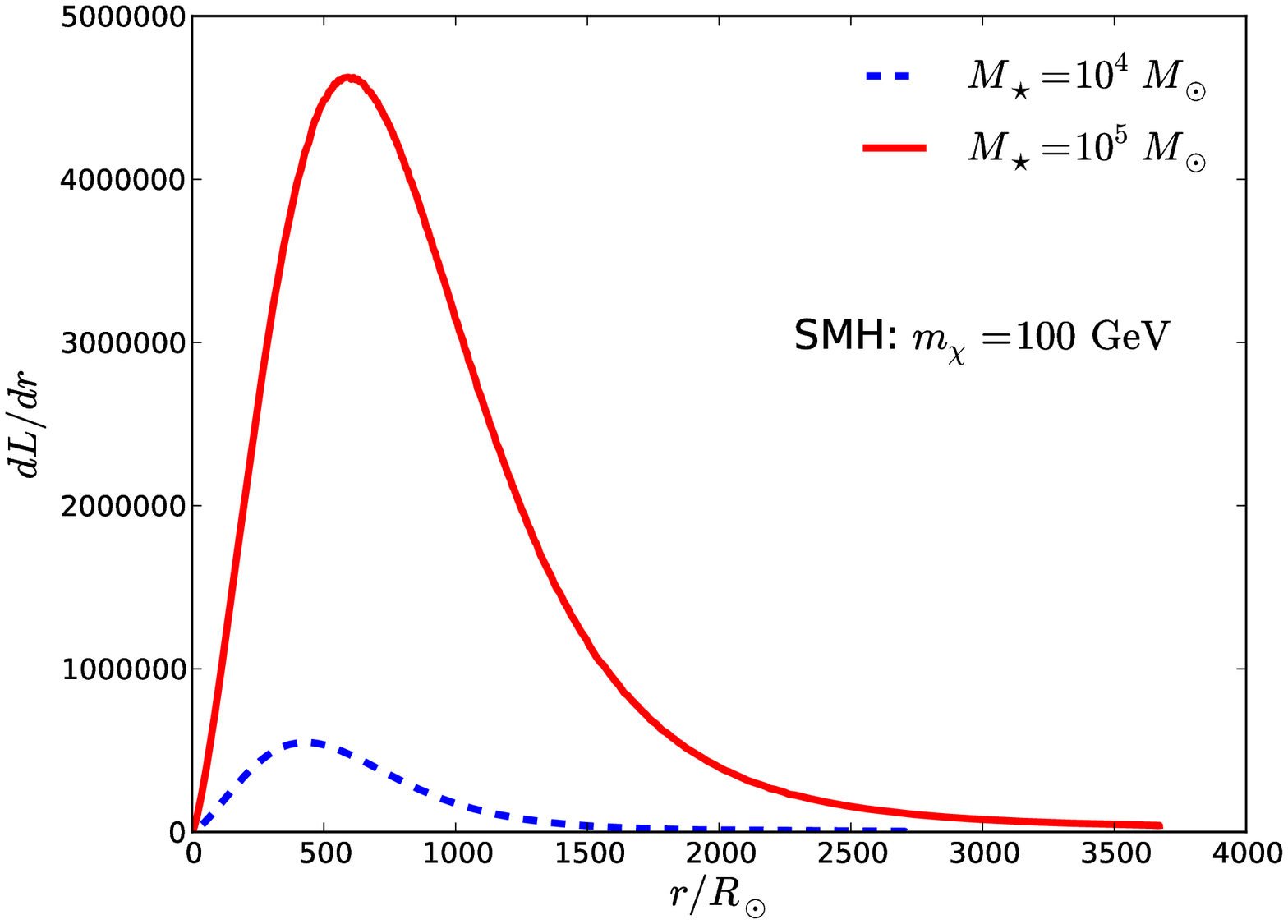}
     \vspace{0.05cm}
    \end{minipage}
    \begin{minipage}{0.5\linewidth}
      \centering\includegraphics[width=7.5cm]{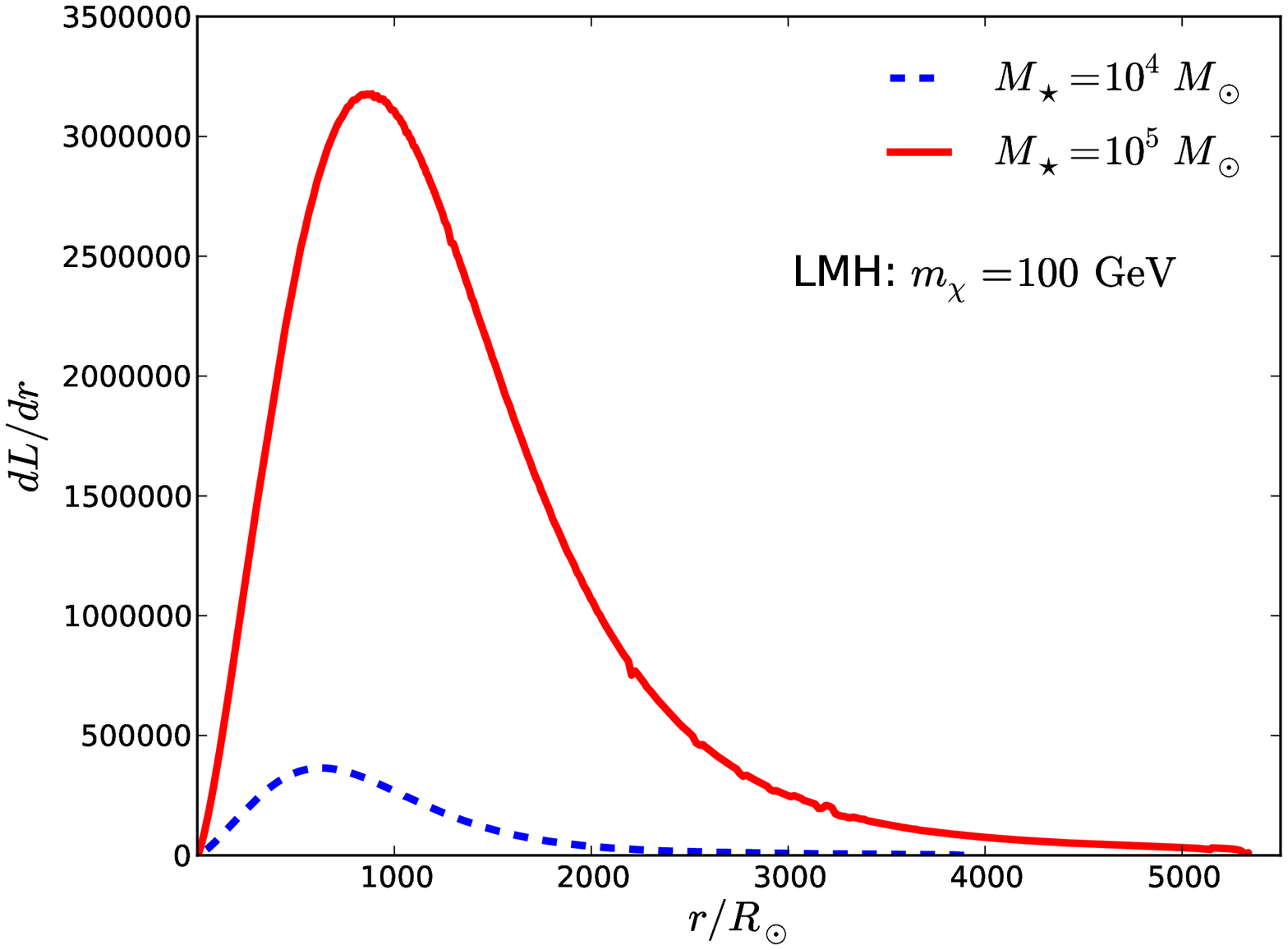}
     \hspace{0.05cm}
    \end{minipage}
    \begin{minipage}{0.5\linewidth}
     \centering
     \includegraphics[width=7.5cm]{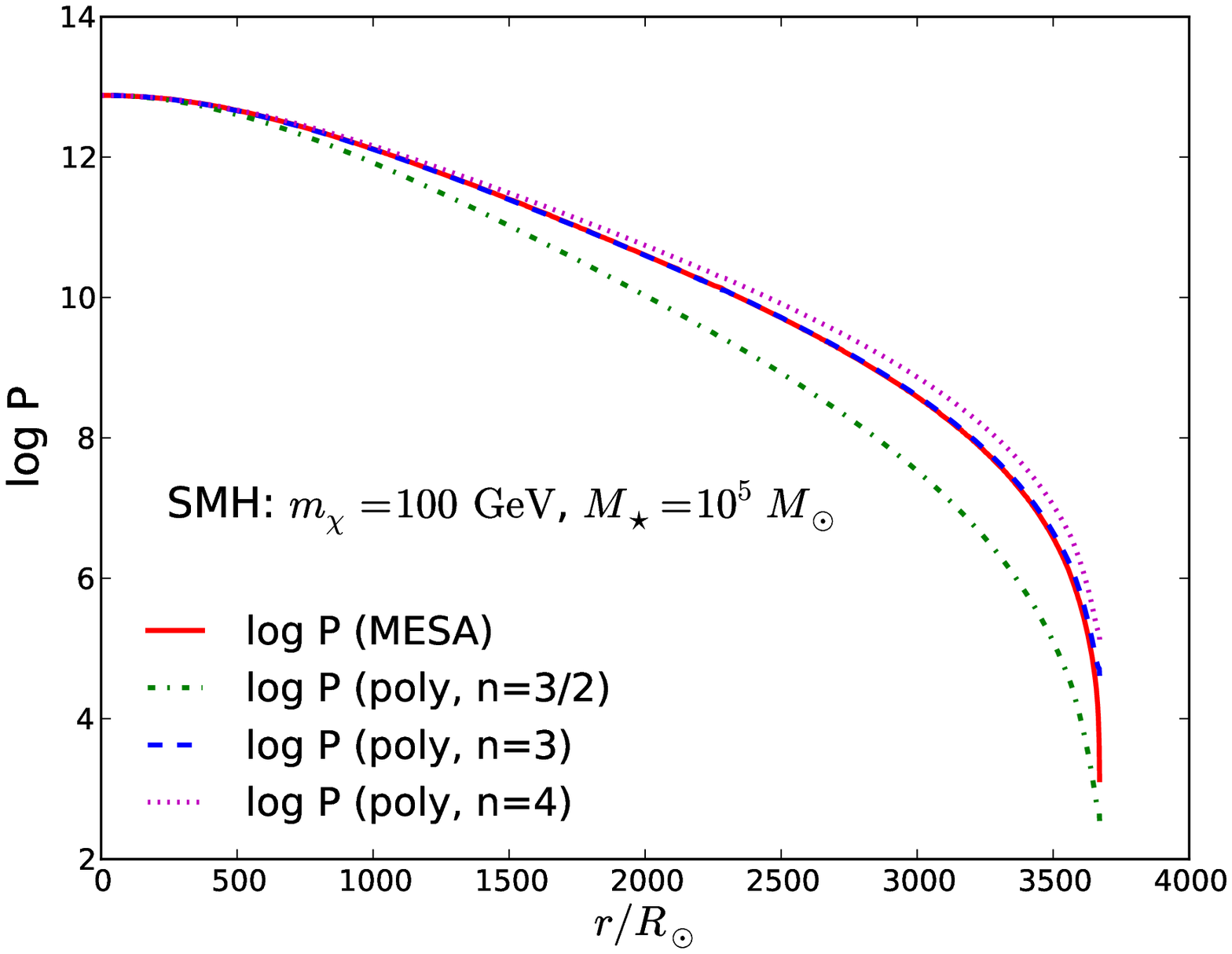}
     \vspace{0.05cm}
    \end{minipage}
    \begin{minipage}{0.5\linewidth}
      \centering\includegraphics[width=7.5cm]{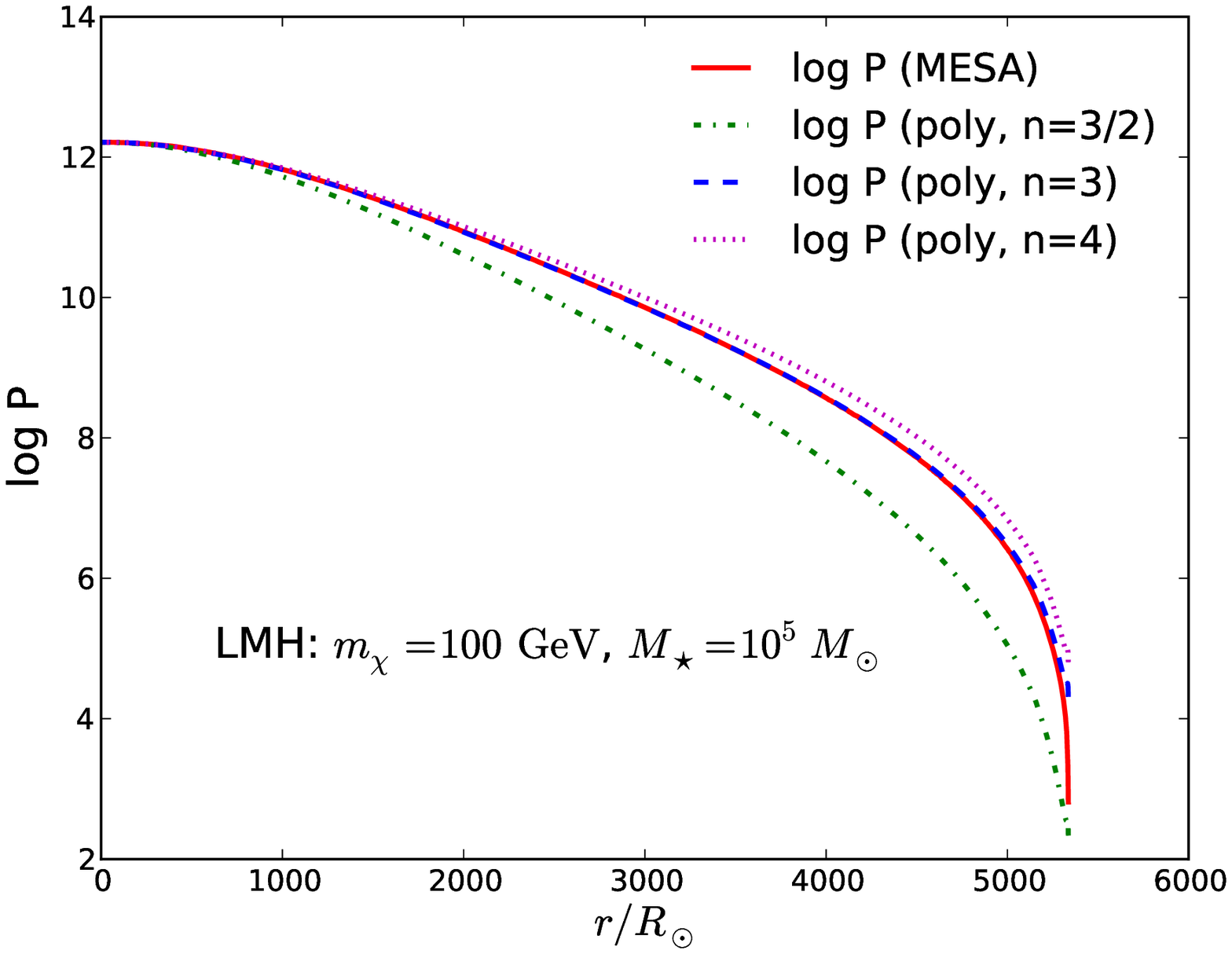}
     \hspace{0.05cm}
    \end{minipage}
 \caption{\tx{Top row:} Effective polytropic index of MESA models, according to Equ.(\ref{neff}), as a function of radius.
 The sharp drop at high $r/R_{\odot}$ corresponds to the surface of the star. The spikes close to the center are caused by large luminosity 
 gradients, see middle row. \tx{Middle row:} Luminosity gradient due to dark matter heating, as a function of radius. 
 \tx{Bottom row:} Dark star pressure as a function
 of radius: comparison of MESA's results with polytropes of index $n = 3/2$, $n=3$ and $n=4$, assuming the same central pressure
 and density. The respective halo environments (SMH, LMH) and DS masses ($10^4, 10^5 M_{\sun}$) are indicated in the legends of the 
 plots. In each case, the DM particle mass is $m_{\chi} = 100$ GeV. We can see that supermassive dark stars
 can be very well approximated by $(n=3)$-polytropes. The same plots for different WIMP masses can be found in
 Appendix A, Fig.\ref{polycomp2} and \ref{polycomp3}.}
 \label{polycomp}
\end{figure*}

A closer examination of the stellar structure reveals further
interesting comparison between the results of MESA and those assuming
polytropes of \cite{Freese10}. First, we focus on the pressure
distribution within the star. Assuming a polytropic law, $P/P_c =
(\rho/\rho_c)^{1+1/n_{\rm{eff}}}$ with the central pressure $P_c$ and
central density $\rho_c$, we solve for $n_{\rm{eff}}$,
\beq \label{neff} n_{\rm{eff}} = \left[\f{\log (P/P_c)}{\log
    (\rho/\rho_c)} - 1 \right]^{-1}.  \eeq In the case of exact
polytropic models, $n_{\rm{eff}}$ is simply the usual polytropic
index, ranging between $n=0,..,5$.  Radiative stars are well
approximated by ($n$=3)-polytropes, and dark stars more massive than a
few hundred solar masses have been found to follow ($n$=3)-polytropes
to a good extent for much of their stellar interior
\cite[see][]{Spolyar09,Freese10}. It is thus instructive to plot
relationship (\ref{neff}) for our models, i.e. for $P(r), P_c,
  \rho(r), \rho_c$, as calculated by MESA. This way, we can
  see where departures from a polytropic law are most pronounced, and
  which parts of the interior of a given supermassive DS are well
  approximated by ($n$=3)-polytropes.  For $m_{\chi} = 100$ GeV,
those results can be found in the top row of Figure~\ref{polycomp}.
We can see that for supermassive DSs the effective polytropic index
is, indeed, remarkably close to $n=3$ for most of the stellar
interior, in accordance with \cite{Freese10}. Again, the ``bump''
close to the surface for the $10^4 \,\M$ model is caused by the high
superadiabaticity gradients, which develop at masses above $100 \,\M$.
Without using MLT++, this bump would have been more pronounced
(see also Sec.3).
The impact of superadiabaticity decreases again for higher stellar masses, as is evident in the
curve for the $10^5 \,\M$ model.  In summary, in the mass range 100 -- $1000\,\M$,
very inefficient convection with large superadiabatic gradients
develop in the envelopes of these models.
While the effective polytropic index of low-mass DSs with about
$10-20\, \M$ is close to $n_{\rm{eff}}=3/2$, this value steadily
increases to above $n_{\rm{eff}}=2$ for more than $100 \,\M$. The
numerical signature of our MLT++ prescription limiting the
superadiabaticity happens when $L_c \sim L_{\rm{rad}}$, while
$n_{\rm{eff}}$ continues to approach the value of $3$ for increasing
DS mass. Then, $L_{\rm{rad}}$ becomes increasingly important.  The
corresponding proximity to ($n$=3)-polytropes is further illustrated
in the bottom row of Figure~\ref{polycomp}, where we compare the
radial run of the DS total pressure of our MESA results with
polytropes of indices $n = 3/2$, $n=3$ and $n=4$. Except for close to
the surface, the pressure of our MESA models lies basically on top of
the ($n$=3)-polytrope, showing again that supermassive DSs can be very
well approximated by ($n$=3)-polytropes. We note that also in
\cite{Freese10}, an interpolation between $n=1.5$ and $n=3$ was
necessary for DSs beyond a few $100 \,\M$. It is remarkable that we find
the same behavior using MESA, given its capability to draw on
elaborate equation-of-state tables.
 
Focusing again on the top row of Figure~\ref{polycomp}, we can see
that the curve for the $10^5 \,\M$ model has a ``spike'' near the
center.  This deviation from a clean polytropic model is caused by the
substantial release of heat due to the DM annihilation. In the middle
row of Figure~\ref{polycomp}, we plot the differential change in
luminosity, as a function of radius, i.e.  \beq \f{dL}{dr} = 4\pi r^2
f_Q \hat Q_{DM} \eeq (see Eq.(\ref{DMheating})).  One can see that the
``spike'' in the effective polytropic index in the top panel coincides
with the location of the maximal change in luminosity in the middle
panel of the figure.  The absolute values of $dL/dr$ increase with
increasing WIMP mass, since DSs are smaller and hence denser for
larger WIMP mass (see also Figs.~\ref{HR} and \ref{DSdens}). The
higher DM densities boost $dL/dr$, in turn.


In order to illustrate the robustness of the above results, we also show in Appendix A the corresponding plots for the cases of $m_{\chi} = 10$ GeV and 
$m_{\chi} = 1000$ GeV in Figs.~\ref{polycomp2} and \ref{polycomp3}, respectively.

\newpage 

\section{Adiabatic pulsation periods}

An interesting question of dark star astrophysics is the possibility
of pulsations. If DSs were found to pulsate, this could represent yet
another observational distinction to other objects at high redshifts,
such as the first galaxies and quasars.  
In this section, we only examine the representative range of
  pulsation frequencies of our DS models; we do not consider the more
  detailed question of the driving and damping of these pulsation
  modes. This allows us to treat the pulsations adiabatically, which
  greatly simplifies the calculation. The nonadiabatic calculation
  of the driving and damping of modes in these models
  will be presented in a forthcoming paper.  

  While supermassive DSs are mostly dominated by radiative transfer,
  as described in Section 3 and 4, we can see from Fig.~\ref{lum1}
  that $L_{c}(r)/L(r) \gtrsim 0.1$, i.e. convection is not completely
  absent in supermassive DSs. Equivalently, looking at the square of
  the Brunt-V\"ais\"al\"a frequency defined as \beq N^2 =
  \f{g}{r}\left[\f{1}{\Gamma_1}\f{d \ln P}{d \ln r} - \f{d \ln \rho}{d
      \ln r}\right], \eeq with $\Gamma_1 \equiv (d \ln P/d \ln
  \rho)_{\rm{ad}}$, we find that it is negative throughout most of the
  DS interior, implying convective instability. We therefore do not
  expect to find gravity modes (``g-modes'') in supermassive DSs.

\begin{figure*}
  \begin{minipage}{0.5\linewidth}
     \centering
     \includegraphics[width=7.5cm]{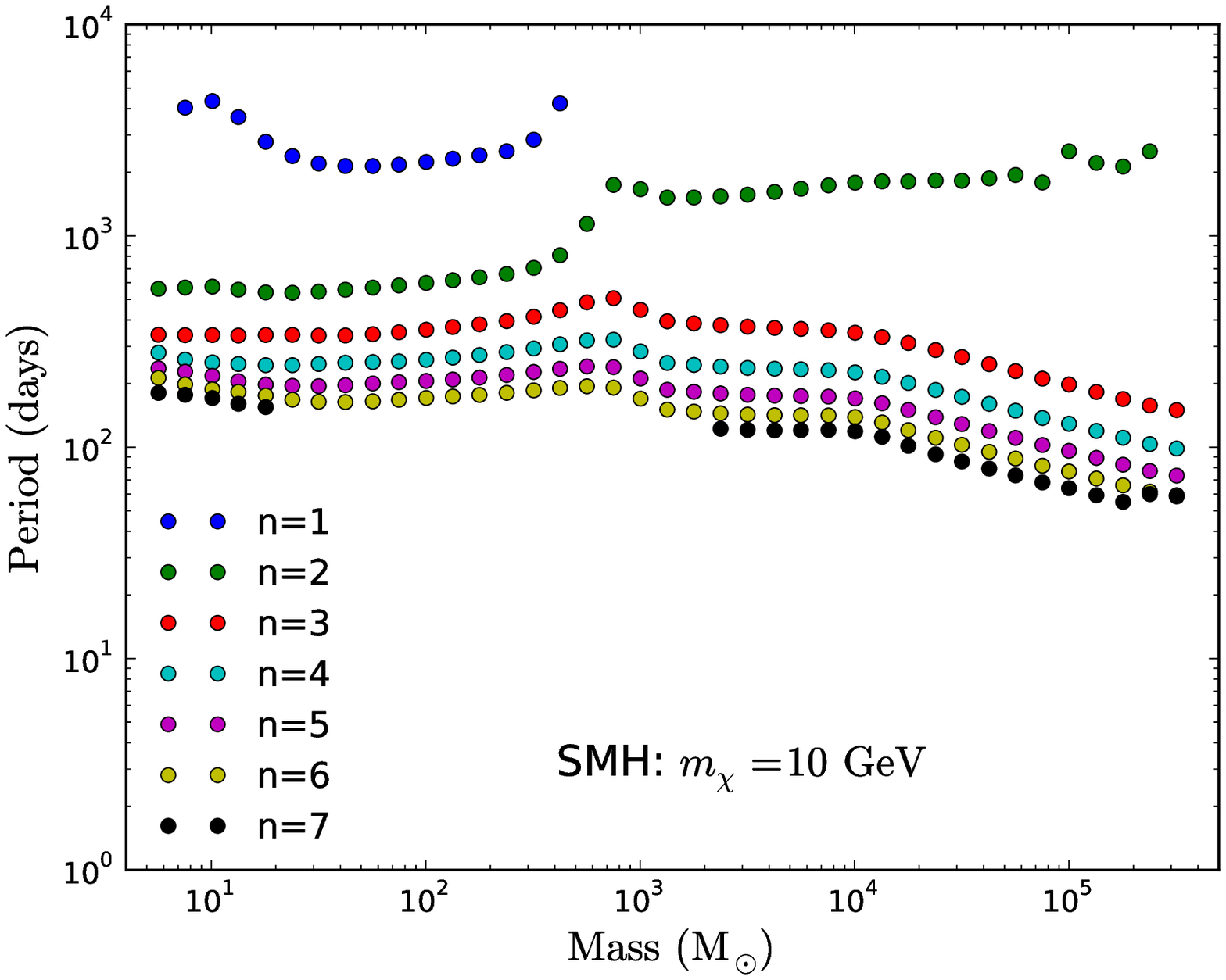}
     \vspace{0.05cm}
    \end{minipage}
    \begin{minipage}{0.5\linewidth}
      \centering\includegraphics[width=7.5cm]{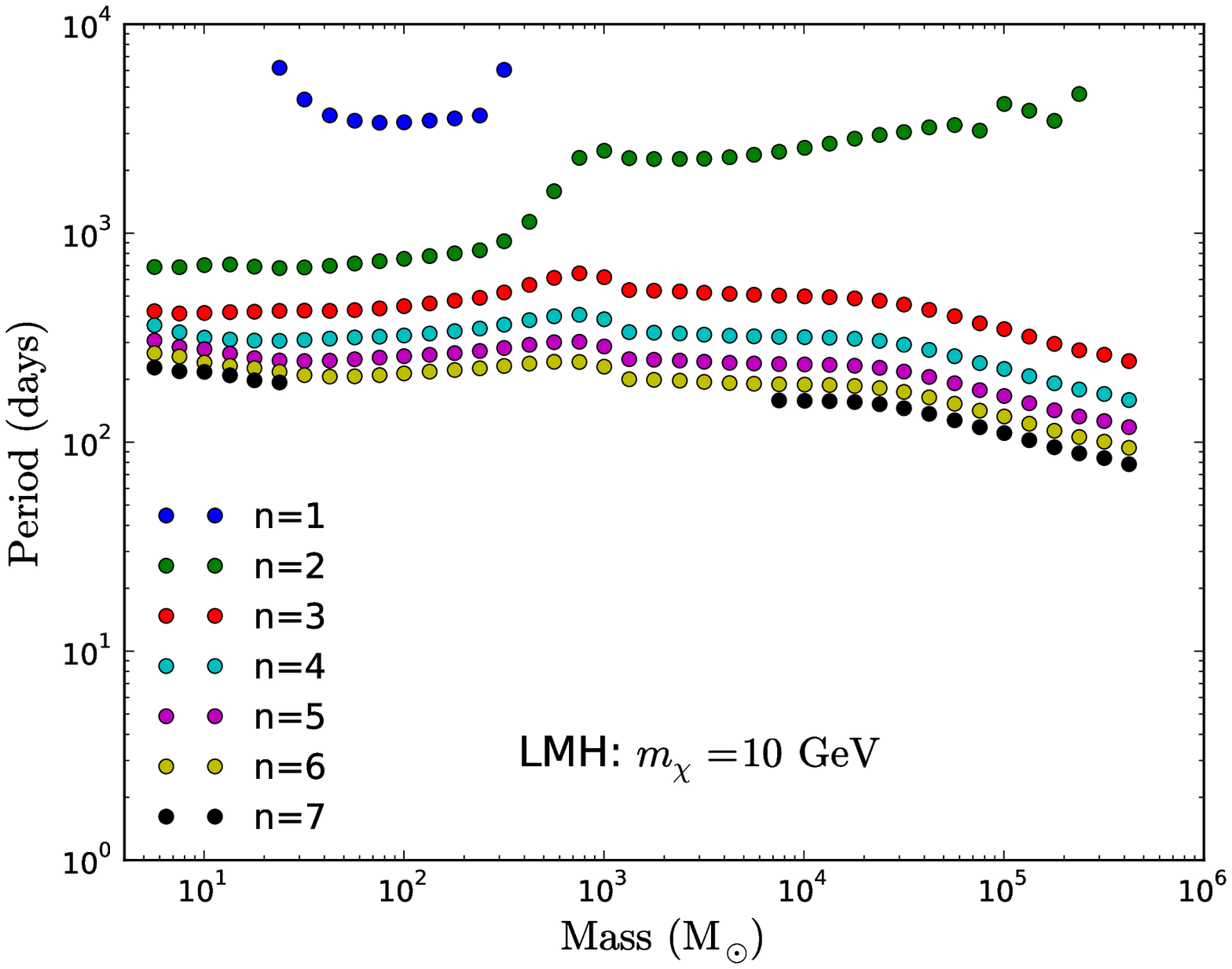}
     \hspace{0.05cm}
    \end{minipage}
 \begin{minipage}{0.5\linewidth}
     \centering
     \includegraphics[width=7.5cm]{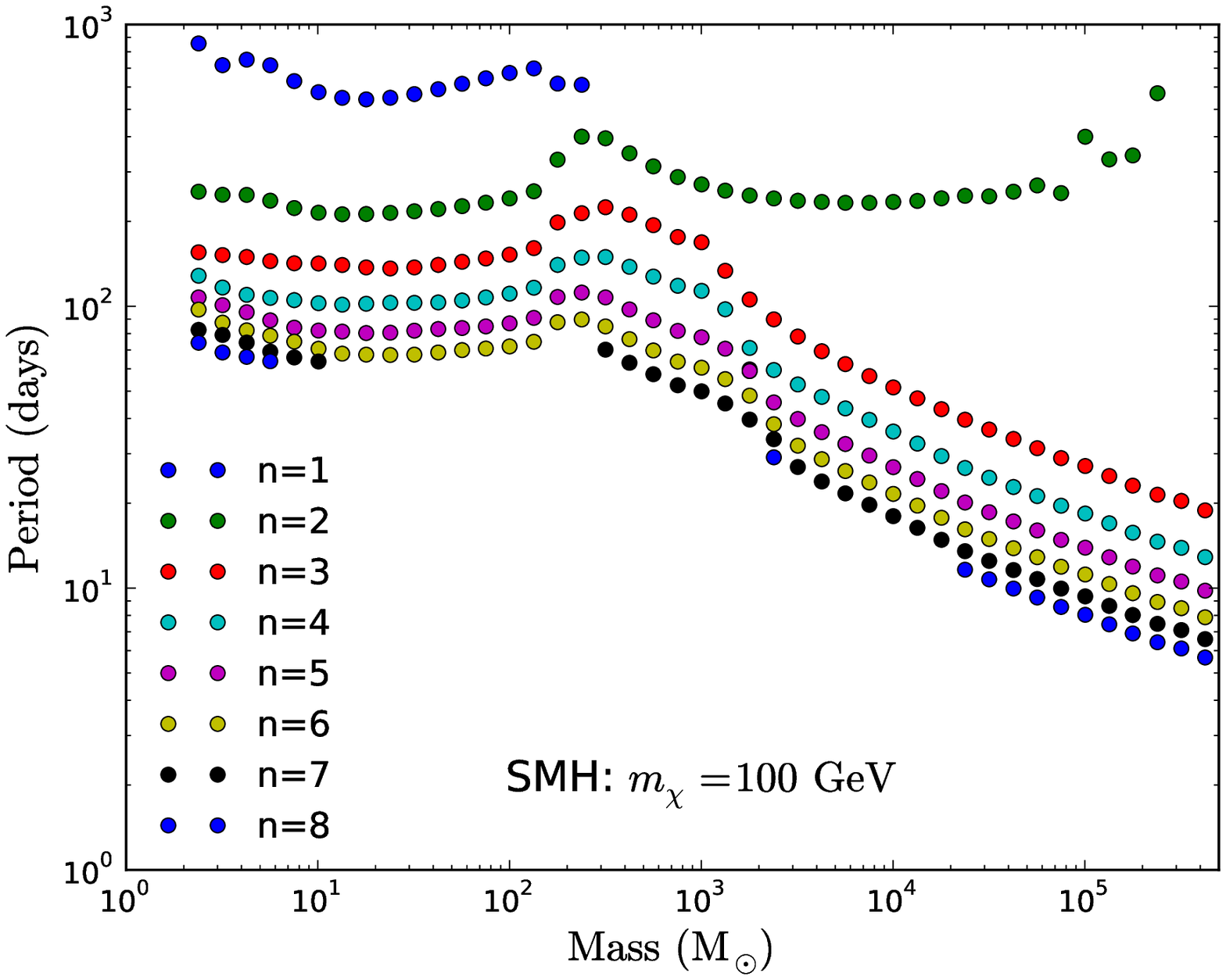}
     \vspace{0.05cm}
    \end{minipage}
    \begin{minipage}{0.5\linewidth}
      \centering\includegraphics[width=7.5cm]{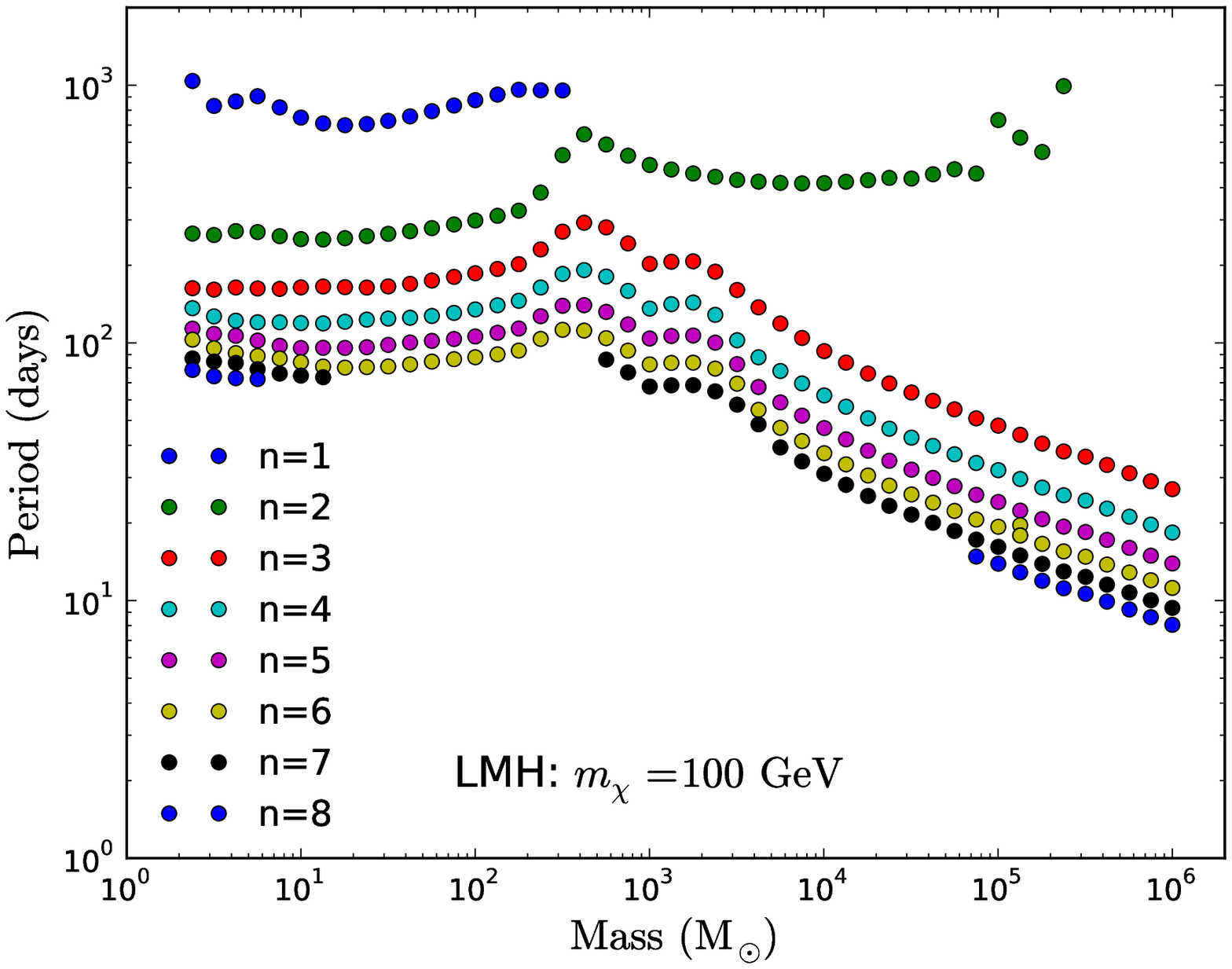}
     \hspace{0.05cm}
    \end{minipage}
    \begin{minipage}{0.5\linewidth}
     \centering
     \includegraphics[width=7.5cm]{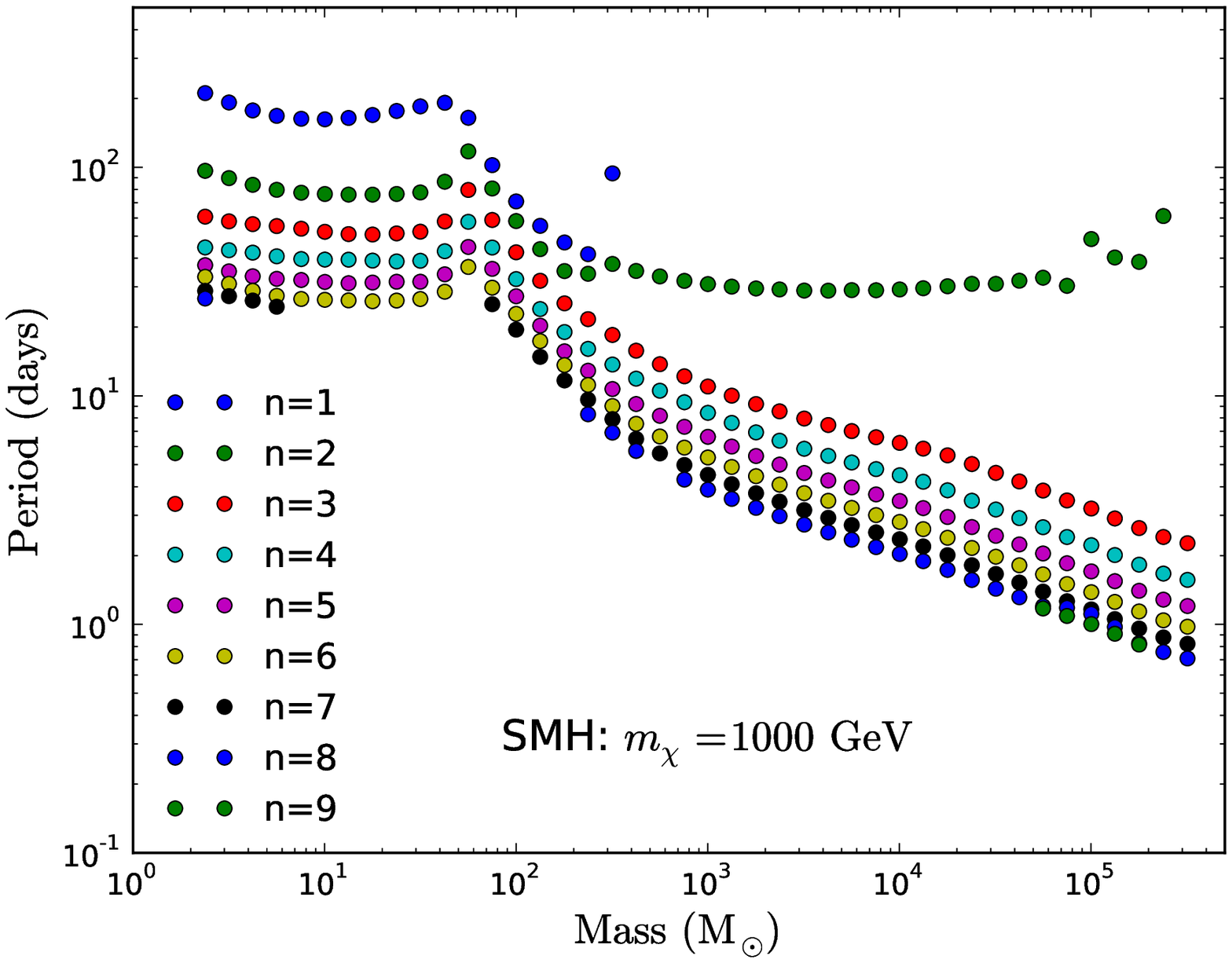}
     \vspace{0.05cm}
    \end{minipage}
    \begin{minipage}{0.5\linewidth}
      \centering\includegraphics[width=7.5cm]{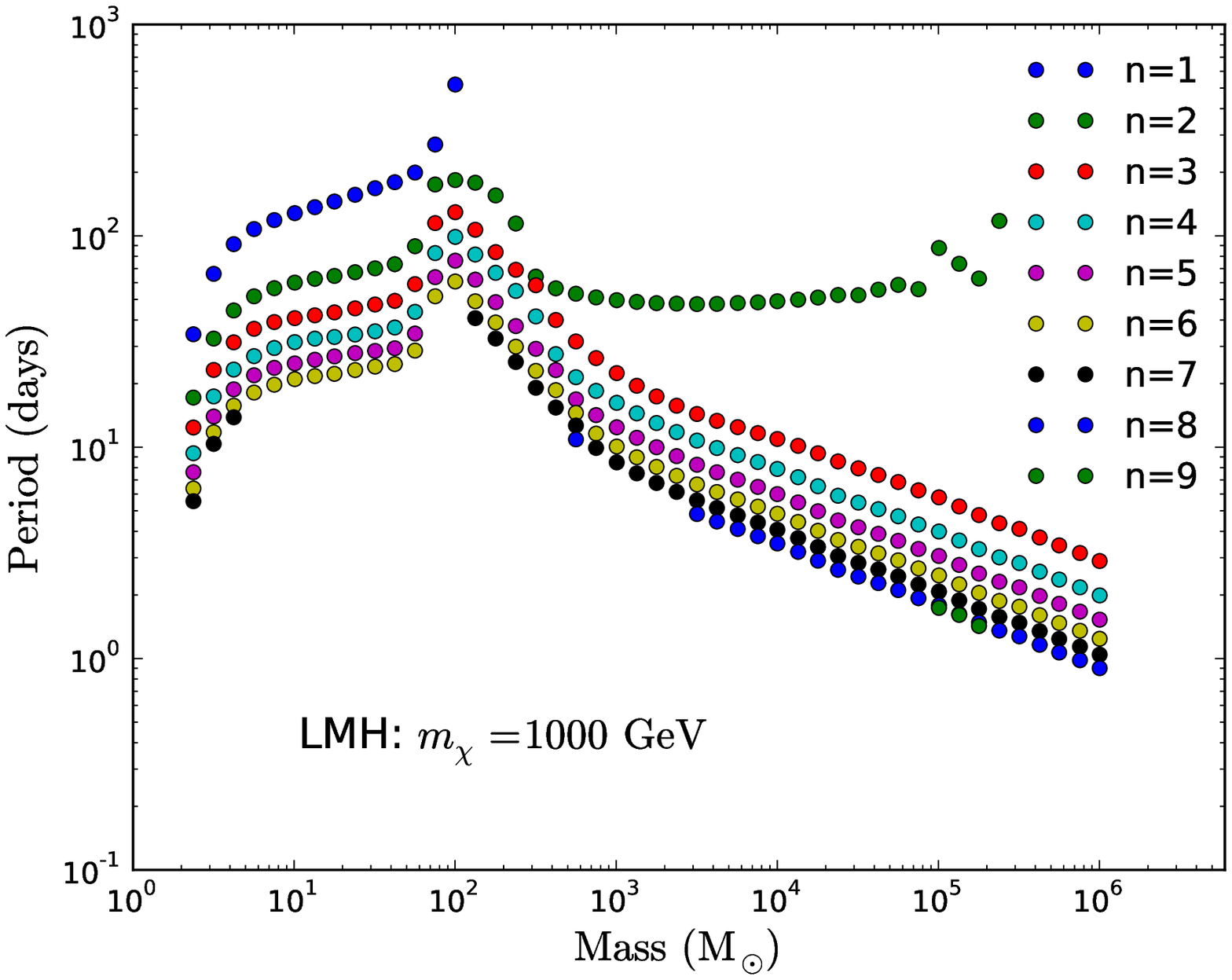}
     \hspace{0.05cm}
    \end{minipage}
 \caption{Radial, adiabatic pulsation periods as a function of DS mass for models with
 $m_{\chi} = 10, 100, 1000$ GeV (\tx{top, middle, bottom row}). Halo environments are indicated in the
 legends of the plots. The curves are for different overtone number, from $n=1$ (upper-most curve; the fundamental radial oscillation)
 to the highest overtone number in the respective plots (lower-most curve; the dots for $n=9$ lie close to the ones for $n=8$
 in the bottom panel). 
 The sharp feature at $\sim 100 \,M_\odot$ is a signature of the onset of
 superadiabaticity in the envelope, as discussed in the text. }
 \label{freqs}
\end{figure*}

On the other hand, acoustic modes, or p-modes, could be present in DSs. We calculated the adiabatic pulsation periods of radial
modes (i.e. those for which $l=0$) with different overtone number $n$,
where $n=1$ is the fundamental (``breathing'') mode, and
$n > 1$ are higher overtone modes\footnote{The pulsations were calculated using the MESA implementation of
the ADIPLS code. We note that ADIPLS defines the order of the mode such that the lowest-order radial oscillation
has order $n=1$, contrary to the commonly used convention of assigning order $0$ to the fundamental radial oscillation.
For $p$-modes, $n>0$.}. Figure~\ref{freqs} shows the rest-frame pulsation periods for our dark star models, covering the
whole range of $\sim 10 - 10^6 M_{\sun}$. We see that the $n=1$ modes
disappear above DS masses of around $100 ~\M$. We believe this 
is due to the change in energy transport from convection to
radiation-domination above that mass range. Similarly, we note the change in slope in the 
plots for the periods around that same mass, again caused by the sharp
transition to superadiabaticity and its suppression, which was also
responsible for the ``bump''
in radius in Figure~\ref{evol3}; as a reminder, this suppression is
put in by hand to limit the growth of the superadiabatic 
gradients. As before, the transition is milder for small WIMP mass. As
expected,
the periods are much shorter for higher overtone number $n$. We also see that the periods are shorter for higher 
WIMP mass: while periods for high $n$ can span a range of 60--400 days for the 10 GeV case, they are of order of days or less than a day
for the 1000 GeV case, in the rest frame. Also, more modes with high $n$ are excited for large WIMP mass. 

In Table \ref{tab3} and Table \ref{tab4}, we give a more detailed list of the corresponding pulsation 
frequencies and periods for supermassive DSs in the range $10^4-10^6
\,\M$ for the $m_{\chi} = 100$ GeV WIMP case. 
In this case, the pulsation periods lie in a range
between 8 days and more than 2 years in the rest frame of the DSs.  
When converting to the observer's frame, one needs to
multiply the periods by a factor of $(1+z_{\star})$, where $z_{\star}$ denotes the redshift at which the DS under consideration
has acquired its final mass. The time frame for this can vary
tremendously, depending on the accretion rate (see
left-hand upper plot in Figs.~\ref{evol}, \ref{evol2} and \ref{evol3},
respectively). This is independent of WIMP mass, however. A DS with
$10^5 \,\M$ has an
age of about $10^8$ yrs in a low-accretion rate environment (SMH), in contrast to an age of about $10^6$ yrs in the high-accretion
rate environment (LMH). For our adopted $\Lambda$CDM cosmology,
this corresponds to redshifts of $z_{\star} \simeq 14.82$ (SMH) and $z_{\star} \simeq 14.95$ (LMH), respectively
(compare these to the halo formation redshifts of $z=20$ (SMH) and $z=15$ (LMH)). We include the converted periods for the case of
a $10^5 \,\M$ DS in the tables, as well.
The shortest periods to be expected
in the observer's frame are given by the 1000 GeV case, amounting to less than about 
50 days, for modes with $n>6$. 

Work is in progress to study the pulsations of these objects in more
detail.  In particular, we defer the nonadiabatic calculation of the
possible driving and damping mechanisms for the pulsation modes to a
future paper. Preliminary results suggest that the traditional
$\kappa-\gamma$ mechanism could operate in these stars \citep[see
e.g.][]{Unno89}.  A further source of driving could come from the dark
matter itself.  As the dark star undergoes small perturbations, local
changes in its baryonic density could lead to local changes in the
dark matter density. This in turn would modulate the local dark matter
heating rate. Depending on the size and relative phasing of these
effects, this could be a source of driving for the pulsations. Of
course, a much more quantitative approach is needed to assess the
viability of this mechanism.  We will study this and other
possibilities in future work.

\begin{table}
\caption{Radial ($l=0$) pulsation periods with overtone number $n$, 
for dark stars of different mass with $m_{\chi} = 100$ GeV and 
$\dot M = 10^{-3} M_{\sun}/yr$ (SMH); periods are in the rest frame of the respective DS, unless otherwise indicated} 
\label{tab3}
\begin{center}
\hspace{0.02cm}
$M_{\star} = 10^4 ~M_{\sun}$: \\
$T_{\rm{eff}} = 15228.50$ K, $L_{\star} = 3.423 \cdot 10^8 ~L_{\sun}$, $R_{\star} = 2660.48 ~R_{\sun}$
\hspace{0.02cm}
\begin{tabular}{r|r|r|r}

          & $n$  & $f~ [\mu \rm{Hz}]$ & periods [days]  \\
\hline

          & $2$        & $0.04932$   & $234.65$    \\   
          & $3$        & $0.22455$   & $51.54$    \\
          & $4$        & $0.32198$   & $35.94$   \\
          & $5$        & $0.43076$   & $26.87$   \\
          & $6$        & $0.53603$   & $21.59$   \\
          & $7$        & $0.64288$   & $18.00$   \\
\hline
\end{tabular}
\end{center}
\begin{center}
\hspace{0.02cm}
$M_{\star} = 10^5 ~M_{\sun}$: \\
$T_{\rm{eff}} = 24463.62$ K, $L_{\star} = 4.150 \cdot 10^9 ~L_{\sun}$, $R_{\star} = 3589.72 ~R_{\sun}$
\hspace{0.02cm}
\begin{tabular}{r|r|r|r|r}

          & $n$  & $f~ [\mu \rm{Hz}]$ & periods [days] & observer's frame [days]   \\
\hline

          & $2$        & $0.02896$   & $399.69$        & $6323.10$      \\   
          & $3$        & $0.42679$   & $27.12$         & $429.04$      \\
          & $4$        & $0.62926$   & $18.39$         & $290.93$      \\
          & $5$        & $0.83194$   & $13.91$         & $220.06$      \\
          & $6$        & $1.03558$   & $11.18$         & $176.87$     \\
          & $7$        & $1.23825$   & $9.35$          & $147.92$      \\
          & $8$        & $1.43942$   & $8.04$          & $127.19$     \\
\hline
\end{tabular}
\end{center}
\end{table}

\begin{table}
\caption{Radial ($l=0$) pulsation periods with overtone number $n$, 
for dark stars of different mass with $m_{\chi} = 100$ GeV and 
$\dot M = 10^{-1} M_{\sun}/yr$ (LMH); periods are in the rest frame of the respective DS,
unless otherwise indicated} 
\label{tab4}
\begin{center}
\hspace{0.02cm}
$M_{\star} = 10^4 ~M_{\sun}$: \\
$T_{\rm{eff}} = 12538.20$ K, $L_{\star} = 3.408 \cdot 10^8 ~L_{\sun}$, $R_{\star} = 3916.37 ~R_{\sun}$
\hspace{0.02cm}
\begin{tabular}{r|r|r|r}

          & $n$  & $f~ [\mu \rm{Hz}]$ & periods [days]  \\
\hline

          & $2$        & $0.02776$   & $416.88$    \\   
          & $3$        & $0.12461$   & $92.88$    \\
          & $4$        & $0.18514$   & $62.51$   \\
          & $5$        & $0.24784$   & $46.70$   \\
          & $6$        & $0.31034$   & $37.29$   \\
          & $7$        & $0.37280$   & $31.05$   \\
\hline
\end{tabular}
\end{center}
\begin{center}
\hspace{0.02cm}
$M_{\star} = 10^5 ~M_{\sun}$: \\
$T_{\rm{eff}} = 20311.98$ K, $L_{\star} = 4.169 \cdot 10^9 ~L_{\sun}$, $R_{\star} = 5218.99 ~R_{\sun}$
\hspace{0.02cm}
\begin{tabular}{r|r|r|r|r}

          & $n$  & $f~ [\mu \rm{Hz}]$ & periods [days]     & observer's frame [days]  \\
\hline

          & $2$        & $0.01581$   & $732.13$            & $11677.47$     \\   
          & $3$        & $0.24257$   & $47.71$             & $760.97$       \\
          & $4$        & $0.36104$   & $32.06$             & $511.36$      \\
          & $5$        & $0.47971$   & $24.13$             & $384.87$       \\
          & $6$        & $0.59866$   & $19.33$             & $308.31$       \\
          & $7$        & $0.71624$   & $16.16$             & $257.75$        \\
          & $8$        & $0.83278$   & $13.90$             & $221.70$        \\
\hline
\end{tabular}
\end{center}
\begin{center}
\hspace{0.02cm}
$M_{\star} = 10^6 ~M_{\sun}$: \\
$T_{\rm{eff}} = 30976.24$ K, $L_{\star} = 4.860 \cdot 10^{10} ~L_{\sun}$, $R_{\star} = 7661.68 ~R_{\sun}$
\hspace{0.02cm}
\begin{tabular}{r|r|r|r}

          & $n$  & $f~ [\mu \rm{Hz}]$ & periods [days]  \\
\hline
  
          & $3$        & $0.42777$   & $27.06$    \\
          & $4$        & $0.63034$   & $18.36$   \\
          & $5$        & $0.83106$   & $13.93$   \\
          & $6$        & $1.03377$   & $11.20$   \\
          & $7$        & $1.23580$   & $9.36$   \\
          & $8$        & $1.43755$   & $8.05$    \\
\hline
\end{tabular}
\end{center}
\end{table}

\section{Conclusions}

The bulk of this paper has been devoted to studying the properties of dark stars using MESA, 
a fully-fledged 1D stellar evolution code which allows us to solve the stellar structure equations self-consistently, without any a priori assumptions
on the equation of state, or other stellar characteristics.
We were quite surprised how well the previous results using polytropes match the more accurate results using MESA.
We have seen that supermassive DSs are extended, fluffy and cool
objects, and in contrast to ``normal'' stars on the red giant branch,
their low-density ``envelopes'' do not host an ultra-dense core. In fact, as we found, supermassive DSs can be very well approximated
by ($n$=3)-polytropes, so their ratio of central to average density is not much different from a factor of about 54.  

However, there are some differences between the results of MESA and previous polytropic models
 in the details, with positive implications for observability of dark stars.   We found that, 
 in the mass range of $10^4 -10^5 M_\odot$, our DSs are hotter by a factor of 1.5 than those in \cite{Freese10}, 
 are smaller in radius by a factor of 0.6, denser by a factor of 3 -- 4, and more luminous by a factor of 2.   This increased luminosity should of course
 help in searches for dark stars using the James Webb Space Telescope (see \cite{Zack10a, Zack10b}).
 
We also performed a first study of dark star pulsations.
While g-modes are excluded by the presence of convection in these
models, radial and p-modes are permitted.
We find that models of these stars pulsate on timescales which range from less than a day to more than two years in their rest frames,
at a redshift of about $15$, depending on the DM particle mass and
overtone number. 
The pulsation periods are significantly shorter for modes with high
overtone number. 
In general, periods are also significantly shorter for higher WIMP mass:
converting to the observer's frame, we find that the shortest periods are less than about 50 days, in the 1000 GeV case 
for modes with $n>6$.

Work is in progress to study pulsations in more detail, including the novel idea of dark-matter driven pulsations.  
As the
DS undergoes small perturbations, changes in the local baryonic
density could lead to changes in the local dark matter density, in
turn
modulating the local dark matter heating rate. Depending
on the size and relative phasing of these effects, this could be a
source of driving for the pulsations; a much more quantitative
approach is needed to assess the viability of this mechanism.
We will study this and other possibilities
in future work.  
DS pulsations could someday be used to identify bright, cool objects uniquely as dark stars.
If the pulsations are detectable, then dark stars could, in principle, also provide novel standard candles for
cosmological studies.





\acknowledgments
We are grateful for helpful discussions with Douglas Spolyar, Peter Bodenheimer, Paul Shapiro and Matthew Turk.
TRD thanks Josiah Schwab for valuable help with programming issues.
TRD and KF acknowledge the support by the Department of Energy under grant DOE-FG02-95ER40899 and the Michigan Center for
Theoretical Physics at the University of Michigan, Ann Arbor. 
MHM and DEW acknowledge the support of the National Science Foundation
under grants AST-0909107 and AST-1312983, and MHM acknowledges the
support of NASA under grant
NNX12AC96G. This work was supported in part by the National Science Foundation under grant PHYS-1066293 and the
hospitality of the Aspen Center for Physics.
The stellar evolution calculations presented in this paper were made
using the MESA code \citep{Paxton11,Paxton13}. The
adiabatic pulsation calculations were made using the MESA
implementation of the ADIPLS code \citep{Chris08}.





\appendix

\section{Dark star properties for different WIMP masses}

For better readability of the paper, we collect some of the plots of Sec.3-4, 
for WIMP masses different from 100 GeV, in this section.

\begin{figure*} 
\begin{minipage}{0.5\linewidth}
     \centering
     \includegraphics[width=7cm]{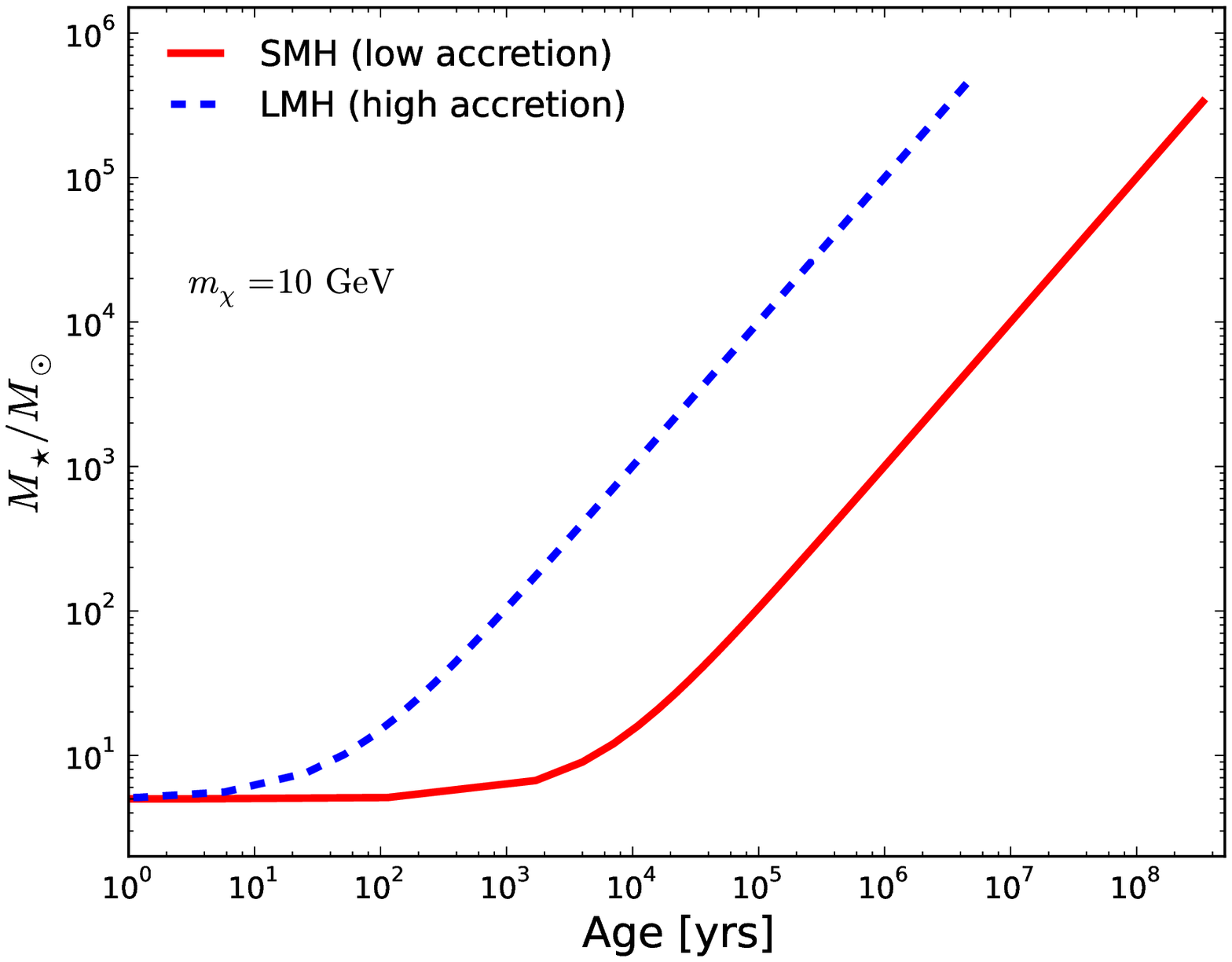}
     \vspace{0.05cm}
    \end{minipage}
    \begin{minipage}{0.5\linewidth}
      \centering\includegraphics[width=7cm]{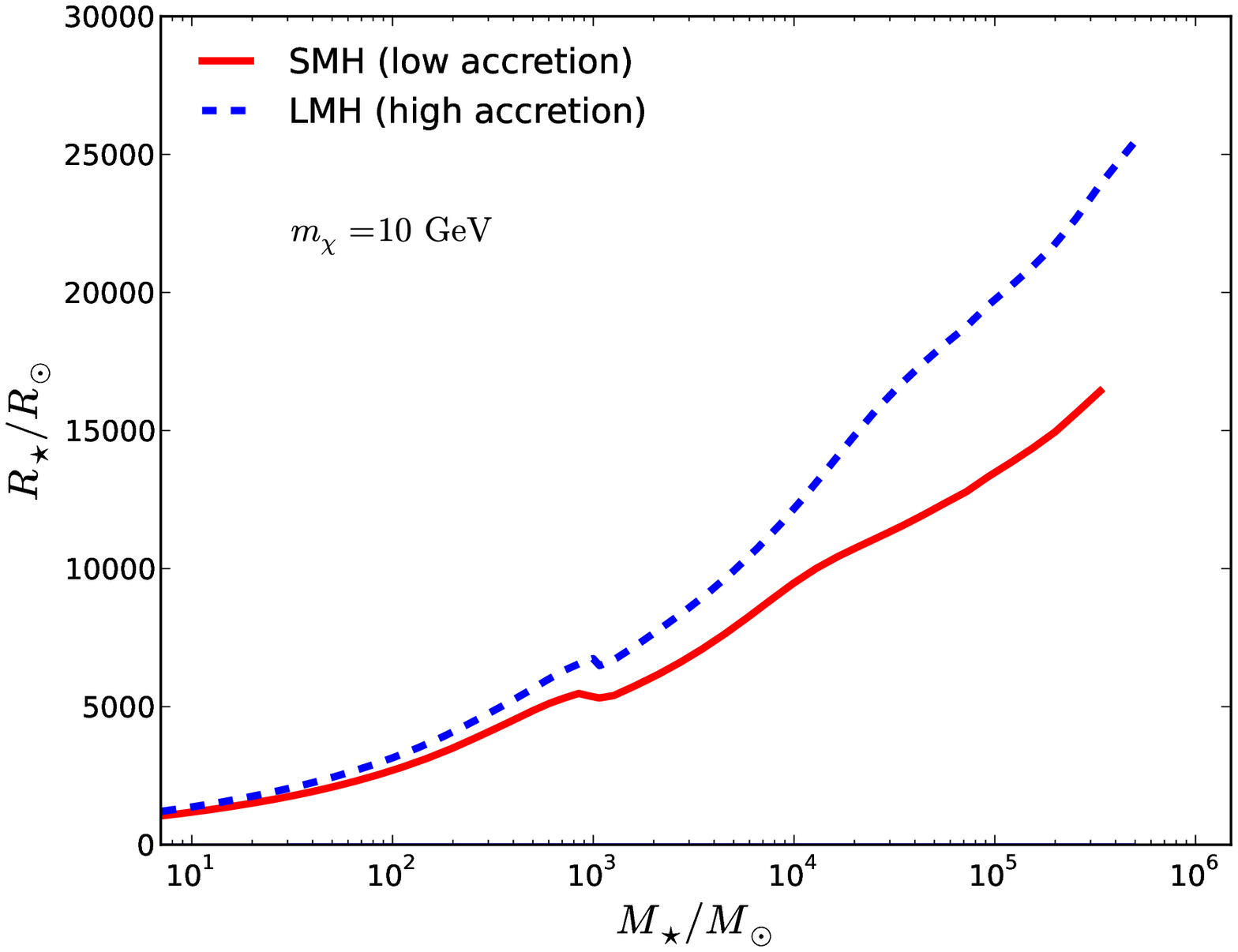}
     \hspace{0.05cm}
    \end{minipage}
 \begin{minipage}{0.5\linewidth}
     \centering
     \includegraphics[width=7cm]{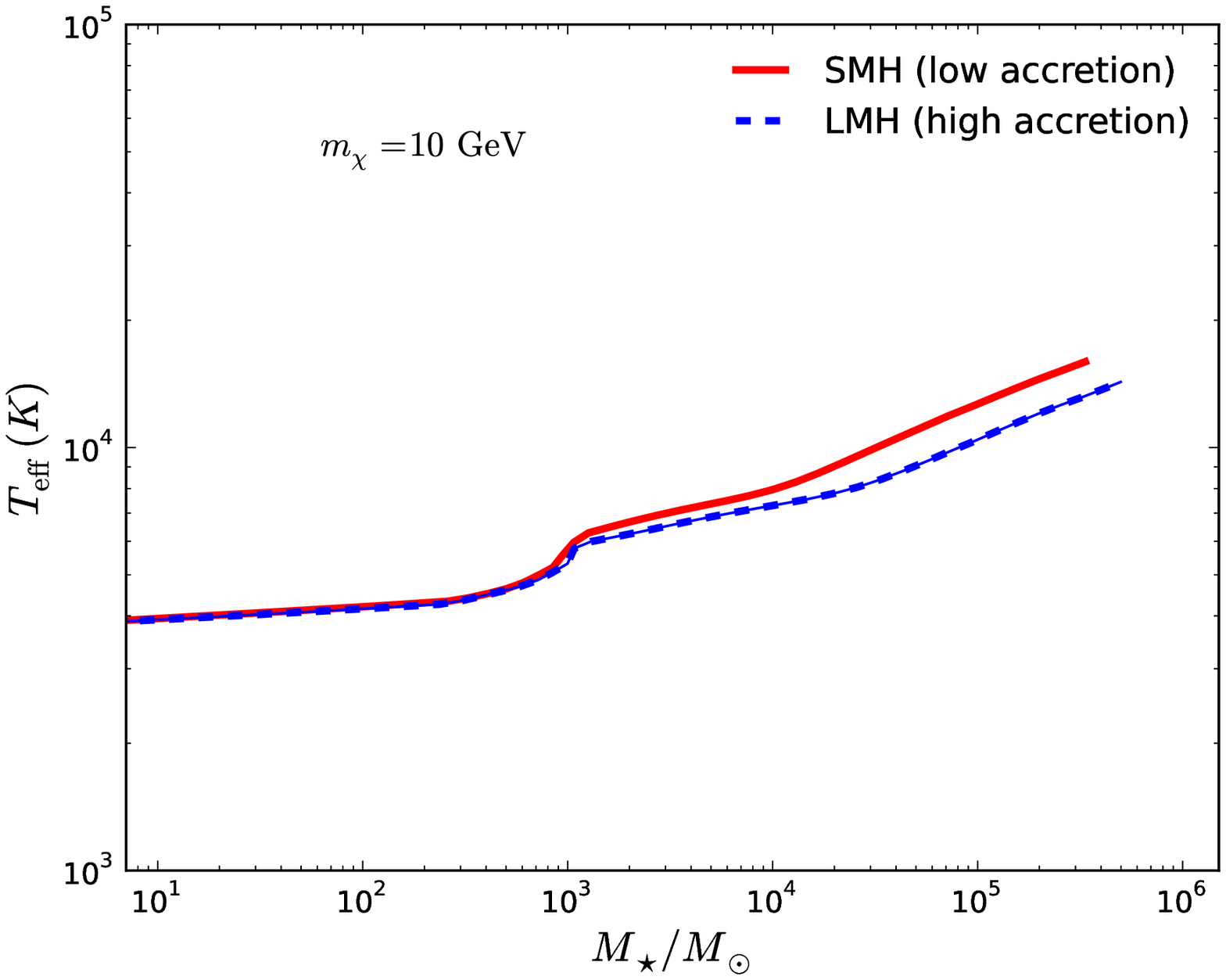}
     \vspace{0.05cm}
    \end{minipage}%
    \begin{minipage}{0.5\linewidth}
      \centering\includegraphics[width=7cm]{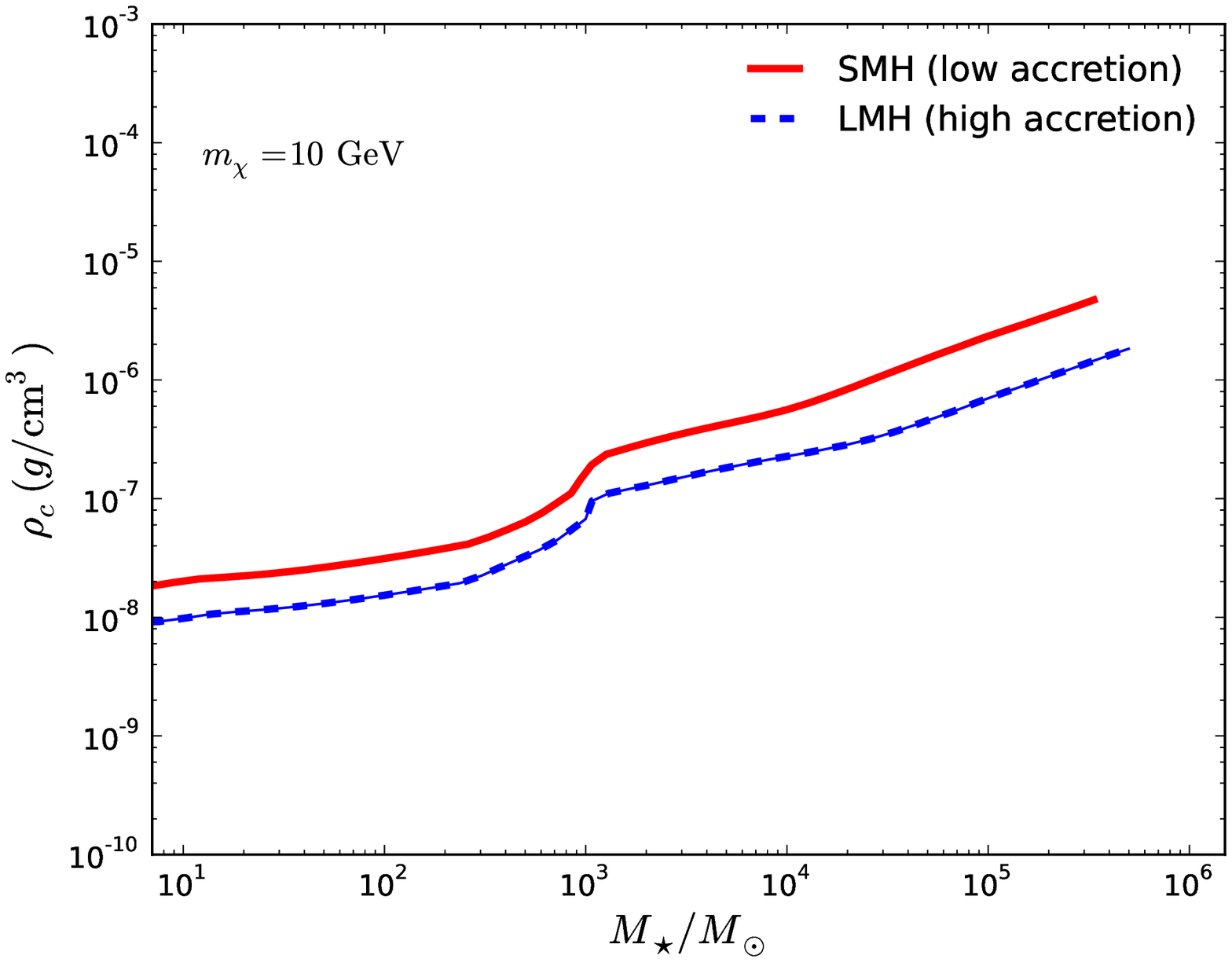}
     \hspace{0.05cm}
    \end{minipage}
 \caption{The same as Figure~\ref{evol}, but for $m_{\chi} = 10$ GeV.}
 \label{evol2}
\end{figure*}

\begin{figure*} 
\begin{minipage}{0.5\linewidth}
     \centering
     \includegraphics[width=7cm]{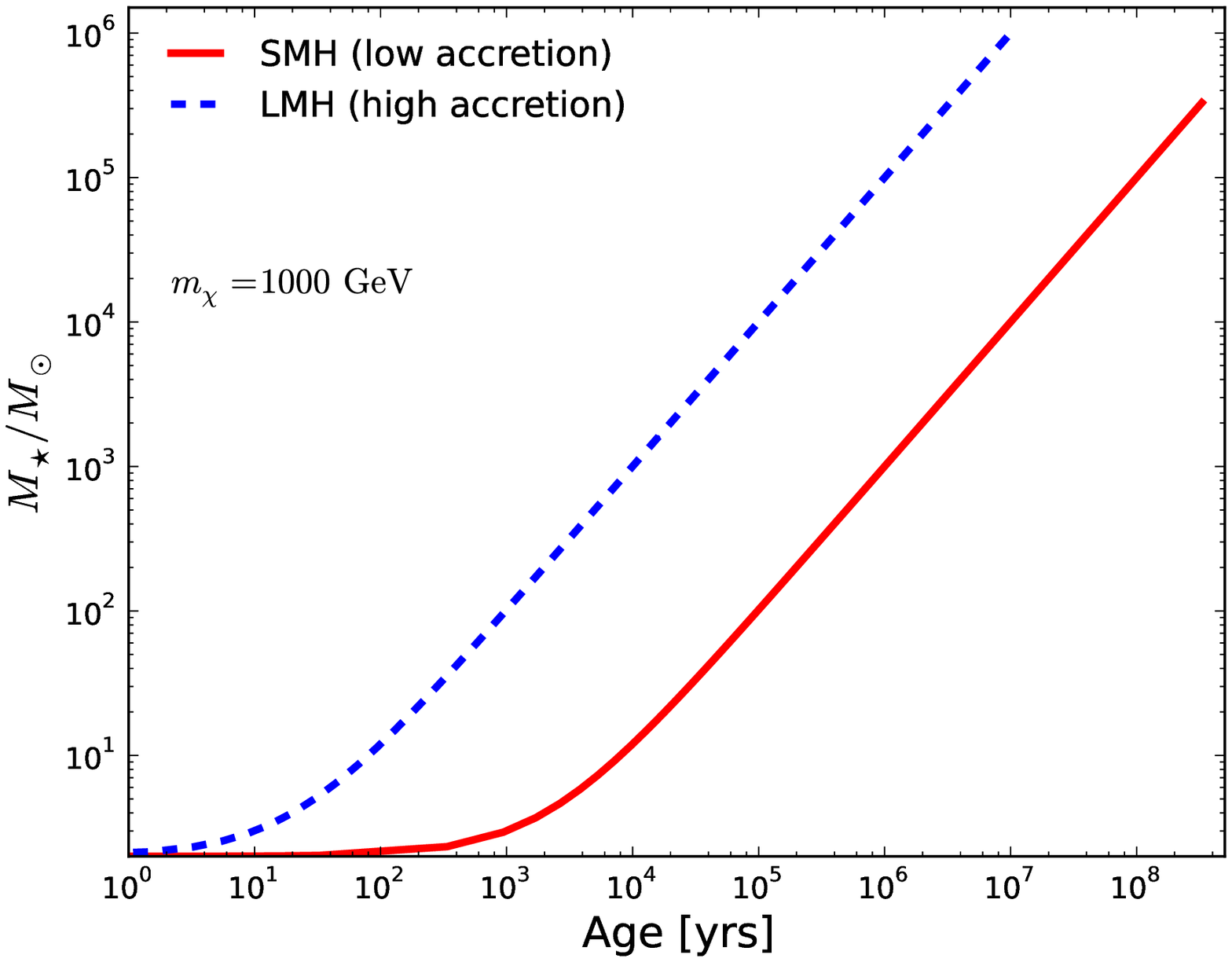}
     \vspace{0.05cm}
    \end{minipage}
    \begin{minipage}{0.5\linewidth}
      \centering\includegraphics[width=7cm]{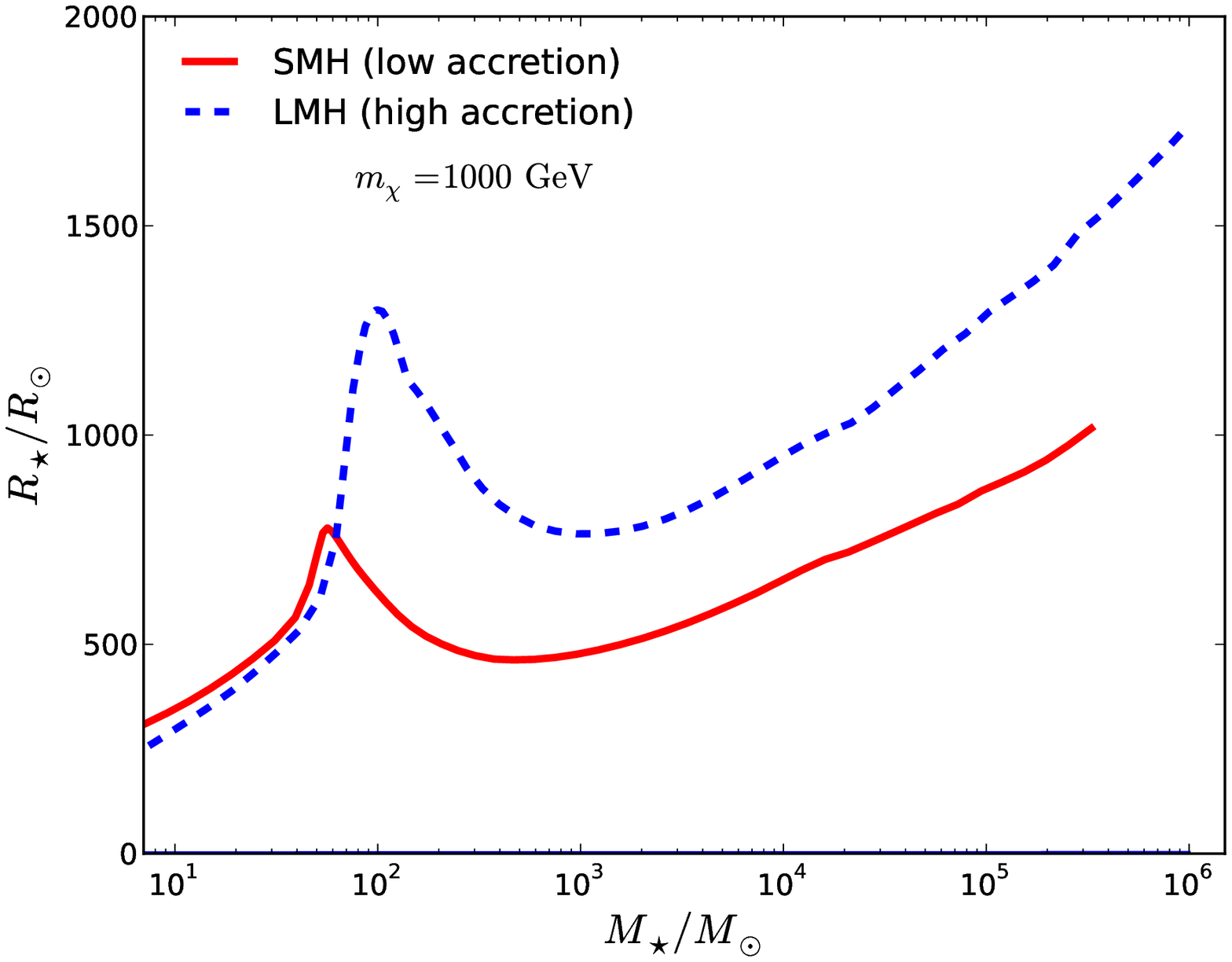}
     \hspace{0.05cm}
    \end{minipage}
 \begin{minipage}{0.5\linewidth}
     \centering
     \includegraphics[width=7cm]{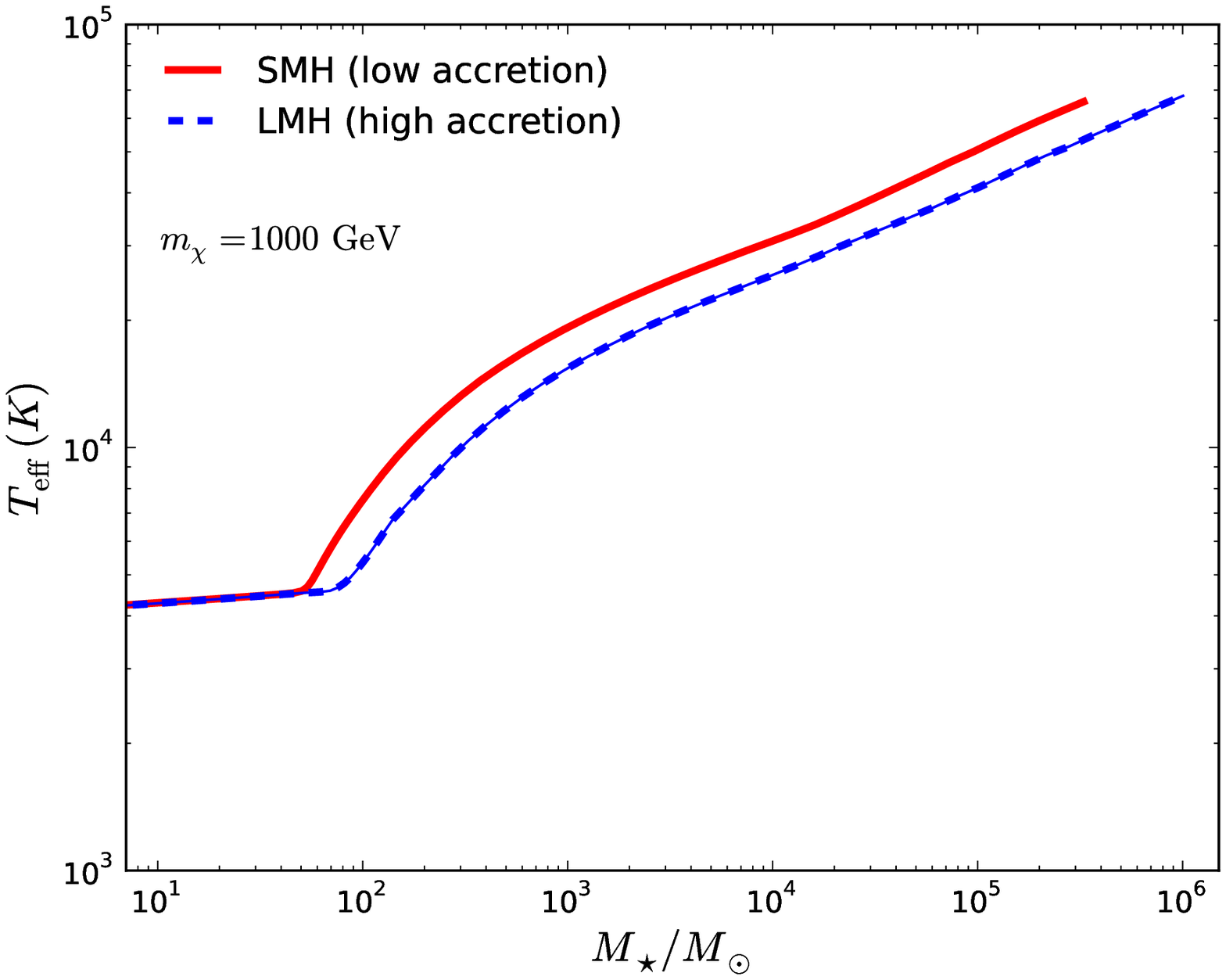}
     \vspace{0.05cm}
    \end{minipage}%
    \begin{minipage}{0.5\linewidth}
      \centering\includegraphics[width=7cm]{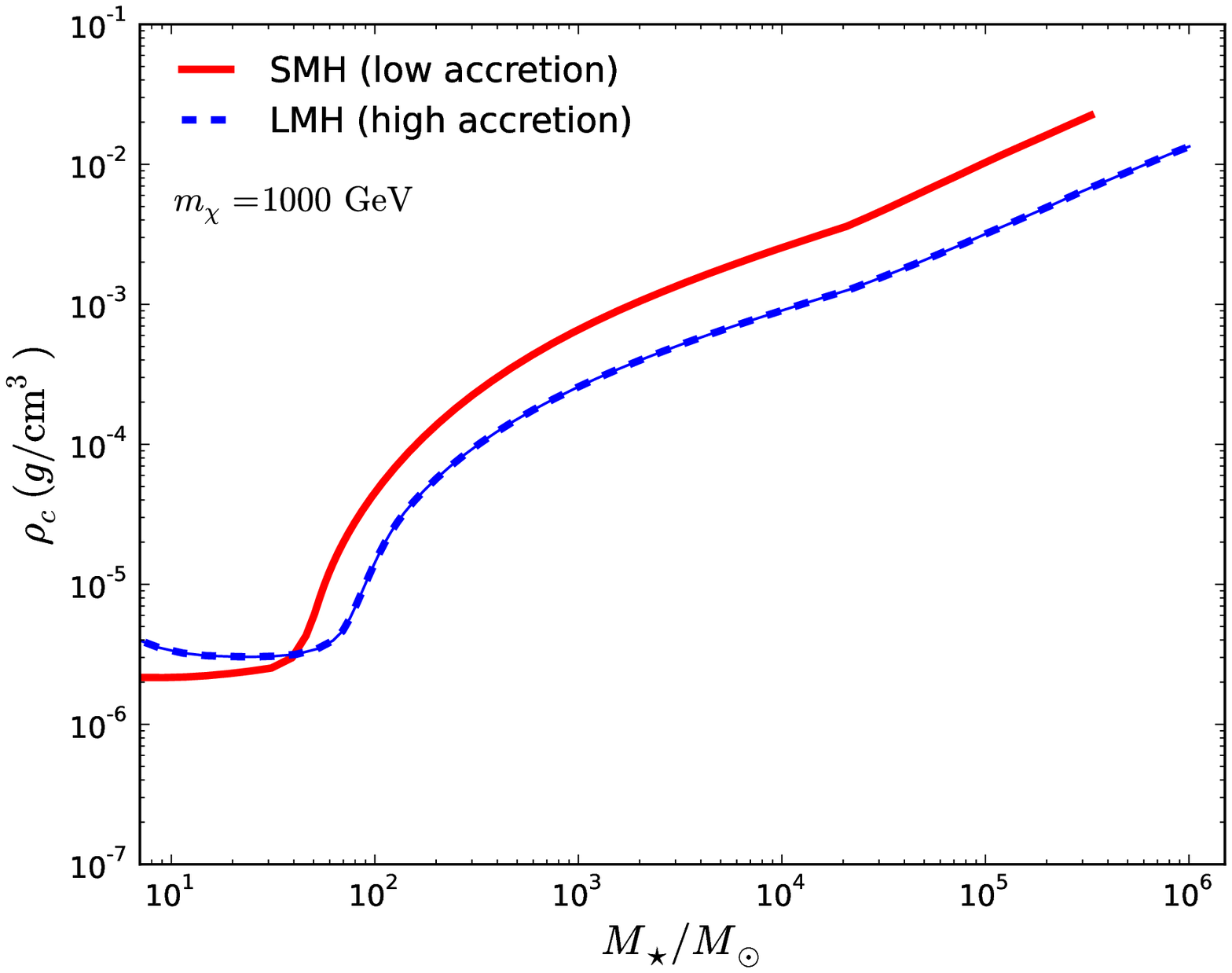}
     \hspace{0.05cm}
    \end{minipage}
 \caption{The same as Figure~\ref{evol}, but for $m_{\chi} = 1000$ GeV. The ``bump'' in the radius at $\sim 100 M_\odot$  
 (top right plot)
 is due to the onset of superadiabaticity in the stellar
 envelope, as explained in the text.}
 \label{evol3}
\end{figure*}

\begin{figure*}[t]
\begin{minipage}{0.5\linewidth}
     \centering
     \includegraphics[width=7.5cm]{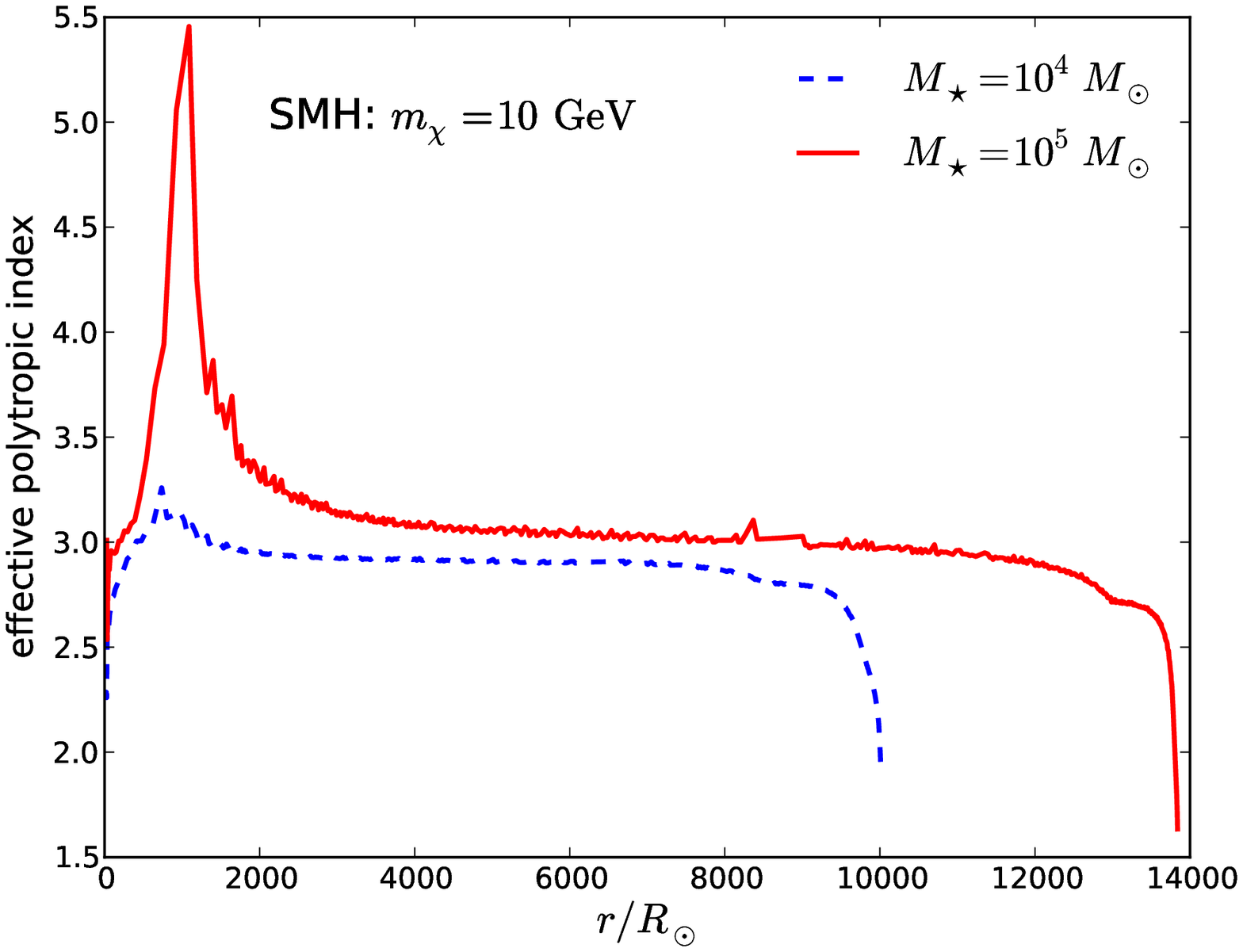}
     \vspace{0.05cm}
    \end{minipage}
    \begin{minipage}{0.5\linewidth}
      \centering\includegraphics[width=7.5cm]{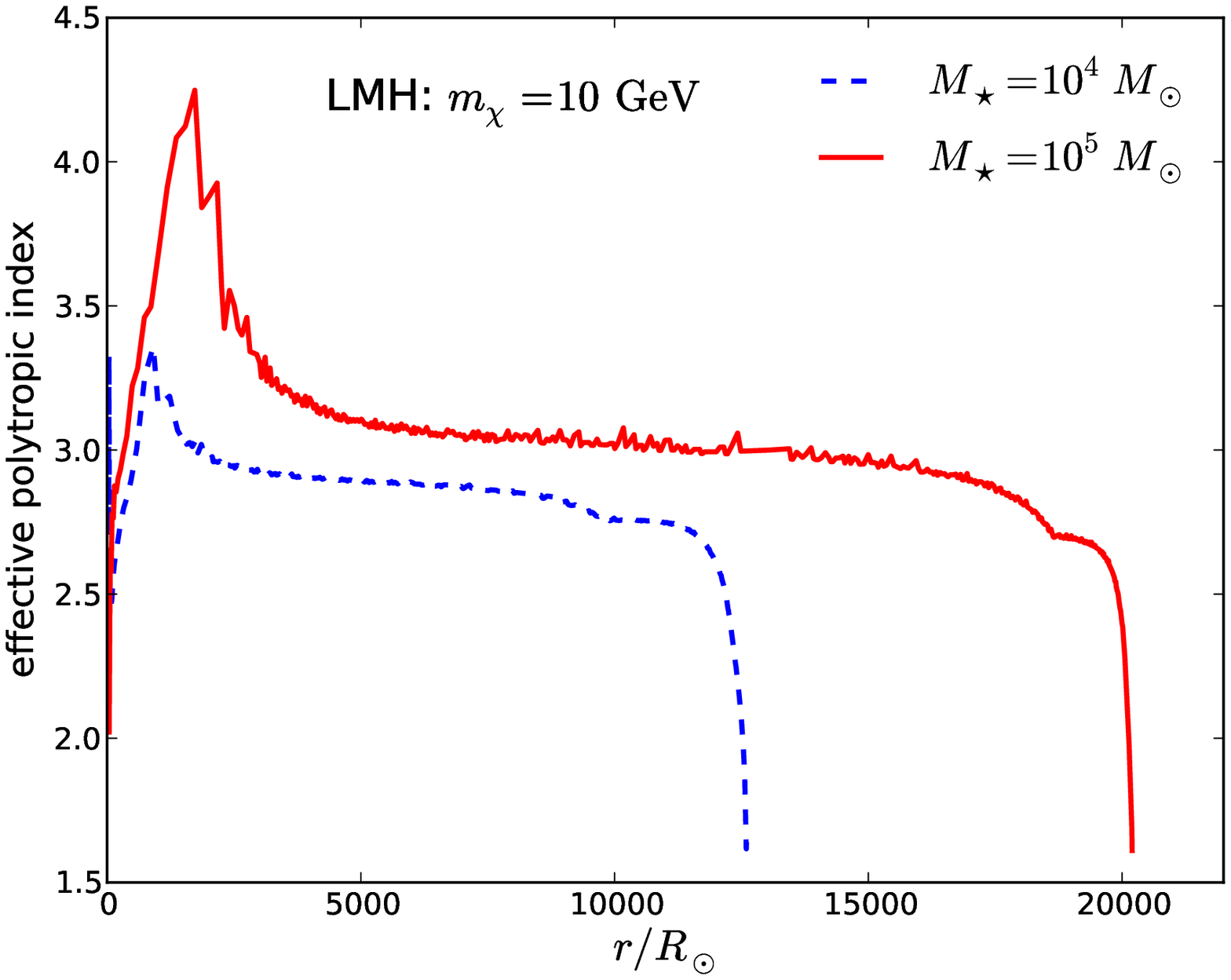}
     \hspace{0.05cm}
    \end{minipage}
\begin{minipage}{0.5\linewidth}
     \centering
     \includegraphics[width=7.5cm]{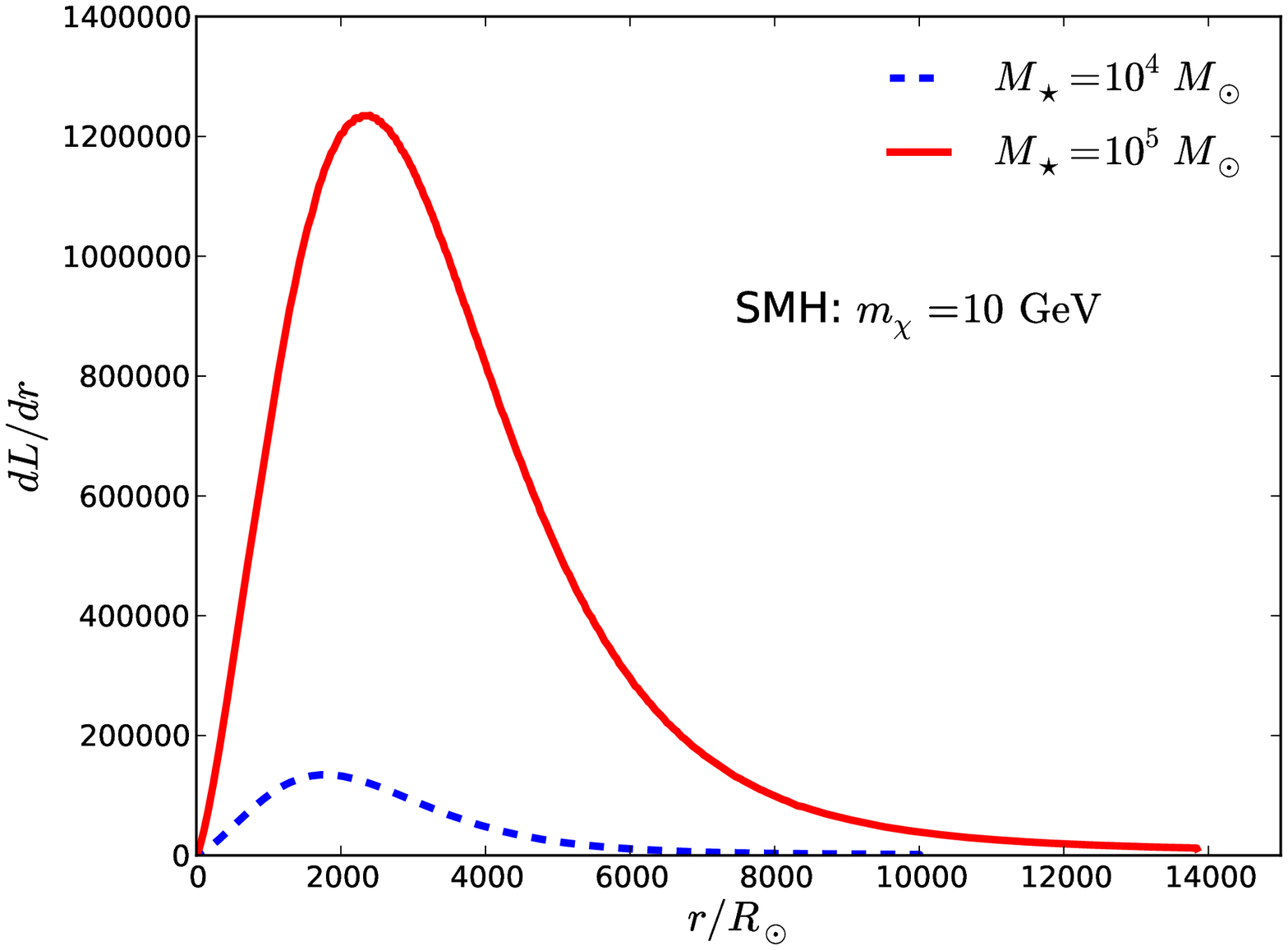}
     \vspace{0.05cm}
    \end{minipage}
    \begin{minipage}{0.5\linewidth}
      \centering\includegraphics[width=7.5cm]{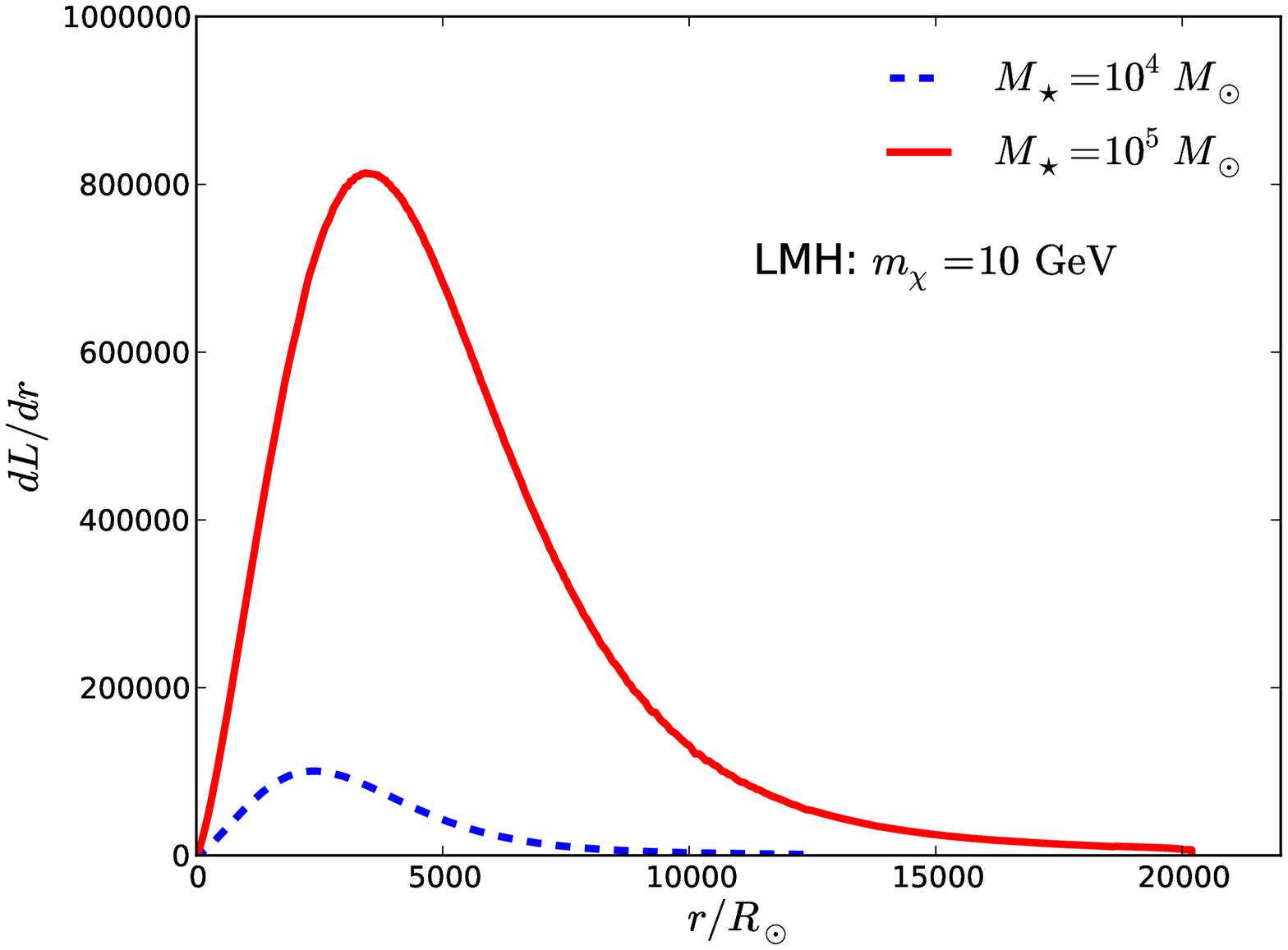}
     \hspace{0.05cm}
    \end{minipage}
    \begin{minipage}{0.5\linewidth}
     \centering
     \includegraphics[width=7.5cm]{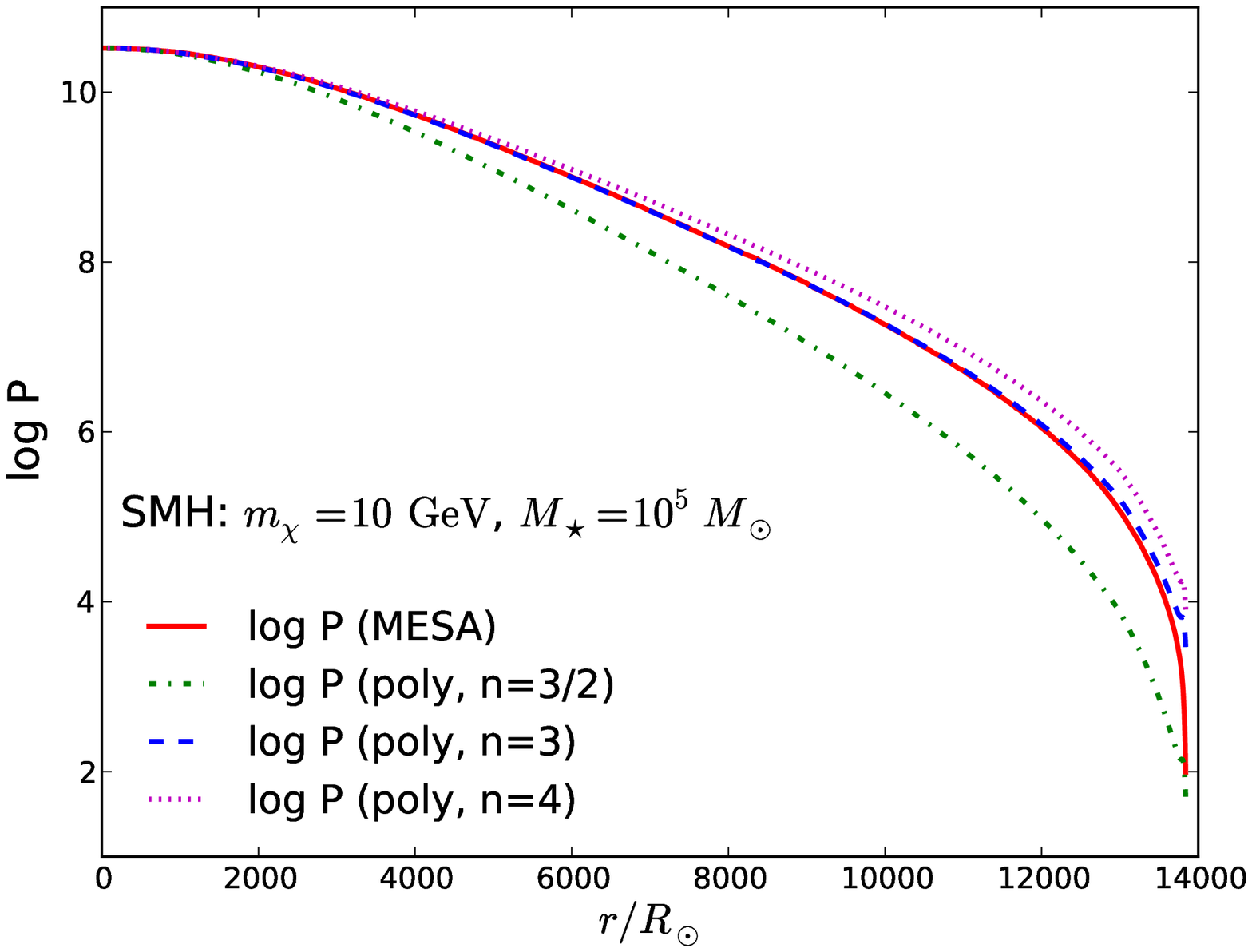}
     \vspace{0.05cm}
    \end{minipage}
    \begin{minipage}{0.5\linewidth}
      \centering\includegraphics[width=7.5cm]{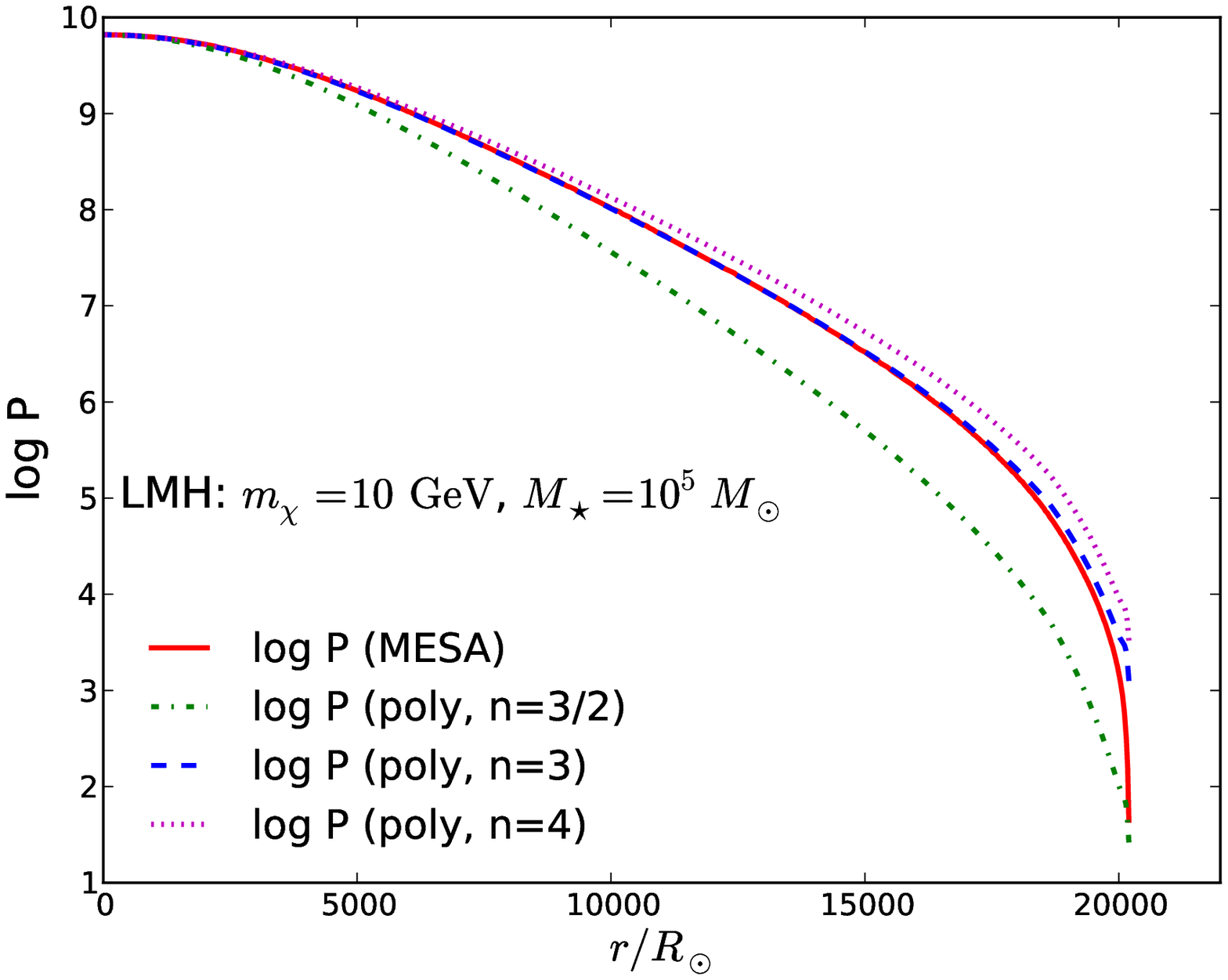}
     \hspace{0.05cm}
    \end{minipage}
 \caption{The same as Figure \ref{polycomp}, but for $m_{\chi} = 10$ GeV.}
 \label{polycomp2}
\end{figure*}

\begin{figure*}[t] 
\begin{minipage}{0.5\linewidth}
     \centering
     \includegraphics[width=7.5cm]{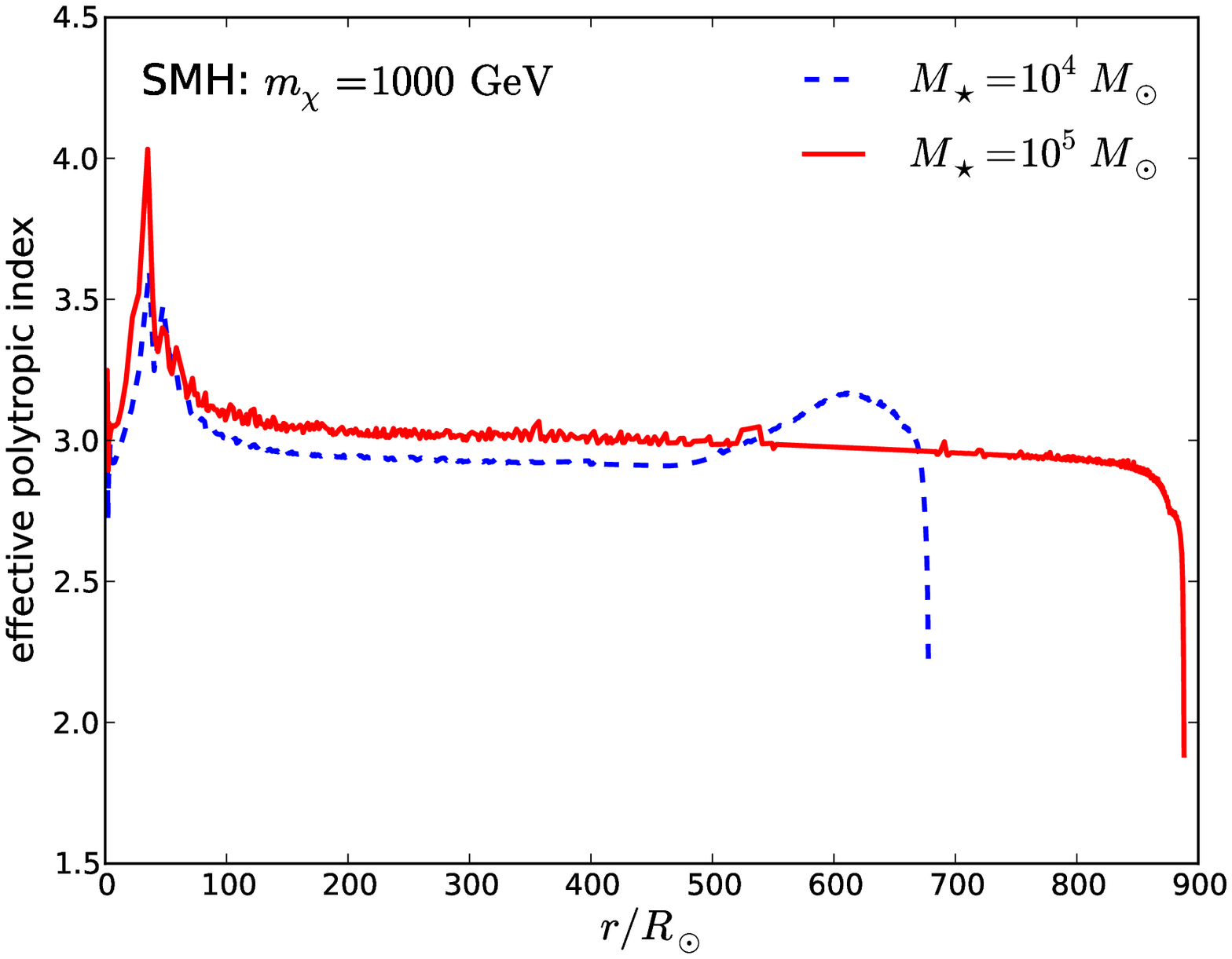}
     \vspace{0.05cm}
    \end{minipage}
    \begin{minipage}{0.5\linewidth}
      \centering\includegraphics[width=7.5cm]{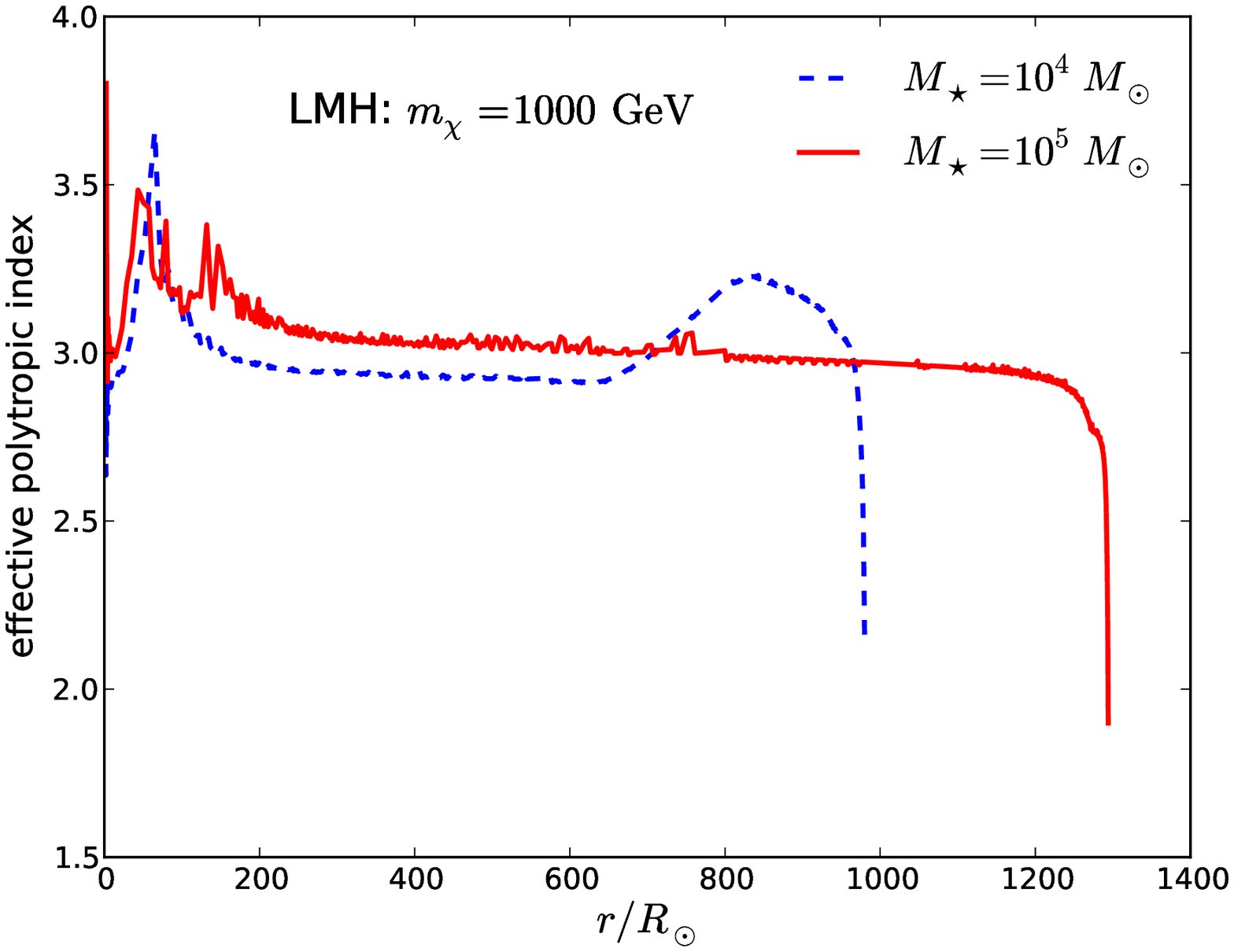}
     \hspace{0.05cm}
    \end{minipage}
\begin{minipage}{0.5\linewidth}
     \centering
     \includegraphics[width=7.5cm]{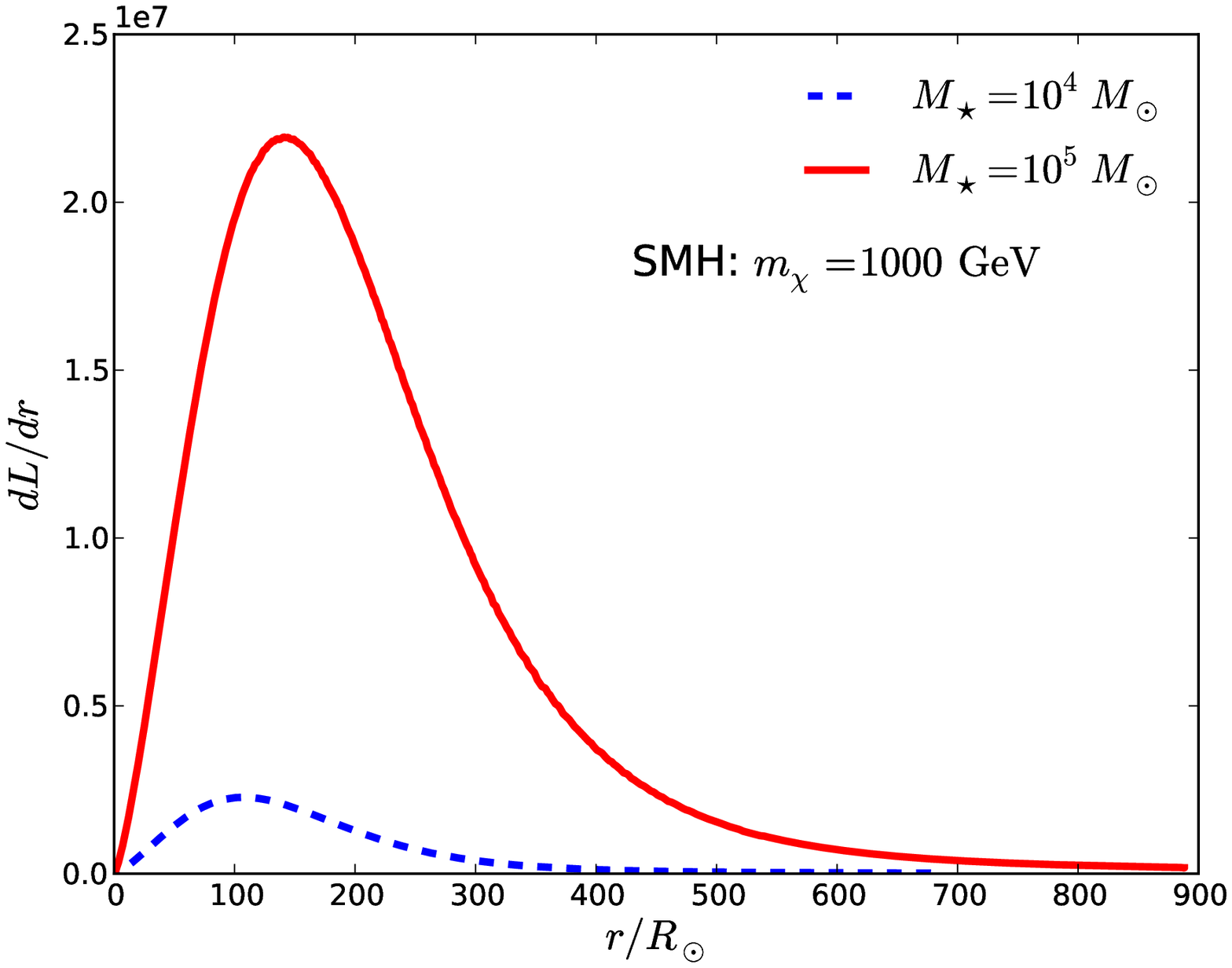}
     \vspace{0.05cm}
    \end{minipage}
    \begin{minipage}{0.5\linewidth}
      \centering\includegraphics[width=7.5cm]{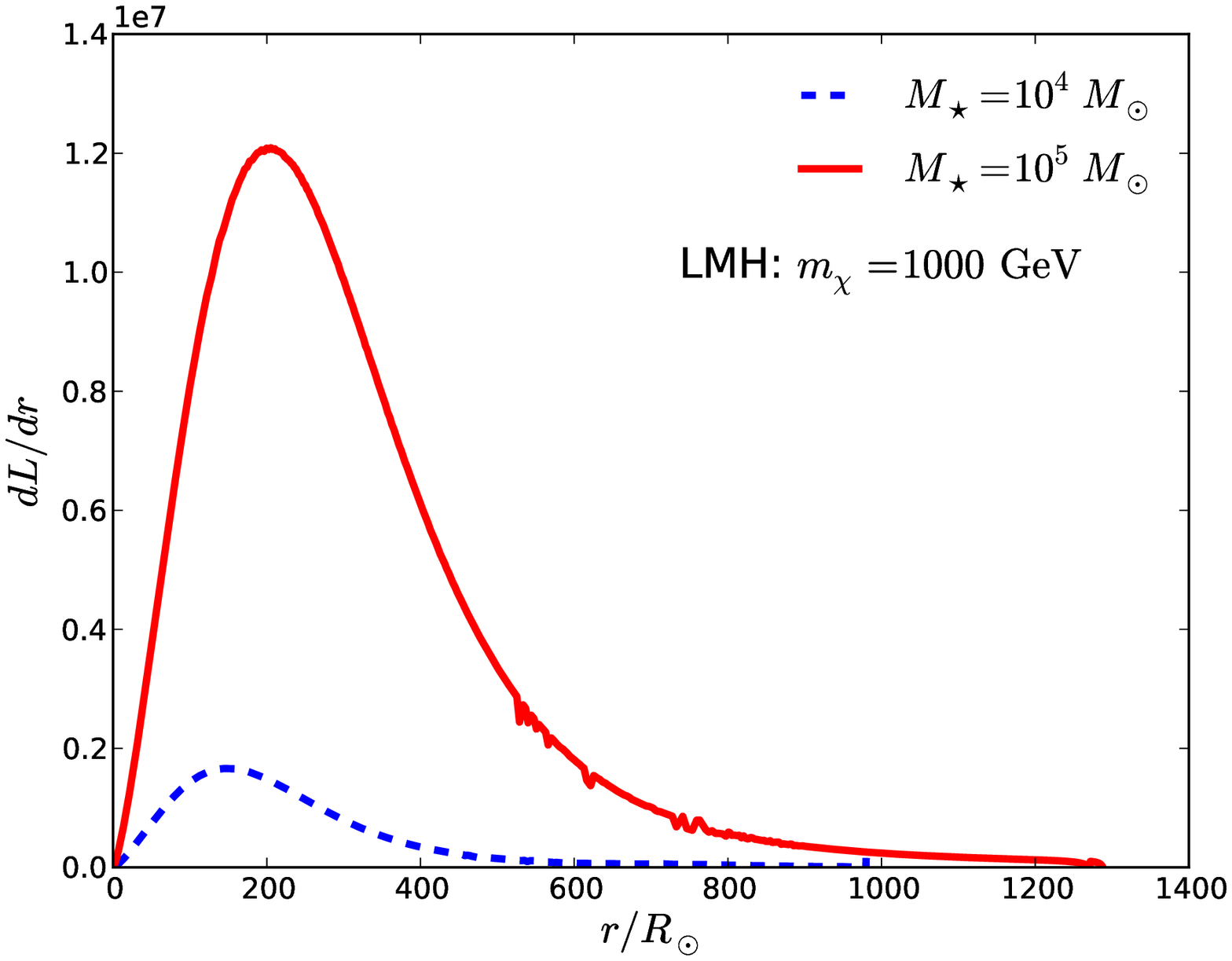}
     \hspace{0.05cm}
    \end{minipage}
    \begin{minipage}{0.5\linewidth}
     \centering
     \includegraphics[width=7.5cm]{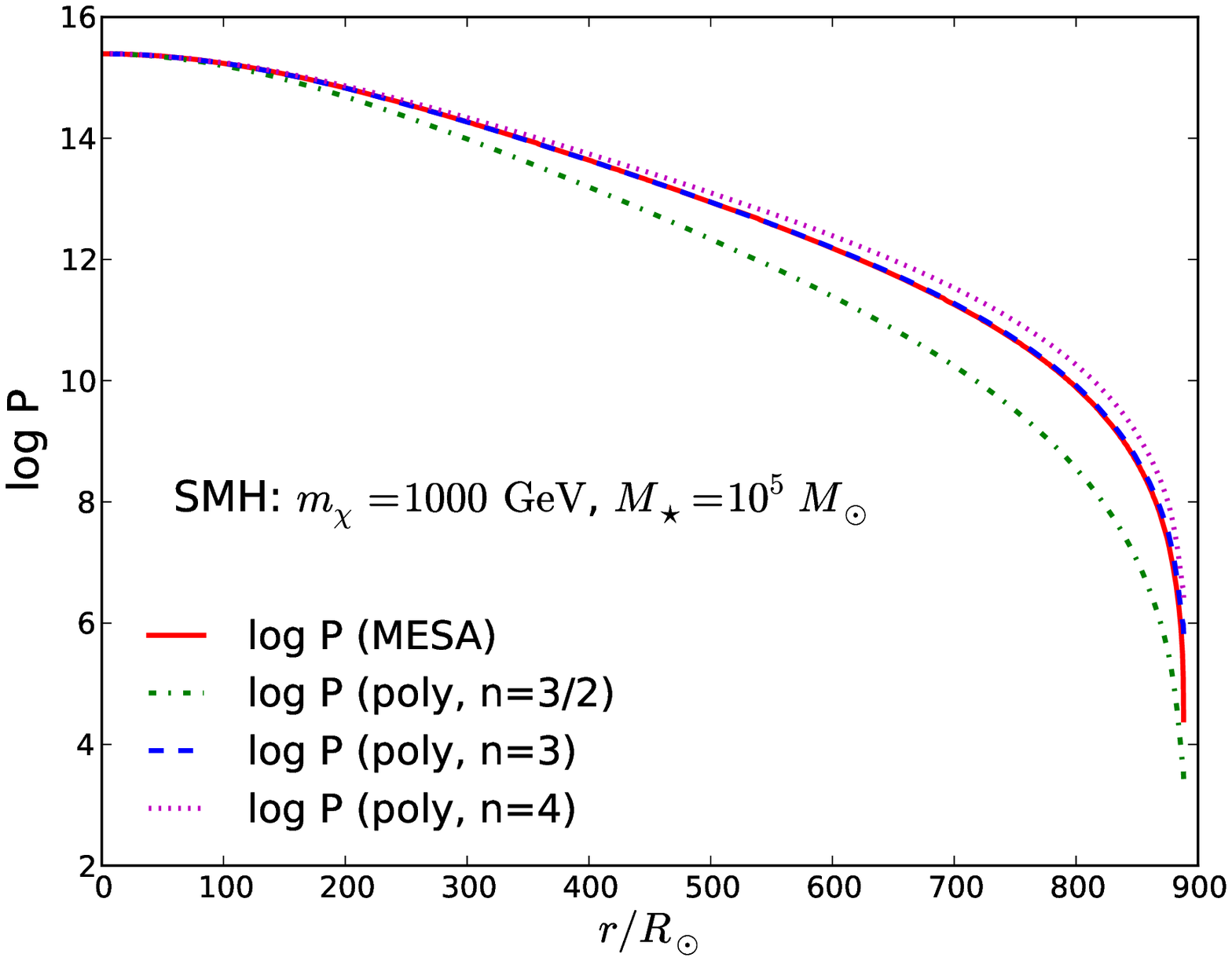}
     \vspace{0.05cm}
    \end{minipage}
    \begin{minipage}{0.5\linewidth}
      \centering\includegraphics[width=7.5cm]{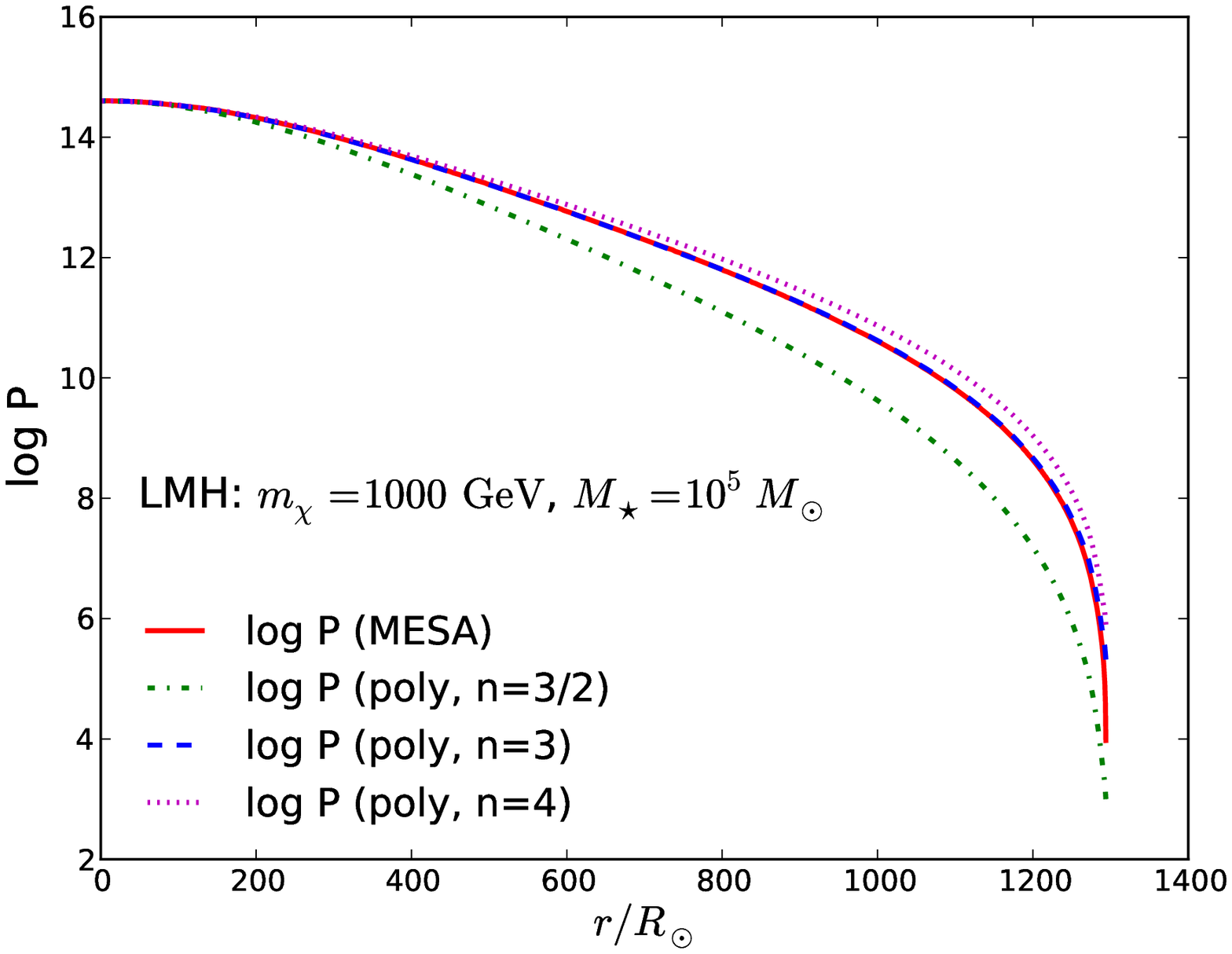}
     \hspace{0.05cm}
    \end{minipage}
 \caption{The same as Figure \ref{polycomp}, but for $m_{\chi} = 1000$ GeV.}
 \label{polycomp3}
\end{figure*}
 
\clearpage
 
\section{Dark matter heating in MESA}

The stellar evolution project MESA \citep{Paxton11,Paxton13} provides
a number of tools that allow one to go beyond the standard picture of
stellar evolution. 
For this study, we have used the \verb+other_energy_implicit+ module
in MESA to include the energy deposited in the model due to DM
annihilation. During a time step, this extra energy is added
self-consistently to the model, in the same way that energy due to
nuclear reactions would be. This code is included in a MESA
\verb+run_star_extras+ file, which itself includes a Fortran program
for calculating the DM heating rate, and the adiabatic contraction. The latter program has been used in 
\cite{Spolyar09} and \cite{Freese10}, and
has been provided to us from the authors for use in MESA. These files,
along with the \verb+inlist+ file used for the stellar evolution
calcuations, can be found on \verb+http://mesastar.org/results+.

We note that while the heating mechanism we investigate is due to the
annihilation of DM particles, we do not assume that this leads to a
depletion of DM.  Rather, we assume that this DM is replenished due to
a continuous infall of DM on centrophilic orbits within the minihalo. Thus, we adopt the same assumption as in
\cite{Freese10}, the reference on which we have based our comparison between MESA's results and polytropes (see Sec.4).
In the code, we accomplish this by not removing (annihilated) DM from the reservoir. 

Work is in progress to include the effects of DM capture via nuclei in the dark star, as well as nuclear fusion. These mechanisms
become important, once the DM 'fuel' from adiabatic contraction and infall runs out. Then, the dark star shrinks and stellar
densities increase to high enough values for these mechanisms to set in, as shown in \cite{Spolyar09}. The results of this
analysis will be presented in a future publication.

\end{document}